\documentclass[12pt]{article}
\usepackage[latin1]{inputenc}

\usepackage{lipsum}
\usepackage{amsmath}
\usepackage{color}
\usepackage{amsfonts}
\usepackage[mathscr]{euscript}
\usepackage{graphicx}
\usepackage{geometry}
\usepackage{amssymb,epsfig}
\usepackage{subfigure}
\usepackage{comment}
\usepackage{tabularx}
\usepackage{bm}
\usepackage[dvipsnames]{xcolor}
\usepackage{exscale}
\usepackage{amsbsy}
\usepackage{textcomp}
\usepackage{hyperref}
\usepackage{slashed}
\usepackage{authblk}
\usepackage{tabularx}
\usepackage{color}
\usepackage{exscale}
\usepackage{doi}
\usepackage{tensor}
\usepackage{wrapfig}
\usepackage[font=footnotesize,labelfont=bf]{caption}
\usepackage{ulem} 
\usepackage{makecell}

\def\ba#1\ea{\begin{align}#1\end{align}}
\def\bg#1\eg{\begin{gather}#1\end{gather}}
\def\bm#1\em{\begin{multline}#1\end{multline}}
\def\bmd#1\emd{\begin{multlined}#1\end{multlined}}
\newcommand{\s}[1]{\mathcal{#1}} 
\newcommand{\bb}[1]{\mathbb{#1}} 

\newcommand{\be}{\begin{equation}}
	\newcommand{\ee}{\end{equation}}
\newcommand{\bea}{\begin{eqnarray}}
	\newcommand{\eea}{\end{eqnarray}}

\newcommand{\bs}{\boldsymbol}
\newcommand{\matleft}{\left(\begin{array}}
	\newcommand{\matright}{\end{array}\right)}

\usepackage[numbers,sort&compress]{natbib}
\def\simge{
	\mathrel{\rlap{\raise 0.511ex 
			\hbox{$>$}}{\lower 0.511ex \hbox{$\sim$}}}}

\def\simle{
	\mathrel{\rlap{\raise 0.511ex 
			\hbox{$<$}}{\lower 0.511ex \hbox{$\sim$}}}}

\makeatletter
\renewcommand\section{\@startsection {section}{1}{\z@}%
	{-3.5ex \@plus -1ex \@minus -.2ex}
	{2.3ex \@plus.2ex}%
	{\normalfont\large\bfseries}}
\renewcommand\subsection{\@startsection{subsection}{2}{\z@}%
	{-3.25ex\@plus -1ex \@minus -.2ex}%
	{1.5ex \@plus .2ex}%
	{\normalfont\bfseries}}
\renewcommand\subsubsection{\@startsection{subsubsection}{3}{\z@}%
	{-3.25ex\@plus -1ex \@minus -.2ex}%
	{1.5ex \@plus .2ex}%
	{\normalfont\itshape}}
\makeatother

\def\pplogo{\vbox{\kern-\headheight\kern -29pt
		\halign{##&##\hfil\cr&{\ppnumber}\cr\rule{0pt}{2.5ex}&\ppdate\cr}}}
\makeatletter
\def\ps@firstpage{\ps@empty \def\@oddhead{\hss\pplogo}%
	\let\@evenhead\@oddhead 
}
\hypersetup{
	unicode=false,          
	pdftoolbar=true,        
	pdfmenubar=true,        
	pdffitwindow=false,     
	pdfstartview={FitH},    
	pdftitle={Hall viscosity of rotating BECs},    
	pdfauthor={},     
	pdfsubject={Subject},   
	pdfcreator={},   
	pdfproducer={}, 
	pdfkeywords={keyword1} {key2} {key3}, 
	pdfnewwindow=true,      
	colorlinks=true,       
	linkcolor=OliveGreen, 
	citecolor=NavyBlue,        
	filecolor=magneta,      
	urlcolor=cyan           
}

\numberwithin{equation}{section}

\newcommand*\samethanks[1][\value{footnote}]{\footnotemark}

\newcommand\beal{\begin{equation}\begin{aligned}}
		\newcommand\eeal{\end{aligned}\end{equation}}

\begin{document}

\normalem

\setcounter{page}0
\def\ppnumber{\vbox{\baselineskip14pt
}}

\def\ppdate{
} 
\date{}

\title{\Large\bf Observable signatures of Hall viscosity in lowest Landau level superfluids}
\author{Seth Musser$^1$, Hart Goldman$^{1,2}$, and T. Senthil$^1$}
\affil{\it\small $^1$ Department of Physics, Massachusetts Institute of Technology, Cambridge, MA 02139, USA}
\affil{\it\small $^2$ Kadanoff Center for Theoretical Physics, University of Chicago, Chicago, IL 60637, USA}
\maketitle\thispagestyle{firstpage}
\begin{abstract}

Hall viscosity is a nondissipative viscosity occurring in systems with broken time-reversal symmetry, such as quantum Hall phases and $p+ip$ superfluids. Despite Hall viscosity's expected ubiquity and past observations in: classical soft matter, optical, and graphene systems, it has yet to be measured experimentally in any macroscopic quantum state of matter. Toward this end, we describe the observable effects of Hall viscosity in a simple family of rotating Bose-Einstein condensates of electrically neutral bosons, in which 
all of the bosons condense into a single lowest Landau level (LLL) orbital. Such phases are accessible to current cold atom experiments, and we dub them LLL superfluids. We demonstrate that LLL superfluids possess a nonuniversal Hall viscosity, leading to a range of observable consequences such as rotation of vortex-antivortex dipoles and wave-vector dependent corrections to the speed of sound. Furthermore, using a coherent state path integral approach, we present a microscopic derivation of the Landau-Ginzburg equations of a LLL superfluid, showing explicitly how Hall viscosity enters.

\end{abstract}

\pagebreak
{
\hypersetup{linkcolor=black}
\tableofcontents
}
\pagebreak

\section{Introduction}

The Hall viscosity, $\eta_H$, is a nondissipative contribution to the viscosity present in any system with broken time-reversal symmetry~\cite{avron_viscosity_1995,avron_odd_1998}. 
In rotationally invariant, incompressible quantum Hall fluids, $\eta_H/\overline{n}$, with $\overline{n}$ the mean number density of particles, is a universal quantum number~\cite{Read2009,Read2011}. More broadly, Hall viscosities can appear in a range of other phases, including superfluids~\cite{Read2009,Read2011,hoyos_effective_2014,Moroz2018,Moroz2019,Nie2020,Rose2020,Furusawa2021}, graphene systems~\cite{berdyugin_measuring_2019}, composite Fermi liquids~\cite{You2014,Son2015,Levin2017,Goldman2018,Pu2020}, and active matter settings~\cite{banerjee_odd_2017, soni_odd_2019, souslov_topological_2019,hargus_time_2020, yamauchi_chirality-driven_2020, han_fluctuating_2021, hosaka_hydrodynamic_2021, banerjee_hydrodynamic_2022, khain_stokes_2022, reynolds_hele-shaw_2022,  de_wit_pattern_2024, fruchart_odd_2023}. 
In both theory and experiment, the existence of a Hall viscosity can have broad consequences that continue to be sought after. Most famously, it  leads to a finite wave-vector correction to the Hall conductivity~\cite{bradlyn_kubo_2012,hoyos_hall_2012} that may be experimentally measurable in realistic systems~\cite{delacretaz_transport_2017}.  Yet remarkably, Hall viscosity has not been experimentally observed in macroscopic quantum states, though it has been measured in the realm of classical soft matter~\cite{soni_odd_2019}, optical systems~\cite{schine_synthetic_2016}, and graphene~\cite{berdyugin_measuring_2019}. 

One natural family of systems exhibiting Hall viscosity are rotating Bose-Einstein condensates (BECs) composed of electrically neutral atoms. Rotating BECs have long been proposed as settings for bosonic fractional quantum Hall (FQH) physics~\cite{Cooper2020}, with the system's rotation playing the same role as the perpendicular magnetic field in conventional electronic quantum Hall systems. While incompressible bosonic FQH phases have yet to be achieved experimentally, recent  advances have allowed for the preparation of rotating BECs where the constituent bosons reside in a single lowest Landau level (LLL) orbital~\cite{fletcher_geometric_2021,mukherjee_crystallization_2022}, which we dub  LLL superfluids. Unlike bosonic FQH phases, LLL superfluids are compressible states with gapless Goldstone excitations~\cite{sinha_two-dimensional_2005,mukherjee_crystallization_2022}. In contrast to the well studied Hall viscosity of $p+ip$ superfluids, where time-reversal is broken spontaneously and $\eta_H/\overline{n}$ takes quantized values, LLL superfluids have explicitly broken time-reversal symmetry and display a universal Hall viscosity without quantization. 

\begin{figure}
    \centering
    \includegraphics[width=0.7\textwidth]{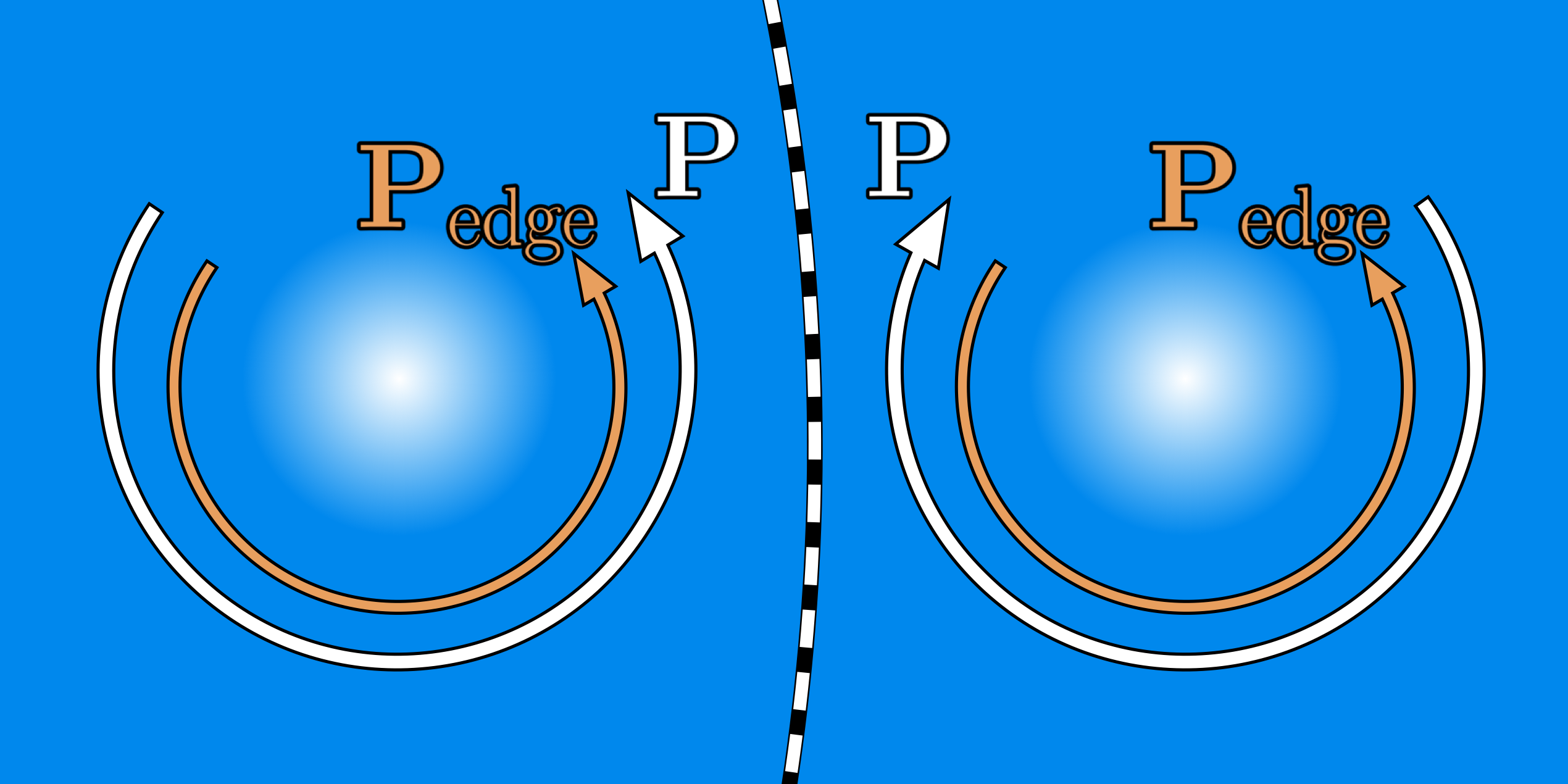}
    \caption{A cartoon of the experimental protocol. Two vortices with opposite signs, i.e. opposite winding of the phase $\phi$ are pictured. The flow due to the winding of the phase is shown as $\vec{P} = -n\vec{\nabla}\phi$. However, each vortex is accompanied by a density depletion shown as the dark region in the cartoon. This leads to an identical contribution to $\vec{P}_{\mathrm{edge}} = -\eta_H \hat{z} \times \vec{\nabla} \log(n)$ for each vortex. The edge current will enhance the force of the right vortex on the left, as shown in the arrows emanating from the left vortex, while decreasing it on the right. Rotation of the dipole will result. This is indicated via the dotted line, which shows the trajectory of the center of mass of the vortex dipole.} 
    \label{fig:vortex_dipole}
\end{figure}

In this work, we describe how Hall viscosity manifests in LLL superfluids and propose realistic protocols for its observation. 
The primary consequence of Hall viscosity we will exploit is that, unlike superfluids in time-reversal invariant systems, the momentum density of a LLL superfluid has two contributions,
\begin{align}
\label{eq: edge momentum}
\s{P}^i=\s{P}^i_{\mathrm{phase}}+\s{P}^i_{\mathrm{edge}}=-n\,(\partial^i \theta - \s{A}^i)  +\eta_H\,\varepsilon^{ij}\partial_j\log n\,.
\end{align}
The first term is the usual momentum density of a superfluid with local number density, $n(x,t)$, and phase variable, $\theta(x,t)$. The second term appears whenever there is a nonvanishing Hall viscosity~\cite{wiegmann_anomalous_2013, wiegmann_anomalous_2014,abanov_hydrodynamics_2020,markovich_odd_2021,machado_monteiro_hamiltonian_2023}. 
Even though it is a total derivative, this term has observable consequences. In particular, it affects the \emph{total vorticity} of the fluid, i.e. the vorticity arising from the momentum current,
\begin{align}
\Omega_{\mathrm{total}}&= \nabla\times \frac{\s{P}}{\rho_m}=\Omega_{\mathrm{phase}} + \omega_c -\nu_H\,\nabla^2\log n\, \label{eqn:total_vorticity}
\end{align}
where $\rho_m = m_b n$ is the mass density of the superfluid with constituent bosons of mass $m_b$, $\nu_H = \eta_H/\rho_m$ is the so-called kinematic Hall viscosity, $\omega_c$ is the constant cyclotron frequency, and $\Omega_{\mathrm{phase}}$ is the vorticity associated with the superfluid phase winding. The presence of this new term enhances (suppresses) the flow around negative (positive) pointlike vortices, which may be readily imaged experimentally~\cite{mukherjee_crystallization_2022}. For example, if one were to prepare a vortex dipole, then the existence of a nonzero Hall viscosity will cause the dipole to rotate as it propagates through the sample (see Fig.~\ref{fig:vortex_dipole}). 

We support our proposals with a derivation of the Landau-Ginzburg equations for the symmetric gauge LLL superfluid starting from the microscopic Hamiltonian, which in turn leads to Eq.~\eqref{eq: edge momentum}. By constructing the coherent state path integral for this case, we demonstrate that the condensation of bosons to a LLL orbital with single value of the angular momentum leads to an effective hydrodynamic theory at low energies with a nonvanishing Hall viscosity, along the way showing how this theory couples to the spatial curvature of the system. To our knowledge, this simple context is the first example where a derivation of Hall viscosity connecting the microscopic many-body Hamiltonian to the low energy effective field theory has been developed.

We proceed as follows. In Section~\ref{sec: 2} we define LLL superfluids and discuss their essential hydrodynamic properties. 
In Section~\ref{sec: 3}, we present the 
observable consequences of Hall viscosity in LLL superfluids and 
present numerical predictions for the vortex dynamics. 
Finally, in Section~\ref{sec: 4}, we present a microscopic derivation of LLL superfluid hydrodynamics starting from a coherent state path integral. We conclude in Section~\ref{sec: discussion}. 

\section{Hall viscosity in LLL superfluids}
\label{sec: 2}


\subsection{LLL superfluids: Definition and basic properties}
\label{sec:LLL_superfluids}

Consider a two dimensional system of $N_b$ interacting bosonic particles of mass $m_b$ rotating with angular speed $\omega_{\mathrm{trap}}$ in a harmonic trap. Due to the rotation of the trap, time-reversal symmetry is broken, and the bosons behave as if they are experiencing a perpendicular magnetic field, $B_{\mathrm{eff}}\equiv\nabla\times\mathcal{A}=2m_b\,\omega_{\mathrm{trap}}$, arising from the Coriolis effect.
As a result, the bosons form highly degenerate Landau levels with gap set by the effective cyclotron frequency, $\omega_c=2\omega_{\mathrm{trap}}$. The full  Hamiltonian can be written as 
\begin{equation}
H = \,\sum_{\alpha=1}^{N_b}\left[ \frac{1}{2m_b}\left(p^{(\alpha)}_i-\mathcal{A}_i\right)^2+V(r_\alpha)\right]+H_{\mathrm{int}} \label{eqn:BEC_Ham}\,.
\end{equation}
The first term is a sum of single-particle Hamiltonians, with the potential, $V(r_\alpha)$, determining the geometry of the system. The second term defines the boson-boson interactions, which we will take to be repulsive and be characterized by an energy scale, $E_{\mathrm{int}}$.

Recent cold atom experiments \cite{fletcher_geometric_2021,mukherjee_crystallization_2022} have demonstrated that such systems can be prepared in the LLL limit, $\omega_{\mathrm{trap}}>>E_{\mathrm{int}}$, where the kinetic energy is quenched, and the interactions are only capable of scattering within the LLL. The densities of these systems are comparable to the effective magnetic field i.e., they are engineered at fillings where the density of particles is comparable to that of the superfluid vortices\footnote{At large enough fillings, $\nu\gtrsim 8$, the vortices are believed to form an Abrikosov lattice. The properties of rotating superfluids in this regime have been extensively studied in a number of recent works, see Ref.~\cite{fetter_rotating_2009} for a review.}, $\nu=\overline{n}/B_{\mathrm{eff}}\sim 1$. If the interactions are sufficiently strong, then the bosons can then form incompressible bosonic FQH fluids. However, another possibility in weakly interacting systems -- which are currently more realistic in cold atom experiments -- is for each of the bosons to condense into a single LLL orbital. Indeed, by making a suitable choice of harmonic potential, $V(r_\alpha)$, it is possible to ``squeeze'' all (or almost all) of the bosons into the same LLL state, $\psi_{\mathrm{LLL}}$, with wave function\footnote{The role of interactions can be thought of as causing tunneling between ``nearby'' LLL orbitals. See Appendix \ref{app: Hall visc calculations} for treatment of an example where interactions affect long wavelength properties. For the particular harmonic potentials we focus on in the main text, we will argue that corrections to Eq.~\eqref{eq: LLL SF state} are small enough that they can be neglected.}
\begin{align}
\label{eq: LLL SF state}
\Psi_{\mathrm{SF}}(r_1,\dots,r_{N_b})=\prod_{\alpha=1}^{N_b}\psi_{\mathrm{LLL}}(r_\alpha)\,.
\end{align}
We dub such states \emph{lowest Landau level superfluids}, and their hydrodynamics will be a major focus of this work. 

Depending on the choice of potential, $V(r_\alpha)$, it is possible to select the precise LLL orbital, $\psi_{\mathrm{LLL}}$, the bosons condense into. For example, for a rotationally invariant 
potential of the schematic form, $V(r_\alpha)=-a |r^i_\alpha|^2+b|r^i_\alpha|^4$, $a,b>0$, one can select $\psi_{\mathrm{LLL}}$ to be any symmetric gauge LL orbital. However, choosing a parabolic potential extending along a single spatial direction, $V(r_\alpha)=a (r^1_\alpha)^2$, leads to a Landau gauge orbital. We note that while of course the Hamiltonian is gauge invariant, a LL basis orbital in one gauge is a linear combination of basis orbitals in another. To state that the harmonic potential ``chooses'' a gauge is simply to say that it picks out a wave function that can be expressed as a single basis orbital in that gauge.

Because LLL superfluids have broken time-reversal symmetry, they generically exhibit Hall viscosity.
The viscosity tensor $\eta^{ijkl}$ is defined to be the linear response of the stress tensor $T^{ij}$ to a strain rate $\dot{u}_{kl}$,
\begin{equation}
\delta T^{ij} = \eta^{ijkl}\,\dot{u}_{kl} \label{eqn:defn_visc_tensor}\,.
\end{equation}
We may separate the viscosity tensor into a sum of two components that are respectively symmetric and antisymmetric under $ij\leftrightarrow kl$.
The symmetric term is even under time reversal and dissipates energy, while the antisymmetric term is odd and nondissipative~\cite{avron_odd_1998}. In an isotropic system, the odd piece of the viscosity tensor will have a single independent component, $\eta_H$,  which is dubbed the Hall viscosity. The pressure exerted by the Hall viscosity is illustrated in Fig.~\ref{fig:Hall}, where it can be seen to produce a transverse force to any strain rate.

In any system with a many-body energy gap, the Hall viscosity is proportional to the average angular momentum per unit area~\cite{avron_viscosity_1995},
\begin{equation}
\eta_H = - \frac{1}{2} \frac{\langle\Psi_{\mathrm{SF}}|\hat{L}_z|\Psi_{\mathrm{SF}}\rangle}{{\mathrm{Area}}}\label{eqn:avg_ang_mom}\,.
\end{equation}
Intuitively, this result is the statement that if the bosons comprising the superfluid possess an angular momentum, then they will respond to a time-dependent strain like a gyroscope; namely, by shearing transverse to the direction of the strain. A particularly elegant way of justifying Eq.~\eqref{eqn:avg_ang_mom} is through adiabatic response theory~\cite{avron_viscosity_1995, park_guiding-center_2014}, in which it can be seen that the Hall viscosity is a consequence of a generalized Berry curvature. More generally, this relationship can be established through a Kubo linear response formalism~\cite{bradlyn_kubo_2012}.

In contrast to incompressible quantum Hall fluids and $p+ip$ superfluids, the Hall viscosity of a LLL superfluid is nonuniversal, in the sense that it depends on the choice of interaction Hamiltonian and harmonic potential. We therefore devote most of our attention to the case of a LLL superfluid in which all of the bosons condense into a single symmetric gauge LLL orbital (we treat the Landau gauge case in Appendix~\ref{app: Hall visc calculations}). If $\Psi_{\mathrm{SF}}$ is a product of symmetric gauge wave functions each with angular momentum, $\ell\in\mathbb{Z}$, then one obtains a result that is quantized in units of the mean number density, $\overline n$,
\begin{align}
\frac{\eta_H}{\overline{n}}=-\frac{\ell}{2}\,.
\end{align}
Note that in other LLL superfluid states the Hall viscosity need not be quantized and can even diverge (interactions can regulate this divergence). This occurs for example if $\psi_{LLL}$ is a Landau gauge orbital. We address this case in detail in Appendix \ref{app: Hall visc calculations}. For simplicity, we choose focus on the quantized, symmetric gauge example.

Before proceeding to the hydrodynamics of these systems, we note that, superficially, the physics of LLL superfluids is very similar to that of the $p+ip$ superfluids studied in earlier works \cite{Read2009,Read2011,hoyos_effective_2014,Moroz2019,Rose2020,Furusawa2021}. However, we emphasize that LLL superfluids have \emph{explicitly broken} time-reversal symmetry, whereas in the most commonly studied theories of $p+ip$ superfluids time-reversal is taken to be broken spontaneously. The nonuniversal nature of the LLL superfluid Hall viscosity is a consequence of this distinction, while the Hall viscosity of a $p+ip$ superfluid  is generally universal and quantized. Importantly, however, explicit breaking of time-reversal \emph{is sufficient} for the Hall viscosity to be nonuniversal, meaning that a $p+ip$ superfluid in the presence of explicit time-reversal breaking will have both universal and nonuniversal contributions to its Hall viscosity.

\subsection{LLL superfluid hydrodynamics}
\label{sec:hydro_conseq}

In this section our aim is to explore the consequences of viscosity via the equations of motion obeyed by a fluid's density and velocity in its presence. We will demonstrate and give intuition for how Hall viscosity leads to the generation of vorticity under compression. We will further show that a system with a magnetic field and Hall viscosity can be rewritten as a system without Hall viscosity, but with an additional ``edge'' contribution to the current which propagates tangentially to any density fluctuation. 

Hydrodynamics characterizes the slow relaxation of microscopically conserved quantities in a given system. At low temperatures this takes the form of two equations of motion: one for mass and one for momentum~\cite{acheson_elementary_1990,lucas_hydrodynamics_2018}. In nonrelativistic systems these two can be written as
\begin{align}
\frac{\partial \rho_m}{\partial t} + \partial_i \s{P}^i &= 0 \label{eqn:cons_num1}\\
\frac{\partial \s{P}^i}{\partial t} + \partial_j T^{ij} &= \s{F}^i \label{eqn:cons_mom1},
\end{align}
where $\rho_m$ is the mass density, $\s{P}^i = \rho_m v^i$ is the nonrelativistic mass current, $T^{ij}$ is the stress-energy tensor, and $\s{F}^i$ is the external force applied, which can include Lorentz forces. While these equations are often analyzed in the classical regime they can be reformulated as operator equations in the quantum regime. In this case $\rho_m/m_b$ and $\s{P}^i$ are the many-body number and momentum operator, respectively~\cite{bradlyn_kubo_2012}. In this section we will use the language of classical hydrodynamics, but many of our conclusions will carry over into many-body equations of motion that will prove useful for a Kubo formulation of viscosity. 

\begin{figure*}
    \centering
    \subfigure[Shear viscosity]{
        \includegraphics[width=0.45\textwidth]{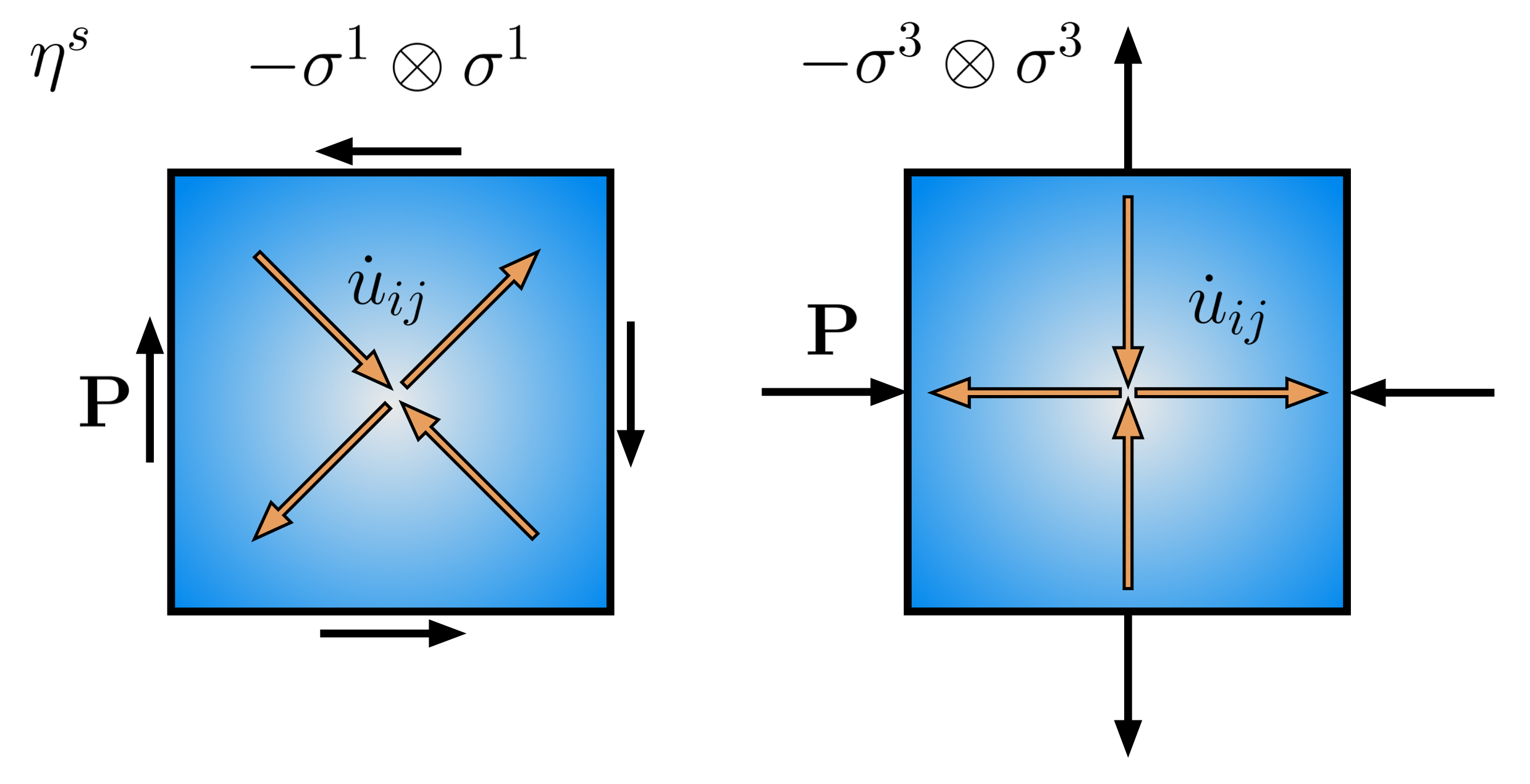}
        \label{fig:shear}
            }
    ~ 
    \subfigure[Hall viscosity]{
        \includegraphics[width=0.45\textwidth]{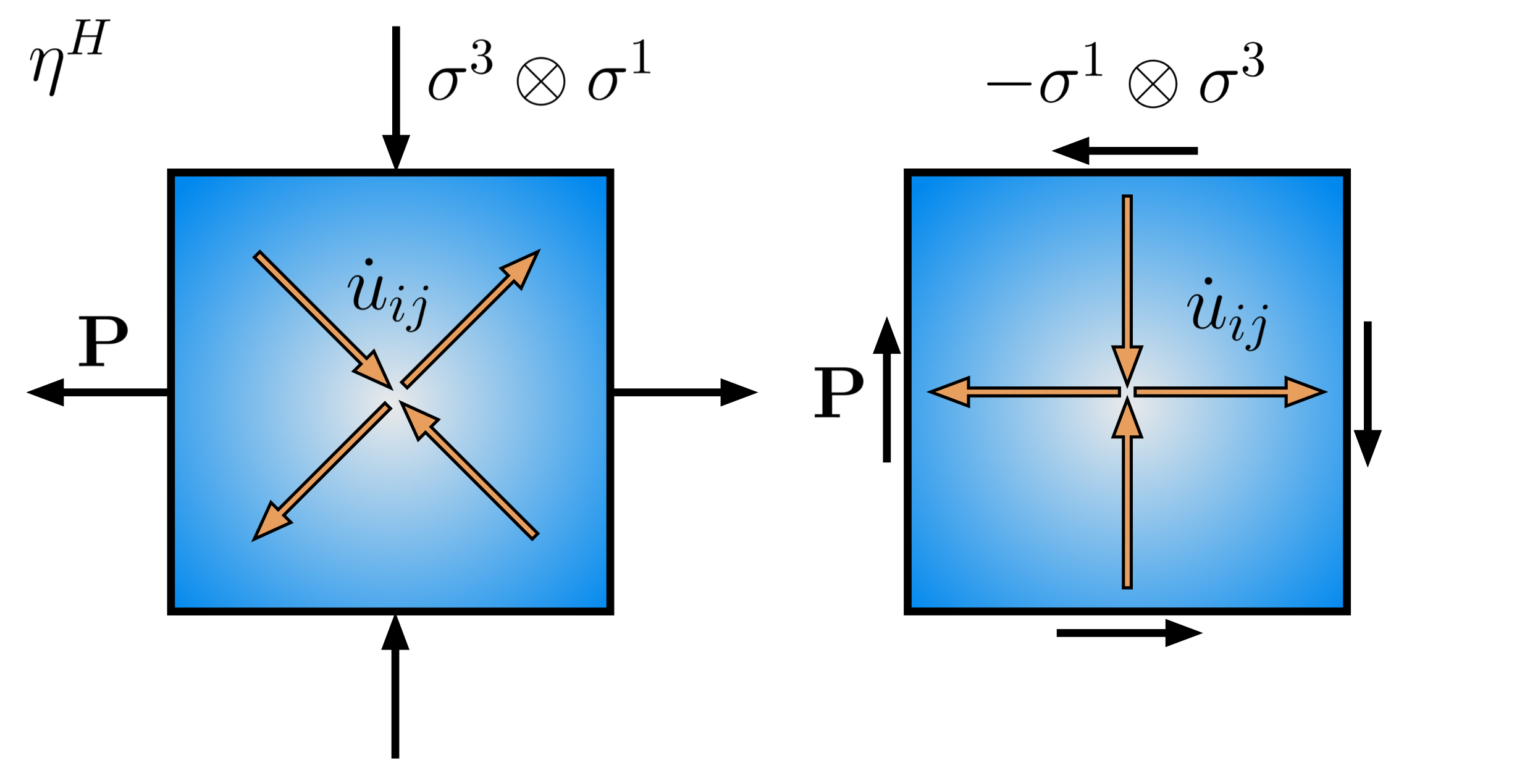}\
        \label{fig:Hall}
            }
    \caption{The pressure due to: \subref{fig:shear} shear and \subref{fig:Hall} Hall viscosities on a fluid element. The strain rate is indicated with orange arrows which show the direction of velocity, while the pressure on the respective edge of the fluid element is indicated with black arrows. The respective tensor element of viscosity is shown above each fluid element.}
	\label{fig:stress_strain}
\end{figure*}

The stress-energy tensor will have three important components and can be written as
\begin{equation}
T^{ij} = \mathfrak{p} \delta^{ij} + \rho_m v^iv^j + \frac{1}{2}\eta_{ijkl}(\partial_k v_l + \partial_l v_k).
\end{equation}
Here $\mathfrak{p}$ is the pressure, $\rho_m v^iv^j$ is due to convection, and the final term is the viscosity contribution. We have used the fact that in a fluid, whose coordinates are advected along with velocity, the strain rate is given by the symmetrized spatial derivative of velocity. If we require isotropy then in two dimensions the viscosity tensor can be written as
\begin{equation}
\eta = -\zeta \sigma^0\otimes \sigma^0 - \eta_s(\sigma^1\otimes \sigma^1 + \sigma^3\otimes \sigma^3) -\eta_H(\sigma^1 \otimes \sigma^3 - \sigma^3 \otimes \sigma^1) \label{eqn:coeffs_defn},
\end{equation}
where $\sigma^0$ is the $2\times 2$ identity matrix and $\sigma^1,\sigma^3$ are the Pauli matrices~\cite{avron_odd_1998}. Here $\zeta, \eta_s$ are the time reversal even components of viscosity which are called the bulk and shear viscosity, respectively. These components lead to dissipation of energy~\cite{acheson_elementary_1990}, while Hall viscosity does not. The forces produced by shear and Hall viscosity are illustrated in Fig.~\ref{fig:stress_strain}.

The conservation of momentum equation of motion is then given by
\begin{equation}
\frac{\partial v^i}{\partial t} + v^j\partial_j v^i = -\frac{1}{\rho_m}\partial^i \mathfrak{p} + \nu_\zeta \partial^i(\partial_j v^j) + \nu_s \nabla^2 v^i - \nu_H \nabla^2 \varepsilon^{ij}v^j + \frac{1}{\rho_m}\s{F}^i. \label{eqn:cons_mom_visc}
\end{equation}
Here we have defined $\nu_s$ to be $\eta_s/\rho_m$; $\nu_\zeta$ and $\nu_H$ are defined likewise \footnote{Note from Eq.~(\ref{eqn:avg_ang_mom}) that $\nu_H$ is a constant in a gapped fluid with an average angular momentum per particle and fixed mass.}. These ratios are called the ``kinematic viscosities'' and have units of $VL$, where $V$ is a velocity and $L$ is a length, regardless of the spatial dimension \footnote{In particular $\nu_s$, the kinematic shear viscosity, is used to define the familiar dimensionless Reynolds number $\mathrm{Re} = VL/\nu_s$, where $V$ is the characteristic velocity of a fluid and $L$ its characteristic length; this number measures the competition between inertial and viscous fluid behavior~\cite{acheson_elementary_1990}.}. The physical content of the various viscosities can then be deduced from Eq.~(\ref{eqn:cons_mom_visc}). Bulk viscosity will induce a force that resists compression, indicated by $\partial_i v^i$. It can be absorbed into the pressure as $\mathfrak{p}\rightarrow \mathfrak{p} - \zeta \partial_i v^i$. Shear viscosity will lead to diffusion of momentum; indeed $\nu_s$ is just the diffusion constant for velocity. The Hall viscosity is less straightforward but can be made clearer if $\s{F}^i$ is taken to be the Lorentz force felt by a particle in a magnetic field pointing out of plane. Then Eq.~(\ref{eqn:cons_mom_visc}) becomes 
\begin{equation}
\frac{\partial v^i}{\partial t} + v^j\partial_j v^i = -\frac{1}{\rho_m}\partial^i \mathfrak{p} + \nu_\zeta \partial^i(\partial_j v^j) + \nu_s \nabla^2 v^i - \nu_H \nabla^2 \varepsilon^{ij}v^j + \omega_c \varepsilon^{ij}v^j, \label{eqn:cons_mom_lorentz}
\end{equation}
where $\omega_c = eB/m_b$ is the cyclotron frequency. The Hall viscosity can thus be understood as a $k^2$ correction to the cyclotron frequency for nonuniform velocities i.e., $\omega_c \rightarrow \omega_c + \nu_H k^2$. Indeed if a relaxation rate $\tau$ is added to Eq.~(\ref{eqn:cons_mom_visc}) and the conductivity tensor is worked out within Drude theory, the contribution of $\nu_H$ will simply be the correction $\nu_H k^2$ to the Hall conductivity. Similar reasoning shows that Hall viscosity will always give a $k^2$ correction to $\sigma_H(\omega,\vec{k})$~\cite{bradlyn_kubo_2012}. 

\begin{figure}
    \centering
    \includegraphics[width=0.5\textwidth]{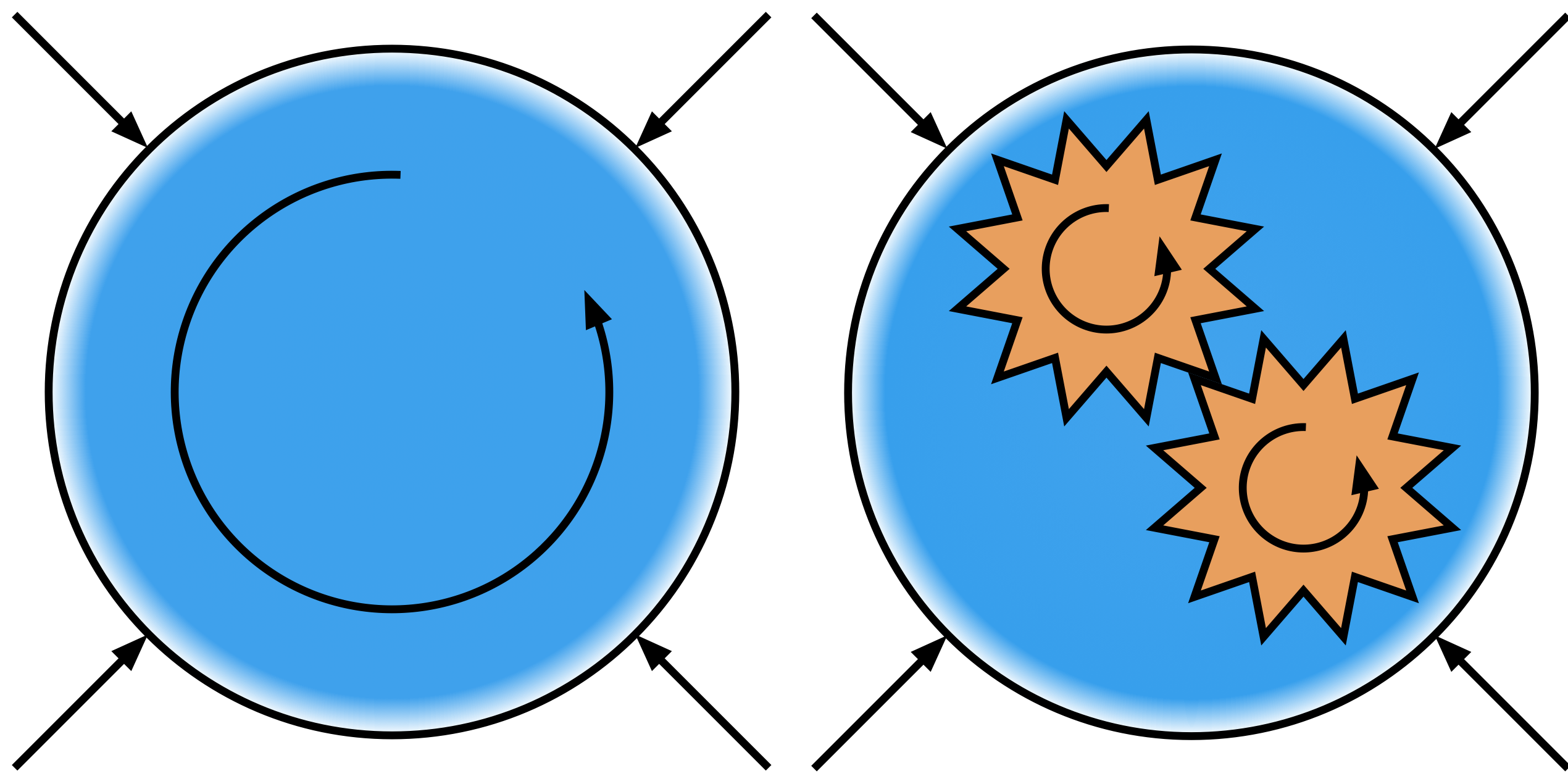}
    \caption{A microscopic realization of Hall viscosity. The left shows a compressible droplet undergoing compression which results in the generation of vorticity via Hall viscosity as in Eq.~(\ref{eqn:vort_EOM}). The right shows a microscopic realization. Rotating gears are suspended in the droplet, so that compression will cause the gears to interlock and turn their internal rotation into vorticity. Note that the gears have an internal angular momentum.}
    \label{fig:gear_fluid}
\end{figure}

An alternative, more hydrodyamically inclined, view of Hall viscosity can be obtained by Helmholtz decomposition of the velocity field into its irrotational and incompressible pieces. When this is done we see that
in an incompressible fluid, where $\partial_i v^i = 0$, the only effect of the Hall viscosity will be to renormalize the pressure with a vorticity dependent term $\mathfrak{p}\rightarrow \mathfrak{p} + \eta_H \Omega$, where $\Omega = \varepsilon_{ik}\partial^i v^k$ is the fluid's vorticity. This term nonetheless can have large impacts because it will alter boundary conditions in the presence of vorticity as in Refs.~\cite{avron_odd_1998,abanov_odd_2018,abanov_hydrodynamics_2020} or in Appendix~\ref{app:Pouiselle_flow}. If the fluid is compressible then Helmholtz decomposition reveals it will exert a force tangential to any compression. A microscopic picture of how such a force might arise was given by Ref.~\cite{han_fluctuating_2021} where a fluid with rotating gears was considered. This is illustrated in Fig.~\ref{fig:gear_fluid}.

This suggests that it is possible to understand Hall viscosity as contributing to vorticity production in compressible fluids. To make this intuition more concrete we will take the curl of Eq.~(\ref{eqn:cons_mom_lorentz}) to derive a conservation equation for vorticity. After some standard manipulations it will be given by
\begin{equation}
\frac{\partial \Omega}{\partial t} + \partial_i (\Omega v^i) = \nu_s \nabla^2 \Omega + \nu_H \nabla^2 (\partial_i v^i) - \omega_c \partial_i v^i, \label{eqn:vort_EOM}
\end{equation}
where we have supposed that the pressure $\mathfrak{p}$ is a function only of the mass density i.e., that it is barotropic~\cite{acheson_elementary_1990}. We stress that this is not a new conservation equation, but is rather a consequence of the equation of motion for momentum conservation. 
Shear viscosity again plays the role of a diffusion constant, now for vorticity, while both the magnetic field and Hall viscosity allow for the generation of vorticity in any region with compression.

We now consider linearizing these equations of motion about a small density fluctuation in a background density $\rho_{m,0}$. This is done in Appendix~\ref{app:lin_hydro}. There we show that in a system without shear or bulk viscosity the Hall viscosity contribution amounts to a correction to the speed of sound and the presence of an ``edge'' contribution to the physical velocity. This allows the full velocity to be written as 
\begin{equation}
v^i = -\partial^i \theta + \s{A}^i + v^i_{\mathrm{edge}}\,,\qquad v^i_{\mathrm{edge}} = \nu_H \varepsilon^{ij}\partial^j \log(\rho_m),
\end{equation}
and where $\s{A}^i$ contains the background vorticity of the fluid and will be proportional to the cyclotron frequency $\omega_c$. With this redefinition the dynamics of the field $\theta$ will be, up to linear order in the density fluctuation, independent of Hall viscosity apart from a momentum dependent correction to the sound velocity. This result is consistent with a recent approach that expressed fluid dynamics in the presence of Hall viscosity in terms of a Hamiltonian framework~\cite{machado_monteiro_hamiltonian_2023} and was also noted in Refs.~\cite{lingam_hall_2015,abanov_hydrodynamics_2020}. In particular, Ref.~\cite{markovich_odd_2021} showed that $v^i_{\mathrm{edge}}$ is necessary to define the correct center-of-mass momentum density in the presence of intrinsic angular momentum.

The presence of this extra edge term means that the total vorticity of the fluid is given by Eq.~(\ref{eqn:total_vorticity}). In particular, a collection of vortices can be described by a field $\theta$ which has nontrivial windings around a discrete set of points. To avoid singularities in the momentum current the density will be required to go to zero at these points, leading to density depletions near the vortices. Far from any vortices, where the linearized analysis applies, we can see that this term will alter the circulation around the vortices and must be considered for any tracer particle embedded in the flow. To determine whether the vortices themselves are advected with the full momentum current, or just the phase piece of the momentum current, it is necessary to go beyond the linearized regime. We will treat this later in our discussion of microscopics, and will show that the vortices are indeed advected with this additional edge current. For now, however, our intuition for Hall viscosity is sufficiently advanced that we can discuss our proposed experimental protocols.

\section{Experimental realization in rotating BECs}
\label{sec: 3}
Recent experiments have demonstrated that it is possible to prepare LLL superfluids by starting with a rotating BEC and ``squeezing'' it into a particular LLL orbital using an external confining potential. These systems are readily tunable by changing the confining potential, allowing the construction of setups with different values of Hall viscosity. The hydrodynamic properties of these systems can be studied by imaging their density profiles, meaning that signatures of Hall viscosity should be readily observable. Here we propose several  experimental protocols, which leverage Hall viscosity's linking of compression with vorticity discussed in the previous Section.

\subsection{Setup: LLL superfluid from rotating BECs}
\label{sec:LLL_superfluid_results}

As discussed in Section~\ref{sec:LLL_superfluids} the Hamiltonian of rotating bosons will be given by Eq.~(\ref{eqn:BEC_Ham}). In particular, the potential $V$ will be given by
\begin{equation}
V(r_\alpha) = V_{\mathrm{original}}(r_\alpha) - \frac{1}{8}m_b\omega_c^2 r_\alpha^2,
\end{equation}
where $\alpha$ labels the particle number and $V_{\mathrm{original}}$ is the potential applied before the centrifugal potential due to rotation is subtracted off. 
As long as the potential $V_{\mathrm{original}}$ is large enough at infinity to compensate for the centrifugal force, the bosons will remain confined. In particular if it is a quadratic plus a quartic, then the total potential will have a minimum at a finite radius, as discussed in the case of the symmetric gauge condensate in Section~\ref{sec:LLL_superfluids}. 
If the condensate is rotated fast enough that it is in the LLL then we may express $r_\alpha^2$ in terms of ladder operators for the guiding center coordinate, projecting out the LL ladder operators.

For simplicity, we will avoid the complication of projecting to the lowest Landau level by focusing on the noninteracting Hamiltonian,
\begin{equation}
H_0 = \sum_\alpha \omega_c\left(a^\dagger_\alpha a_\alpha + \frac{1}{2}\right) + U\left(b^\dagger_\alpha b_\alpha - \ell\right)^2 \label{eqn:H0_Mexican_hat},
\end{equation}
where $U$ is a positive constant, and adding other terms as perturbations. Here $a^\dagger_\alpha$ and $b^\dagger_\alpha$ are the operators that raise the Landau level and guiding center index of the $\alpha$ particle, respectively. This Hamiltonian will corresponds to a minima at $r/l_B = \sqrt{2(\ell + 1)}$, as in Appendix~\ref{app: Hall visc calculations}. If the experiment has control over both the quadratic and quartic terms in $V_{\mathrm{original}}$, both $U$ and $\ell$ are tunable real numbers. 
If $\ell \in \bb{Z}$ is an integer, then $H_0$ has a unique many-body ground state,
\begin{equation}
|\Psi\rangle_0 = \prod_\alpha |0,\ell\rangle \text{ where } \langle z|0,\ell\rangle \propto z^{\ell} e^{-|z|^2/4l_B^2}\,.
\end{equation}
This state is sometimes referred to as the ``giant vortex state''~\cite{roncaglia_rotating_2011}. This state clearly has a well defined angular momentum per particle,
\begin{equation}
L_{z,\alpha}|\Psi\rangle_0 = \ell |\Psi\rangle_0\,.
\end{equation}
Moreover, there is a many-body energy gap of order $\mathrm{min}(U,\omega_c)$ to the nearest state with a different angular momentum, and thus adiabatic response will hold and we should expect that 
\begin{equation}
\eta_H = -\frac{1}{2}\ell n\,,
\end{equation}
giving the system a well-defined Hall viscosity. 

We treat the full interacting theory in Appendix~\ref{app: Hall visc calculations} using the Kubo formalism of Ref.~\cite{bradlyn_kubo_2012}. We find that because the angular momenta are quantized there continues to be a many-body energy gap even when interactions are included; this gap is independent of the number of particles and thus remains finite even in the limit $N_b\rightarrow \infty$. Adiabatic response thus continues to hold and the Hall viscosity will be given by half of the angular momentum per unit area. Interactions will cause fluctuations out of the state with $\hat{L}_z = \ell$. They will tend to increase the angular momentum because they can increase it without bound while only being able to decrease it to zero. We thus find including interactions will renormalize $\ell$ upwards by $\mu/UN_b$ where the chemical potential $\mu$ is proportional to the strength of interactions. Nonetheless the Hall viscosity will continue to be nonzero, though nonuniversal.


\subsection{Observable signatures of Hall viscosity}
\label{sec:observable_signatures}

We now discuss several bulk~\footnote{We note that there is an extensive literature on the effect of Hall viscosity on boundary modes~\cite{abanov_odd_2018,soni_odd_2019,abanov_hydrodynamics_2020}. We will not concern ourselves with that here due to the presence of the confining potential at the edge of the sample, but in principle these effects might be used as a diagnostic.} hydrodynamic signatures of Hall viscosity that are accessible to current atomic physics experiments. 
These signatures rely on the hydrodynamic principles discussed in Section~\ref{sec:hydro_conseq}.

\subsubsection{Behavior near vortices}

Because Hall viscosity converts compression into vorticity, it is natural to seek its signatures in the motion of superfluid vortices. Indeed, it is possible to generate vortices experimentally in a rotating BEC by dragging a laser through the sample. This technique has been used in the past to produce vortex dipoles~\cite{t_w_neely_observation_2010} as well as longer ``streets'' of vortices~\cite{kazuki_sasaki_benard-von_2010,kwon_observation_2016}. 
There have also been theoretical proposals for the controllable nucleation of vortices using airfoil shaped potentials~\cite{musser_starting_2019}. Measurement of vortices is remarkably simple in rotating BEC setups, as the density depletions produced by vortices 
can be imaged directly~\cite{mukherjee_crystallization_2022}. 

\begin{figure}
    \centering
    \includegraphics[width=0.6\textwidth]{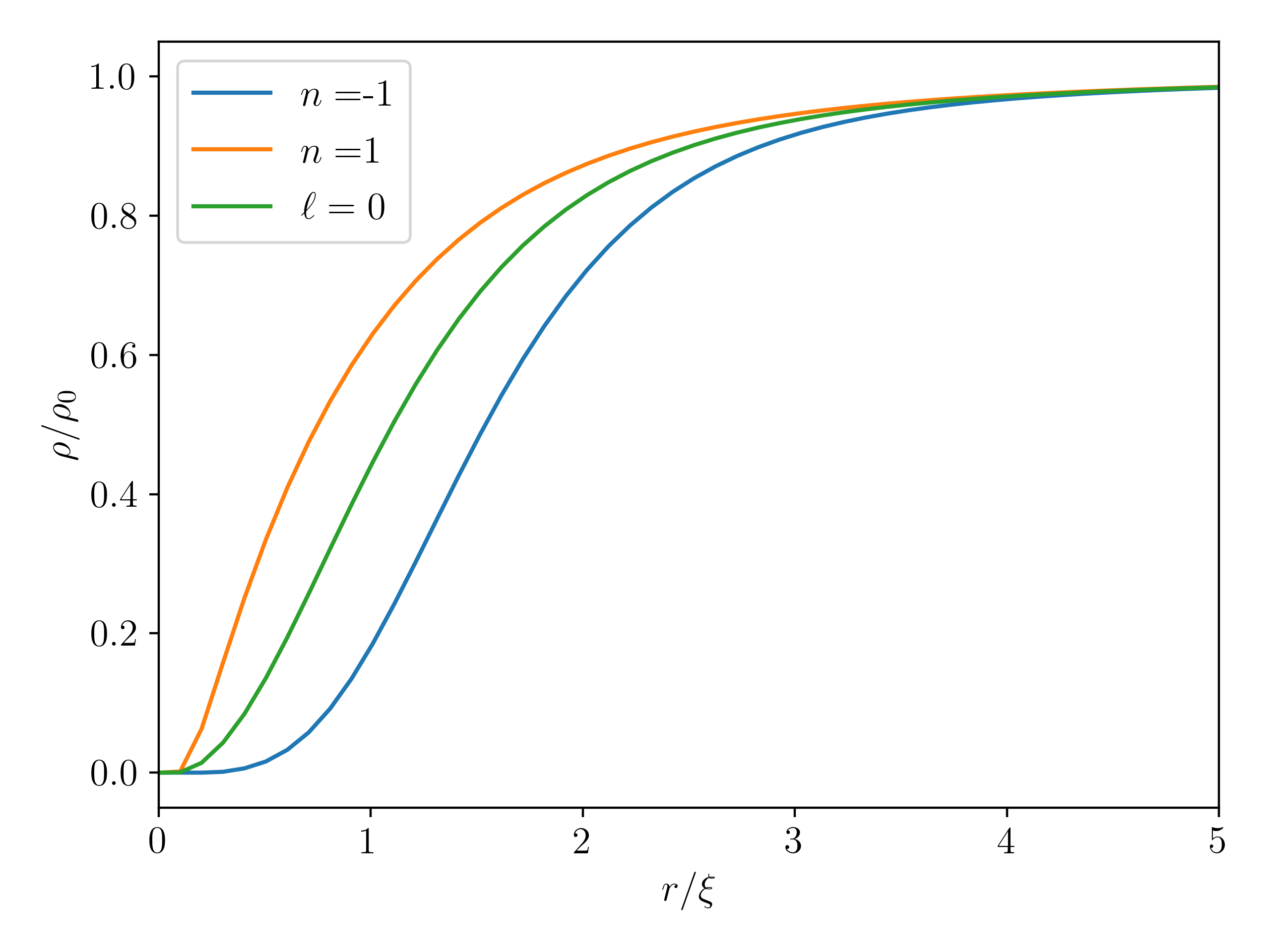}
    \caption{Density profiles of vortices in a superfluid with $\ell = 1$. At the longest distances from the vortex, both the $n=\pm 1$ vortices have a correction to the equilibrium density which decays as $1/r^2$. However, at closer distances their densities begin to differ. In particular, their scaling near the origin is very different. The density of a vortex in a superfluid with $\ell = 0$ is also plotted for reference. It can be seen to interpolate between the two vortices in the superfluid with $\ell = 1$.}
    \label{fig:dens_profiles}
\end{figure}

In this spirit it may be possible to suspend tracer particles that are advected with the condensate's momentum current. These particles would then pick up the extra edge current in the presence of density fluctuations that was discussed in previous sections. As Fig.~\ref{fig:vortex_dipole} reveals the edge current will enhance the circulation around one species of vortex, while decreasing it around the other. In particular, the speed at which a tracer particle embedded in the fluid orbits a single vortex at a given distance will be different depending on the sign of the vortex. 
We give an order of magnitude estimate of this effect. As computed in Appendix~\ref{app:vortex dynamics} and~\cite{hoyos_effective_2014} the density profile of a singly charged vortex a distance $r\gg \xi$ away, where $\xi$ is the coherence length, is given by
\begin{equation}
n(r) = n_0 - \frac{n_0 \xi^2}{2r^2} + \cdots,
\end{equation}
where $n_0$ is the asymptotic value of the superfluid number density. This density profile will continue to hold even when the effects of Hall viscosity are included. A plot of the full density is shown in Fig.~\ref{fig:dens_profiles} obtained with the boundary conditions $n(0) = 0, n(\infty) = n_0, n'(\infty) = 0$. We note that close to the vortex core the scaling of $n(r)$ is very different for the two charges of vortices. This is because in the presence of the extra edge current the fluid circulates much faster around the core of one charge of vortex than around the other. This extra centrifugal force must be balanced by a larger pressure, which requires a sharper change in density, as the pressure near the vortex core is a function of the gradient of density. See Appendix \ref{app:vortex dynamics} for more details.

Nonetheless, far from either vortex the edge contribution to the velocity will be of order:
\begin{equation}
v^\phi_{\mathrm{edge}} \sim -c_s\ell \left(\frac{\xi}{r}\right)^3 \sim \ell \left(\frac{\xi}{r}\right)^2 |v^\phi_{\mathrm{phase}}| \label{eqn:edge_to_phase},
\end{equation}
where $\xi$ is the coherence length and $c_s$ is the speed of sound in the superfluid. Here we used the fact that $v^\phi_{\mathrm{phase}} \sim \pm c_s\xi/r$ due to the phase winding of the vortex. Note that we can deduce this even at the level of the linearized analysis of Section~\ref{sec:hydro_conseq} since we are far from the vortex core. Thus, if we compare the ratio of the time for a tracer particle embedded in the fluid to orbit one sign of a vortex to another at the same distance, then we will find that they differ from one by a term of order
\begin{equation}
\ell\, \frac{\xi^2}{d^2}\,,
\end{equation}
where $d$ is the distance from the vortex.

Adding and tracking such tracer particles to the fluid may be difficult, however. We would thus like to know whether the vortices themselves can act as tracer particles i.e., if the vortices are advected with the momentum current. If this is the case, then we expect a vortex dipole to rotate since the momentum current around one sign of vortex will be enhanced relative to the other. We can address this by numerically integrating the Landau Ginzburg equations of motion of the full theory found in Section~\ref{sec:coherent_states}. 
To do so it is necessary to avoid the divergence of the velocity at the core of the vortex; we thus simulate the motion of vortices whose flow and density profile are identical to quantized vortices outside a core region of size $r_0$, but whose velocity goes to zero at $r=0$. This means that the density will no longer be zero at the core of a vortex. While the physics is altered at distances $r< r_0$ from the vortex core, we find that the vorticity stays concentrated near their core throughout the simulation. The physics for $r>r_0$ should thus approximate the physics of the LLL superfluid whose phase vorticity is a series of delta functions. More details of this simulation and the approximations made are discussed in Appendix~\ref{app:vortex dynamics}. As shown in Fig.~\ref{fig:dipole_rot_sim} we find that a vortex dipole does indeed rotate. We can then conclude that vortices themselves can be used as effective tracer particles to detect whether or not Hall viscosity is zero.

\begin{figure*}
    \centering
    \subfigure[$\ell = 0$]{
        \includegraphics[width=0.45\textwidth]{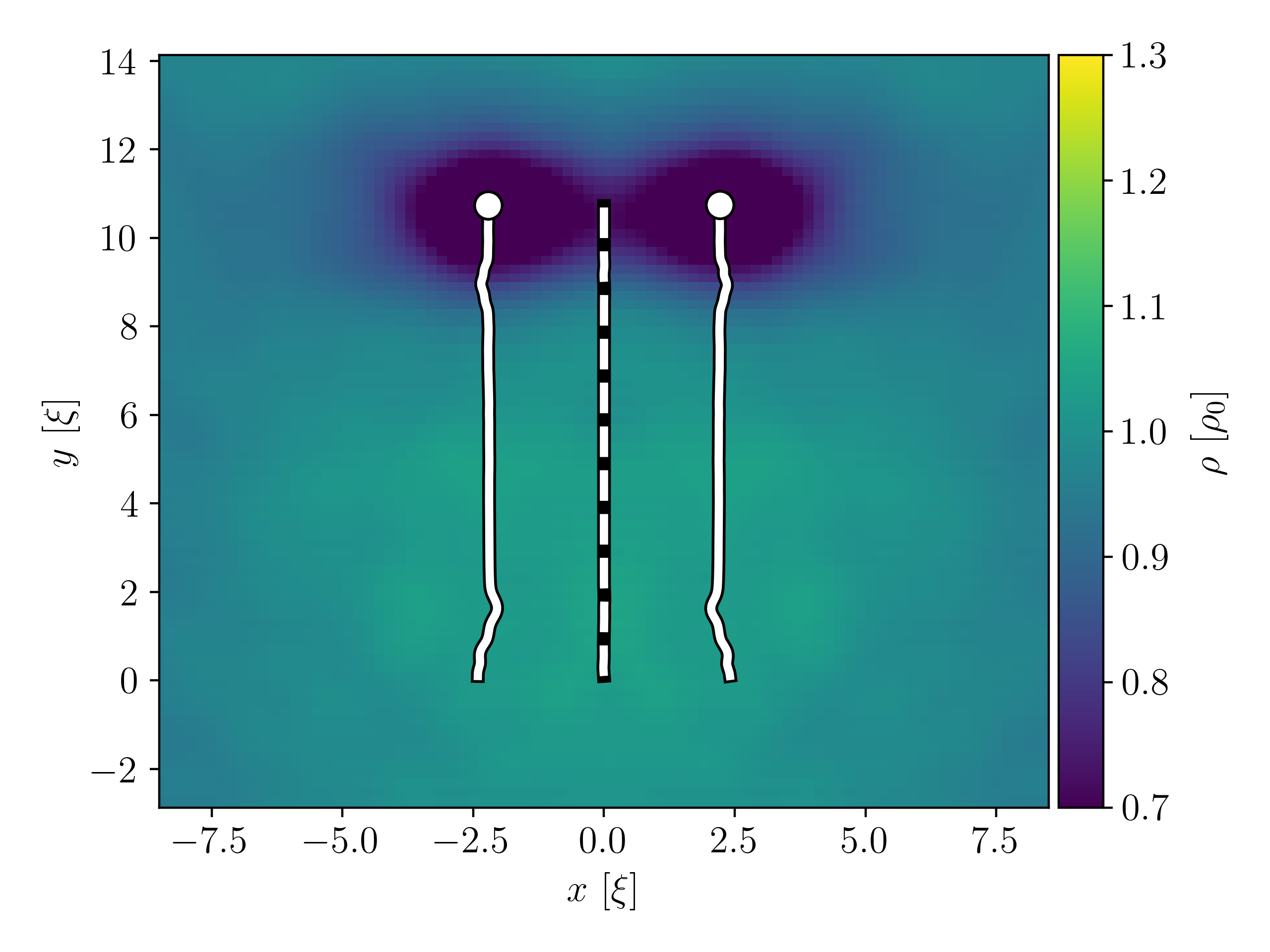}
        \label{fig:no_etaH}
            }
    ~ 
    \subfigure[$\ell = 1$]{
        \includegraphics[width=0.45\textwidth]{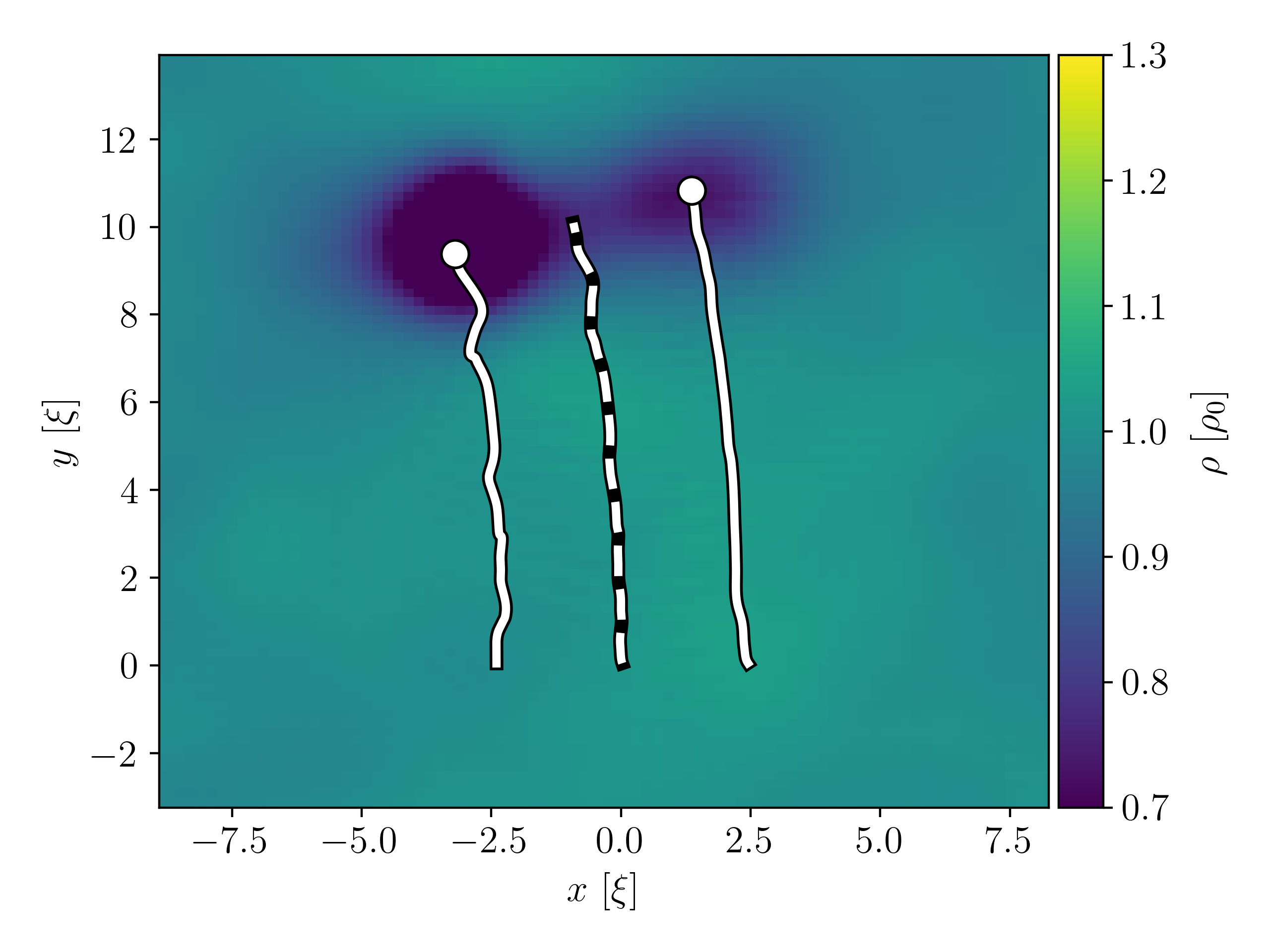}\
        \label{fig:etaH}
            }
    \caption{Behavior of a vortex dipole when Hall viscosity is: (a) zero and (b) nonzero. The tracking of the dipole is discussed in Appendix~\ref{app:vortex dynamics}, here each trajectory is shown as a solid white line, with the centerline shown as a dotted white line. In panel \subref{fig:no_etaH} the vortices have the same density profile. They can be seen to move perpendicular to their dipole. In panel \subref{fig:etaH} the vortices have different density profiles, owing to their different charge in the $\ell \neq 0$ superfluid. While they stay a roughly constant distance from one another their centerline is clearly curved due to their different velocities.}
	\label{fig:dipole_rot_sim}
\end{figure*}

\begin{figure*}
    \centering
    \subfigure[Initial density]{
	    \includegraphics[width=0.45\textwidth]{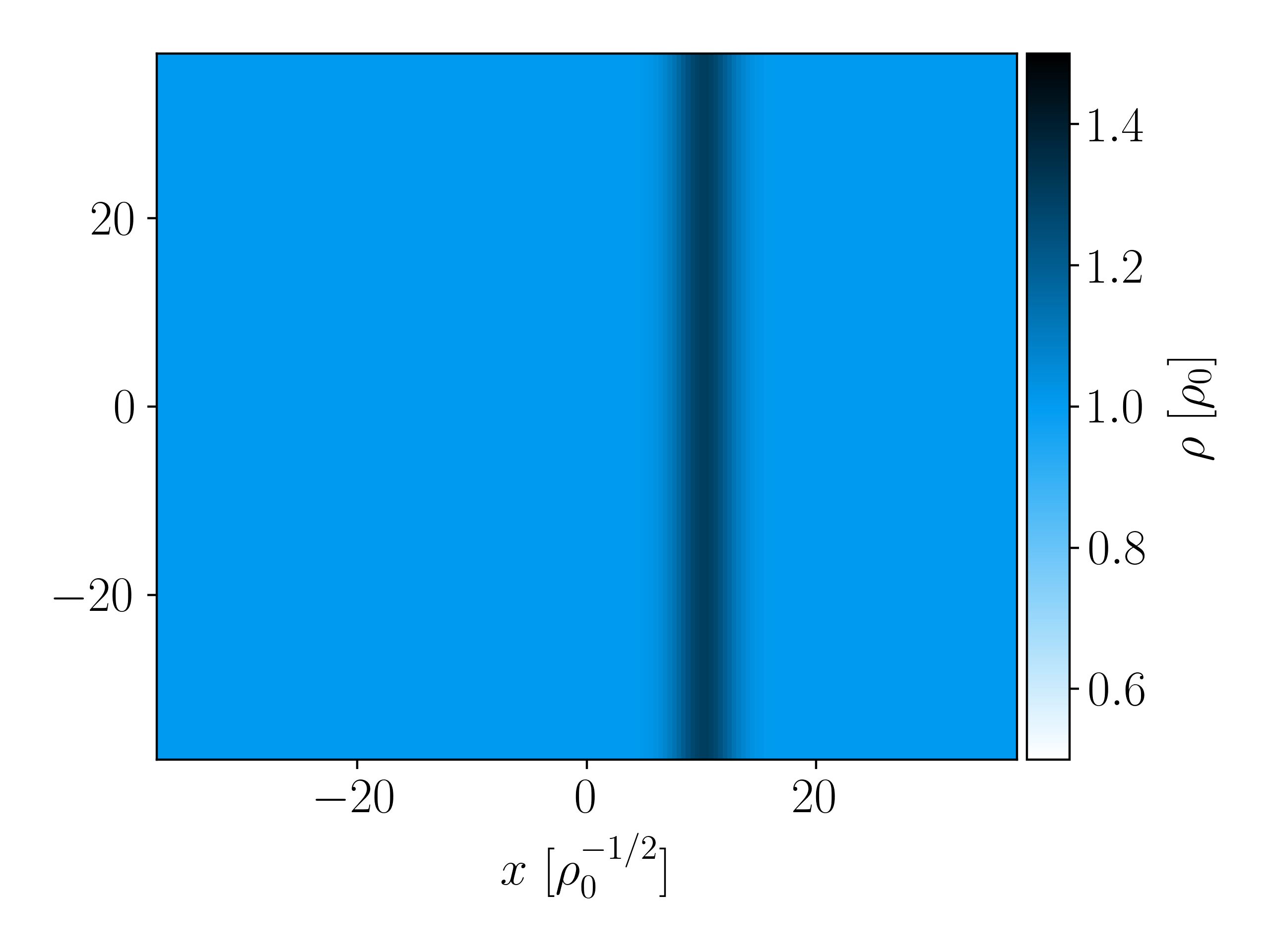}\
	    \label{fig:density_initial}
        	}
    ~ 
    \subfigure[Initial position]{
        \includegraphics[width=0.45\textwidth]{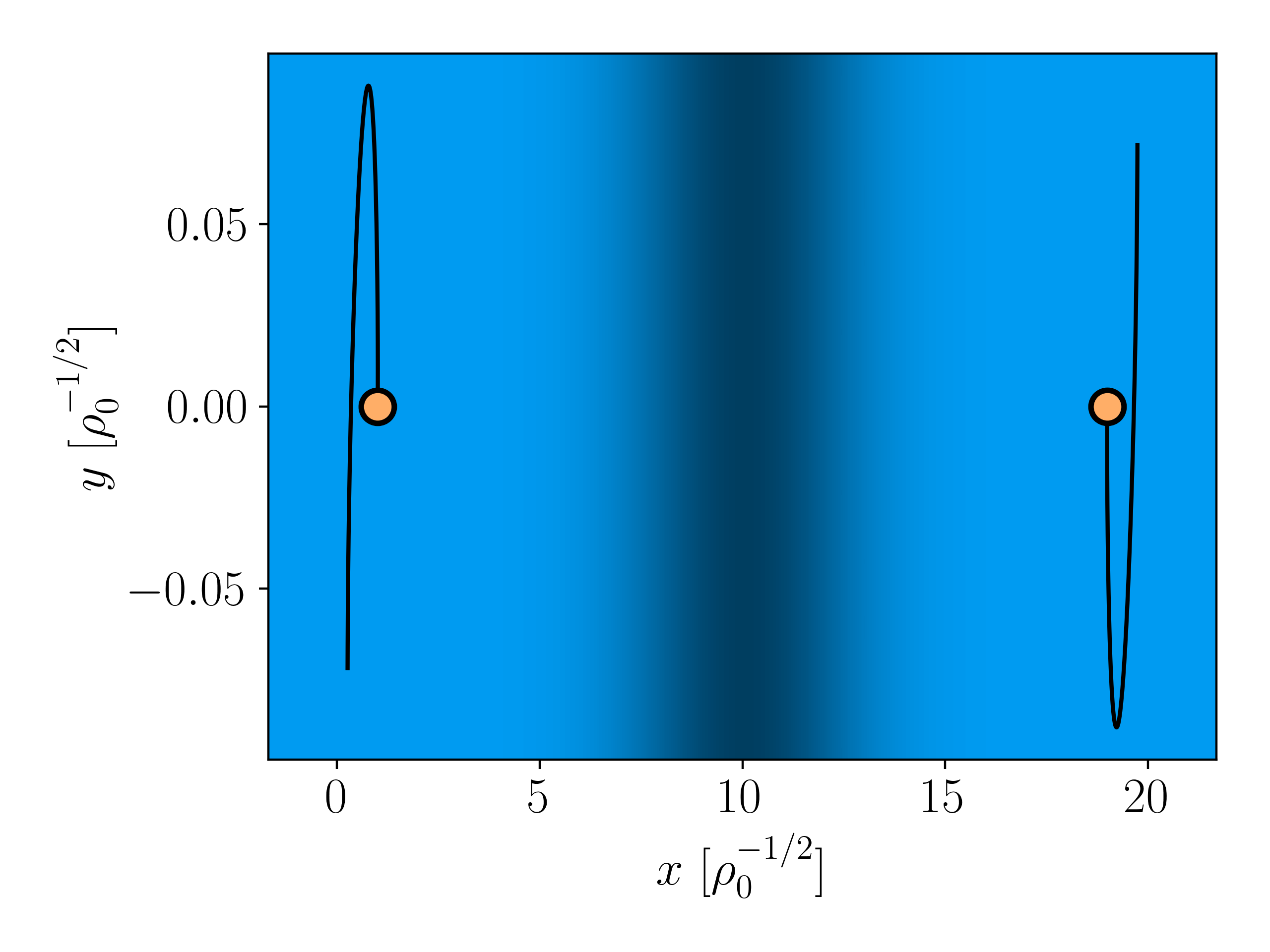}\
        \label{fig:tracer_initial}
            }
    ~ 
    \subfigure[Final density]{
        \includegraphics[width=0.45\textwidth]{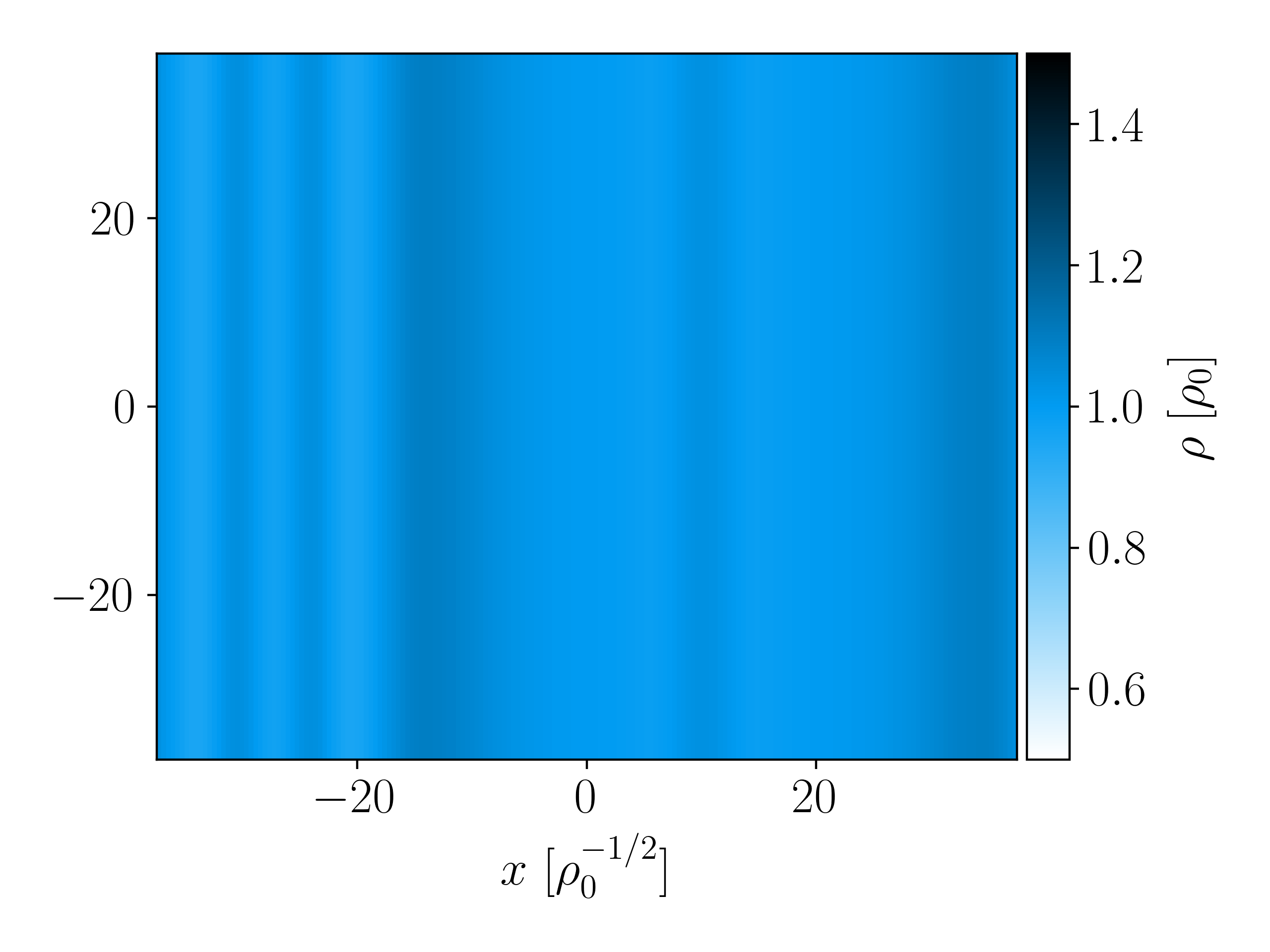}\
        \label{fig:density_final}
            }
    ~ 
    \subfigure[Final position]{
        \includegraphics[width=0.45\textwidth]{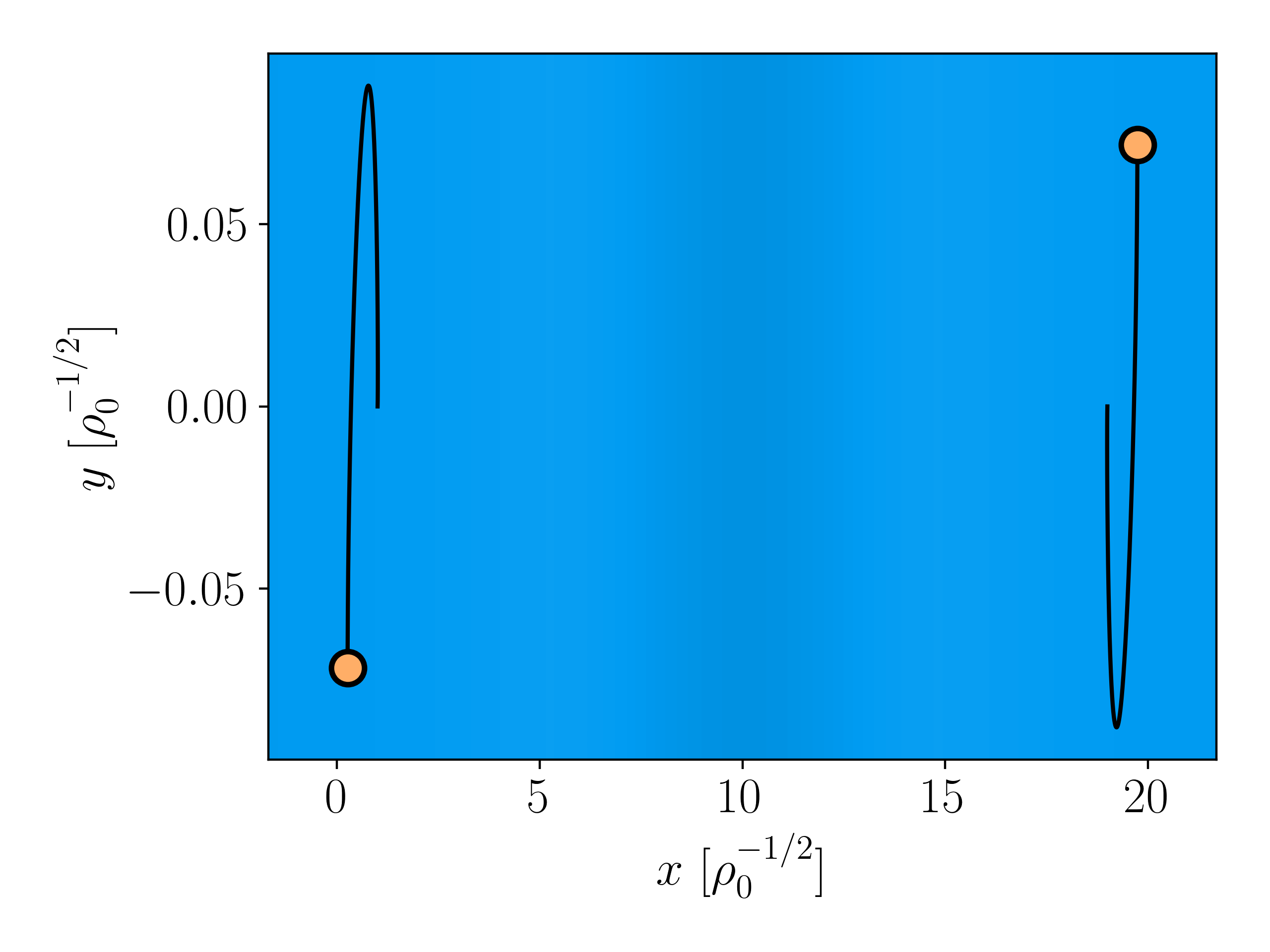}\
        \label{fig:tracer_final}
            }
    \caption{The behavior of tracer particles that are advected with the fluid flow in the presence of a density wave with $\eta_H \neq 0$. The full density is shown in panels \subref{fig:density_initial} and \subref{fig:density_final}, while a zoomed in view is shown in the background of panels \subref{fig:tracer_initial} and \subref{fig:tracer_final}. The tracer particles are indicated as orange circles in panels \subref{fig:tracer_initial} and \subref{fig:tracer_final}, while their full trajectories are shown as black lines. The simulation has periodic boundary conditions and is initialized with a Gaussian density peak at rest. The peak spreads out due to the pressure of the fluid. In panel \subref{fig:density_final} the lack of ballistic propagation is apparent from the multitude of density peaks at the left of image and is due to the $\nu_H^2 k^2$ corrections to the speed of sound. The initial density peak was placed at $x=10/\sqrt{\rho_0}$ so that these would not be obscured by the edges of the simulation.}
	\label{fig:tracer_density}
\end{figure*}

\subsubsection{Effects near sound waves}

The extra edge current in the presence of density fluctuations will also alter the behavior of tracer particles, and thus vortices, near a sound wave. This can be seen in Fig.~\ref{fig:tracer_density} which shows the behavior of tracer particles as a Gaussian density wave passes them. As the density wave approaches the particle on the left is advected upwards and once it passes is advected downwards. The particle on the right is initially advected downwards and then upwards. This is in contrast to a fluid without Hall viscosity, where parity symmetry would forbid such behavior. As in the case of a vortex dipole, we expect that vortices themselves will be similarly affected.

Figure \ref{fig:tracer_density} also demonstrates the $\s{O}(k^2)$ correction to the speed of sound. As discussed in Appendix~\ref{app:lin_hydro} the Hall viscosity will generically contribute a correction,
\begin{equation}
c_s^2 \rightarrow c_s^2 + \nu_H^2 k^2\,,
\end{equation}
where $c_s$ is the speed of sound and again $\nu_H = \eta_H/\rho_m = - \ell/2m_b$ in this system. This correction will alter the ballistic propagation of soundwaves, as can be seen in Fig.~\ref{fig:density_final}. The diffusivity of sound in a strongly interacting Fermi gas has already been studied in the cold atoms context~\cite{patel_universal_2020} via the use of a sinusoidally modulated trap. There absorption imaging was used to image the resulting density fluctuations. A similar procedure could in principle be used to image sound waves in a BEC with a nonzero Hall viscosity to extract this $k^2$ correction. While there is expected to be a $k^2$ correction to the superfluid sound velocity even when $\eta_H = 0$, it is proportional to constants of the superfluid and not tunable \cite{rogel-salazar_grosspitaevskii_2013}. Thus, by tuning the trap radius and hence $\eta_H$ it may be possible to extract the $k^2$ correction unique to the Hall viscosity.

\section{Derivation of LLL superfluid hydrodynamics}
\label{sec: 4}

\subsection{Coherent state path integral in the LLL}
\label{sec:coherent_states}

Although it is possible to calculate the value of $\eta_H$, Eq.~\eqref{eqn:avg_ang_mom}, either using (single particle) adiabatic response or (many body, but linear response) Kubo formulas, it still remains to establish how Hall viscosity enters into the full hydrodynamic equations of the LLL superfluid.  
Remarkably, because of the simple structure of the LLL superfluid's many-body wave function, we are able to derive its Landau-Ginzburg equation by explicitly constructing the coherent state path integral and effective action for the system. This Landau-Ginzburg equation is the analog of the Gross-Pitaevskii equation (GPE) but for a superfluid with a nonzero average angular momentum per particle. In the process, we show that it is possible to  microscopically derive the coupling of the LLL superfluid to spatial curvature (in the form of a background SO$(2)$ spin connection). This coupling determines the Hall viscosity, because the linear stress  response to spatial curvature is equivalent to the linear response to a time-dependent strain, as in Eq.~\eqref{eqn:defn_visc_tensor}~\cite{bradlyn_kubo_2012,hoyos_hall_2012}. 
Our low energy theory mostly coincides with the theory of the chiral superfluid developed in Ref.~\cite{hoyos_effective_2014}, with the primary difference being a nonuniversal value of the Hall viscosity depending on the choice of symmetric gauge orbital.

We consider a LLL superfluid of $N_b$ bosons at positions $\bs{x}_1,\dots,\bs{x}_{N_b}$ (boldface denotes 2D spatial vectors), each occupying the same symmetric gauge orbital,
\begin{align}
|0,\ell;\bs{x}_\alpha\rangle&=(b^\dagger_\alpha)^\ell\,|0\rangle\,.    
\end{align}
The many-body ground state wave function for the noninteracting LLL superfluid is therefore  
\begin{align}
\label{eq: gs}
|\Psi_{\mathrm{SF}}(\bs{x}_1,\dots,\bs{x}_{N_b})\rangle=\prod^{N_b}_\alpha|0,\ell;\bs{x}_\alpha\rangle\,.   
\end{align}
In this ground state, each particle has fixed angular momentum, $\ell$,
\begin{align}
\label{eqn:defn_const_Lz}
\hat{L}_z^{(\alpha)}|0,\ell;\bs{x}_\alpha\rangle&= \ell|0,\ell;\bs{x}_\alpha\rangle\,,
\end{align}
We wish to study the ground state fluctuations of the LLL superfluid, projecting to the many-body Hilbert space of the LLL ($\omega_c\rightarrow\infty$). The noninteracting Hamiltonian, Eq.~\eqref{eqn:H0_Mexican_hat}, is a constant on this subspace, so we introduce an interaction potential, $V[\{\bs{x}_\alpha\}]$, such that the LLL projected Hamiltonian is 
\begin{align}
H_{\mathrm{LLL}}&=\mathcal{P}_{\mathrm{LLL}}\left\{U\sum_\alpha \left(L_z^{(\alpha)}-\ell \right)^2+ V[\{\bs{x}_\alpha\}]\right\}\mathcal{P}_{\mathrm{LLL}}\,.
\end{align}
Here $\mathcal{P}_{\mathrm{LLL}}$ is the LLL projection operator, and we have replaced $b^\dagger_\alpha b_\alpha$ in the squeezing potential term with $\mathcal{P}_{\mathrm{LLL}} L_z^{(\alpha)}\mathcal{P}_{\mathrm{LLL}}=\mathcal{P}_{\mathrm{LLL}}(b^\dagger_\alpha b_\alpha-a_\alpha^\dagger a_\alpha)\mathcal{P}_{\mathrm{LLL}}=\mathcal{P}_{\mathrm{LLL}}b^\dagger_\alpha b_\alpha\mathcal{P}_{\mathrm{LLL}}$.  

The interaction potential, $V[\{x_\alpha\}]$, is chosen to only weakly mix states with different angular momenta within the LLL i.e., $U>>E_{\mathrm{int}}$, where $E_{\mathrm{int}}$ denotes the energy scale associated with the interactions. 
However, we note that in current experimental setups \cite{fletcher_geometric_2021, mukherjee_crystallization_2022}, $E_{\mathrm{int}}/U\sim U/\omega_c\sim\mathcal{O}(1)$, meaning that fluctuations in this system will mix both Landau level and angular momentum sectors. Nevertheless, we choose in this section to work with the scale hierarchy ${\omega_c>> U>> E_{\mathrm{int}}}$ to derive a superfluid ground state solution which could possibly be extended to the regime $\omega_c \sim U\sim E_{\mathrm{int}}$. We leave a detailed analysis of the effects of LL mixing to future work, although we do consider interaction effects within the LLL in  Appendix~\ref{app: Hall visc calculations}. 

To model fluctuations, we work in the basis of coherent states, $|\phi_\alpha\rangle$, 
\begin{align}
\label{eq: coherent state ansatz}
|\psi_{\mathrm{coh}}\rangle =\prod_\alpha |\phi_\alpha\rangle\,,\,b_\alpha|\phi_{\alpha'}\rangle=\delta_{\alpha\alpha'}\,\sqrt{\frac{\ell}{\overline{n}}}\,\phi_{\alpha'}|\phi_{\alpha'}\rangle\,. 
\end{align}
Here $\overline{n}=N_b/\mathrm{Area}$ is the boson number density. The reason for the particular choice of normalization  factor will become apparent below.

Allowing $\phi$ to evolve in time in the Heisenberg picture, one can then describe the system in terms of a coherent state path integral. This is constructed from the normal ordered Hamiltonian and given by:
\begin{align}
Z=\int \mathcal{D}\phi\mathcal{D}\phi^\dagger e^{iS}\,,\,S&=\int dt\left\{-i\sum_\alpha\phi^\dagger_\alpha \partial_t\phi_\alpha-\langle\psi_{\mathrm{coh}}|H_{\mathrm{LLL}}|\psi_{\mathrm{coh}}\rangle\right\}\,\\
&=\int dt\left\{-i\sum_\alpha\phi^\dagger_\alpha \partial_t\phi_\alpha-\frac{U\ell^2}{\overline{n}}\sum_\alpha(\phi^\dagger_\alpha\phi_\alpha-\overline{n})^2-\mathcal{V}_{\mathrm{LLL}}[\{\phi_\alpha\}]\right\}\,,
\end{align}
where $\mathcal{V}_{\mathrm{LLL}}=\langle\psi_{\mathrm{coh}}|\mathcal{P}_{\mathrm{LLL}}V[\{\bs{x}_\alpha\}]\mathcal{P}_{\mathrm{LLL}}|\psi_{\mathrm{coh}}\rangle$.

We now pass from this many-body quantum mechanics problem to a coherent state path integral representation involving fields defined on spacetime. If the bosons have hard-core repulsive interactions preventing them from gathering at a single position, then their locations will tend to be packed around the trap minimum with some characteristic spacing, $\mathfrak{a}>\ell_B$, such that as $N_b\rightarrow\infty$ we can replace $\phi_\alpha$ with a spatially-dependent field variable, $\phi(\bs{x})$. Because $b_\alpha$ increases the angular momentum of the boson labeled by $\alpha$ by 1, we can thus understand $\phi^\dagger(\bs{x})$ as creating an excitation carrying unit angular momentum at the spatial location $\bs{x}$.  Furthermore, sums over particle positions can be replaced with integrals, $\sum_\alpha f_\alpha \approx \overline{n}\int d^2\bs{x}\,f(\bs{x})$.  

The resulting effective action is
\begin{align}
S&=\int dt d^2\bs{x}\left\{-i\phi^\dagger\partial_t\phi-\frac{U\ell^2}{\overline{n}}(\phi^\dagger\phi-\overline{n})^2-\mathcal{V}_{\mathrm{LLL}}[\phi]\right\}\,.
\end{align}
Missing from this expression is a dispersion for $\phi$, which is not present in the microscopic Hamiltonian. However, generically a dispersion will be generated by fluctuations, due to the interactions in $\mathcal{V}_{\mathrm{LLL}}$. For the purposes of this section, we will assume this to be the primary role of $\mathcal{V}_{\mathrm{LLL}}$ at long wavelengths and discard it. We are thus left with the effective action,
\begin{align}
\label{eq: coherent state EFT}
S&=\int dt d^2\bs{x}\left\{-i\phi^\dagger\partial_t\phi+\frac{1}{2m_*}|\partial_i\phi|^2-\frac{U\ell^2}{\overline{n}}(\phi^\dagger\phi-\overline{n})^2\right\}\,,
\end{align}
where the effective mass, $m_*$, is set by $E_{\mathrm{int}}$. As discussed in Appendix~\ref{app: Hall visc calculations}, the above action is valid if the interactions are weak compared to the trapping potential, $E_{\mathrm{int}}<<U$.

Unlike in derivations of the coherent state action for more traditional superfluids, in the symmetric gauge LLL superfluid the conserved charge density corresponds to angular momentum,
\begin{align}
\label{eq: Lz field theory}
L_z =\sum_\alpha \langle\psi_\mathrm{coh}|\mathcal{P}_{\mathrm{LLL}}\, L_z^{(\alpha)}\,\mathcal{P}_{\mathrm{LLL}}|\psi_\mathrm{coh}\rangle=\ell\int d^2\bs{x}\, \phi^\dagger(\bs{x})\,\phi(\bs{x})\,. 
\end{align}
This is a consequence of the fact that excitations of $\phi$ carry angular momentum. Here the reason for the choice of normalization in Eq.~\eqref{eq: coherent state ansatz} becomes clear: On condensing into the LLL superfluid ground state, $|\Psi_{\mathrm{SF}}\rangle$, each boson carries the same angular momentum, $\ell$, which then should be interpreted as the charge under spatial rotations generated by $L_z$, 
\begin{align}
e^{i\xi L_z}|\Psi_{\mathrm{SF}}\rangle=e^{i\xi\,\ell\, N}|\Psi_{\mathrm{SF}}\rangle\,,
\end{align}
where $N$ is the total boson number operator and we note that this transformation is accompanied by a global rotation of the spatial coordinates by the angle $\xi$. The total boson number charge in this state is thus $N=L_z/\ell$, which also must be quantized. 
This connection between charge and angular momentum in the LLL superfluid\footnote{We emphasize that this relationship is only strictly speaking valid in the superfluid ground state, which is a condensate of bosons with a single value of  angular momentum $\ell$. In particular, high energy fluctuations of the angular momentum density, $\ell\,\phi^\dagger\phi$, above the ground state need no longer be equal to $\ell$ times the boson number density.} will have important implications for the hydrodynamic equations: Namely, the existence of a nonvanishing Hall viscosity!

From the point of view of the microscopic Hamiltonian, the presence of the trapping potential causes the bosons to each condense into the symmetric gauge orbital with angular momentum $\ell$. In the effective field theory language, Eq.~\eqref{eq: coherent state EFT}, this is the statement that $\phi$  condenses such that the system has uniform angular momentum density $\ell \phi^\dagger\phi=\ell\overline{n}$. Fluctuations about the superfluid ground state can be represented as
\begin{align}
\phi^\dagger\phi=\overline{n}+\delta n(\bs{x},t)\,&,\qquad\phi=\sqrt{[\overline{n}+\delta n(\bs{x},t)]}\, e^{i\theta(x,t)}\,.
\end{align}
The resulting effective action for the density and phase fluctuations about the ground state is therefore
\begin{align}
\label{eq: superfluid EFT}
S_{\mathrm{eff}}&=\int dt d^2x\left\{(\overline{n}+\delta n)\left(-\,\partial_t\theta+\frac{1}{2m_*}|\partial_i\theta|^2\right)+\frac{1}{2m_*}\Big|\partial_i\sqrt{\overline{n}+\delta n}\Big|^2-\frac{U\ell^2}{\overline{n}}\delta n^2\right\}\,.
\end{align}
Superficially, the physics appears to be no different from an ordinary superfluid in a time-reversal invariant system. Crucially, however, to obtain the full hydrodynamic equations, one must determine how the superfluid variables transform under spatial translations and rotations. More technically, it is necessary to couple the theory to a background metric and determine how the effective Lagrangian transforms under diffeomorphisms. 

As we observed in deriving Eq.~\eqref{eq: Lz field theory}, in the symmetric gauge LLL superfluid, the angular momentum operator simply generates U$(1)$ phase rotations of the superfluid ground state. Indeed, once the bosons each condense into the same angular momentum $\ell$ orbital, the total angular momentum simply counts the total number of bosons. 
If one wishes to perform local rotations by acting with $e^{i\xi(\bs{x},t)L_z}$, then, it will be necessary to introduce a background U$(1)$ gauge field, $\omega_\mu$, which transforms under spatial rotations as 
\begin{equation}
\omega_\mu \rightarrow \omega_\mu - \partial_\mu \xi\,.
\end{equation}
The background field, $\omega_\mu$, should be interpreted as a spin connection (see e.g., Refs.~\cite{Nicolis2011,Abanov2014,Gromov2014,Cho2014}). Indeed, because acting with $L_z$ also rotates the spatial coordinates, minimally coupling Eq.~\eqref{eq: superfluid EFT} to $\omega_\mu$ also means introducing a spatial metric, $g_{ij}=\delta_{ij}+\delta g_{ij}$, where the spin connection will be determined by $\delta g_{ij}$~\cite{green_superstring_2012}. We discuss this more in Appendix~\ref{app:galilean_symm}.

The resulting minimally coupled action is
\begin{align}
S_{\mathrm{minimal}} &= \int d td^2 \bs{x}\, \sqrt{g} \ \biggl[  n \left(-\,\s{D}_t \theta + \frac{g_{ij}}{2m_*}\s{D}^i\theta\, \s{D}^j\theta\right) - \frac{U \ell^2}{\overline{n}}n^2
+ \frac{g_{ij}}{2m_*}(\partial^i\sqrt{n})(\partial^j\sqrt{n})\biggr]\,,
\label{eqn:action_BH_Hl}
\end{align}
where $g$ denotes the determinant of the metric, $n\equiv \overline{n}+\delta n$, and the covariant derivative $\mathcal{D}_\mu$ is given by
\begin{equation}
\s{D}_\mu \theta = \partial_\mu \theta - \mathcal{A}'_\mu - \ell\, \omega_\mu,
\end{equation}
where the fixed charge $\ell$ under rotations follows from the fact that phase fluctuations of the condensate do not affect the  angular momentum. Note that we include an ordinary U$(1)$ gauge field, $\mathcal{A}'_\mu$, carrying unit charge, which couples to the boson number current. The total magnetic field felt by the microscopic bosons is thus $B_{\mathrm{tot}}=\varepsilon_{ij}\partial_i(\mathcal{A}_j+\mathcal{A}'_j)=B+\varepsilon_{ij}\partial_i\mathcal{A}_j'$. 

The effective action, Eq.~\eqref{eqn:action_BH_Hl}, however, remains incomplete. Indeed, it is not invariant under Galilean boosts. As long as the superfluid's coherence length is much smaller than the scale of the trap, we can expect this to be a symmetry of the system, just as it is for ordinary trapped superfluids \cite{fetter_rotating_2009}. 
Repairing this issue requires two modifications to our LLL superfluid effective theory. The first is the introduction of the gyromagnetic term,
\begin{align}
\label{eq: gyro}
S_{\mathrm{gyro}}&=\frac{1}{2m_b}\int dt d^2 \bs{x} \sqrt{g} \ n\,\ell\,\varepsilon_{ij}\,\partial_i\mathcal{A}'_j.
\end{align}
Note that we use the microscopic boson mass, $m_b$, here, as opposed to the effective mass, $m_*$ generated by interactions. 
The need for such a term makes physical sense: The LLL superfluid is, after all, composed of particles with an ``intrinsic'' angular momentum, $\ell$. In a different guise this same term also appeared in Ref.~\cite{hoyos_effective_2014}, where it was absorbed into the definition of $\omega_t$. This is discussed in more detail in Appendix~\ref{app:galilean_symm}. 

The gyromagnetic term in Eq.~\eqref{eq: gyro} can be motivated microscopically if we notice that the full coupling of the bosons to the rotation of the trap takes the form~\cite{fletcher_geometric_2021},
\begin{align}
H_{\mathrm{rot}}=-\omega_{\mathrm{trap}}\sum_\alpha\,L_z^{(\alpha)}=-\frac{B}{2m_b}\sum_\alpha\,L_z^{(\alpha)}\,.
\end{align}
If we wish to add external fields, $B'_{(\alpha)}$, which rotate individual bosons in the fluid (e.g., by applying a laser), then following the coherent state construction above we find that the effective magnetic field is shifted as $B\rightarrow B+\varepsilon_{ij}\partial_i \mathcal{A}'_j(\boldsymbol{x})$. This will then result in the additional term in Eq.~\eqref{eq: gyro}.

A second modification of Eq.~\eqref{eqn:action_BH_Hl} is needed to account for the fact that the microscopic Galilean symmetry requires that the momentum density, $\mathcal{P}^i$, and the number current, $j^i$, to satisfy,
\begin{align}
\label{eq: correct momentum}
\mathcal{P}^i=m_b\,j^i=m_b\,\frac{\delta \mathcal{L}_{\mathrm{eff}}}{\delta\mathcal{A}_i}\,,
\end{align}
with the microscopic boson mass, $m_b$. Because $m_b$ differs from the effective mass, ${m_*=\mathcal{Z}^{-1}m_b}$, generated by interactions, it is necessary to add terms to Eq.~\eqref{eq: coherent state EFT} to enforce Eq.~\eqref{eq: correct momentum}. To do so compactly, we introduce a fluctuating ``velocity field,'' $\upsilon^i$, coupling as~\cite{greiter_hydrodynamic_1989,Levin2017}
\begin{align}
\label{eq: supsilon}
S_{\upsilon}=\int dtd^2\bs{x}\,\sqrt{g}\,(\mathcal{Z}-1)\,\left\{\frac{i}{2}\,g_{ij}\,\upsilon^i(\phi^\dagger D^j\phi-D^j\phi^\dagger\,\phi)+\frac{g_{ij}}{2}m_b\upsilon^i\upsilon^j\,|\phi|^2 \right\}\,,
\end{align}
where $D_i\phi=(\partial_i-i\mathcal{A}'_i-i\ell\omega_i)\phi$. This term is the price one has to pay for working with an low energy theory with the effective mass, $m_*$, included in the action. Because our interest is in the physics of fluctuations about the superfluid ground state, with $\phi=\sqrt{n}\, e^{i\theta}$, the equation of motion for $\upsilon^i$ takes the particularly simple form in terms of the superfluid phase variable alone,
\begin{align}
\upsilon_i&=\frac{1}{m_b}\,\mathcal{D}_i\theta.
\end{align}
Hence, when integrated out the effect of introducing $\upsilon^i$ is to replace $m_*$ with $m_b$ in the second term of Eq.~\eqref{eqn:action_BH_Hl}. The phase stiffness is then $\overline{n}/m_b$, while the density fluctuations are left unchanged.

With the full effective action,
\begin{align}
\label{eq: Sfull}
S_{\mathrm{eff}}&=S_{\mathrm{minimal}}+S_{\mathrm{gyro}}+S_{\upsilon}\,,
\end{align}
we can derive the hydrodynamic equations for the symmetric gauge LLL superfluid. 
We start by computing the momentum density, following the gauge invariant procedure in Ref.~\cite{greiter_hydrodynamic_1989} and along the way invoking the equation of motion for $\upsilon^i$,
\begin{align}
\mathcal{P}^i &= m_b\frac{\delta \mathcal{L}_{\mathrm{eff}}}{\delta \s{A}_i} = -n(\partial^i\theta - \s{A}^i - \ell\, \omega^i) + \ell\,\frac{n}{2}\, \varepsilon^{ij}\partial_j\log(n),
\end{align}
Importantly, the second term -- which we have referred to as the ``edge'' term in earlier sections --  leads to a Hall viscosity, $\eta_H=\ell\,\overline{n}/2$, and is nonvanishing even in the absence of spatial curvature, $\omega_\mu=0$. We have thus found that the microscopic origin of the edge term is (1) the condensation of bosons with angular momentum $\ell\neq 0$ in the LLL, and (2) the existence of the  gyromagnetic term featured in Eq.~\eqref{eq: gyro}. 

Before proceeding, we remark that the edge term can be removed by a judicious choice of local frame. Indeed, if one scales coordinates locally such that $\sqrt{g}\,n \equiv \overline{n}$ everywhere, then density fluctuations will be absorbed into the metric, $g_{ij}$, and the term involving the spin connection will exactly cancel the edge term. Physically, passing to this rather complicated frame amounts to replacing density fluctuations with spatial curvature fluctuations, so the observable physics associated with Hall viscosity remains unchanged even though the edge term is absent. 

\subsection{Landau-Ginzburg equations with Hall viscosity}

We now develop the Landau-Ginzburg equations for the effective theory in Eq.~\eqref{eq: Sfull}. Once $\upsilon_i$ is integrated out,  the equations of motion 
become essentially those worked out for chiral superfluids in Ref.~\cite{hoyos_effective_2014}, but with the inclusion of distinct masses for the phase and amplitude fluctuations of the superfluid. 

First, we take $\rho_m = m_bn$ as before. Then we define
\begin{equation}
V^i \equiv \frac{1}{\rho_m}\s{P}^i = -\frac{1}{m_b}\left(\partial^i \theta - \s{A}^i +\frac{\ell}{2} \varepsilon^{ij}\partial^j \log(\rho_m)\right) \,. \label{eqn:defn_big_V}
\end{equation}
The full momentum density is then ${\mathcal{P}^i=\rho_m V^i}$. 
We will find it useful below to separate $V^i$ into  phase and edge contributions
\begin{equation}
v^i_{\mathrm{phase}} = \frac{1}{\rho_m}\s{P}^i_{\mathrm{phase}} \equiv -\frac{1}{m_b}\left(\partial^i \theta - \s{A}^i\right)\,,\qquad v^i_{\mathrm{edge}} = -\frac{\ell}{2m_b}\varepsilon^{ij}\partial^j \log(\rho_m)\,.
\end{equation}
The 
mass conservation equation can then be written in terms of the mass density as
\begin{equation}
\frac{\partial \rho_m}{\partial t} = -\partial_i(\rho_m V^i) = -\partial_i(\rho_m v^i_{\mathrm{phase}}),
\end{equation}
following from the fact that $\partial_i(\rho_m v^i_{\mathrm{edge}})=0$. 

The 
momentum conservation equation is in turn given by 
\begin{equation}
\frac{\partial}{\partial t}(\rho_m V^i) + \partial_j T^{ij} = \omega_c \rho_m \varepsilon^{ij}V^j,
\label{eq: momentum continuity}
\end{equation}
where $\omega_c = eB/m_b$ is the cyclotron frequency and
\begin{align}
T^{ij} &= \frac{2}{\sqrt{g}}\left.\frac{\delta \s{S}_{\mathrm{eff}}}{\delta g_{ij}}\right|_{g_{ij} = \delta_{ij}} \nonumber\\
&= \delta^{ij}\mathfrak{p} + \frac{1}{m_bm_*}(\partial^i\sqrt{\rho_m})(\partial^j \sqrt{\rho_m}) + \rho_m\left(v^i_{\mathrm{phase}}v^j_{\mathrm{phase}} + v^i_{\mathrm{edge}}v^j_{\mathrm{phase}} + v^i_{\mathrm{phase}}v^j_{\mathrm{edge}}\right),\nonumber\\
&\quad+ \frac{\ell}{4m_b}\rho_m\left(\varepsilon^{ik}\delta^{jl}+\varepsilon^{jk}\delta^{il}\right)\left(\partial_k v_{l,\mathrm{phase}} + \partial_l v_{k, \mathrm{phase}}\right)\,.
\label{eq: Tij}
\end{align}
We now remark that the pressure will have a leading order term given by $U\ell^2 n^2/\overline{n}$. This term is what will give rise to the speed of sound. The appearance of the cross terms proportional to $v^i_{\mathrm{phase}}v^j_{\mathrm{edge}}$ and the final term in Eq.~\eqref{eq: Tij}  come from varying the kinetic term, 
\begin{equation}
\sqrt{g}\frac{\rho_m}{2m_b^2}\, g_{ij}\,\s{D}^i\theta \, \s{D}^j\theta
\end{equation}
with respect to the spin connection, which is performed explicitly in Appendix~\ref{app:galilean_symm}. 

We can then express Eq.~\eqref{eq: momentum continuity} as 
\begin{align}
\partial_j T^{ij} &= \partial^i \mathfrak{p} + \frac{1}{m_b m_*}\partial_j[(\partial^i \sqrt{\rho_m})(\partial^j \sqrt{\rho_m})] \nonumber\\
&\quad +  \partial_j\left(\rho_m V^iV^j - \rho_m v^i_{\mathrm{edge}}v^j_{\mathrm{edge}}\right) - \frac{ \ell}{2m_b}\rho_m \varepsilon^{ij}\nabla^2 v^j_{\mathrm{phase}}.
\end{align}
Using the mass conservation equation, we can reduce the momentum conservation equation to
\begin{align}
\frac{\partial V^i}{\partial t} +V^j\partial_j V^i - v^j_{\mathrm{edge}}\partial^j v^i_{\mathrm{edge}} &= -\frac{1}{\rho_m}\partial^i \mathfrak{p} + \frac{1}{2m_bm_*}\partial^i\left(\frac{1}{\sqrt{\rho_m}}\nabla^2\sqrt{\rho_m}\right) \nonumber\\
&\quad - \left(\frac{ \ell}{2m_b}\right)^2 \partial^i\nabla^2 \log(\rho_m) 
+ \frac{ \ell}{2m_b}\varepsilon^{ij}\nabla^2 V^j + \omega_c \varepsilon^{ij}V^j\label{eqn:cons_mom_coherent},
\end{align}
where we absorbed a density term into the pressure. The left hand side of this equation expresses that the phase velocity 
will advect both itself and the edge terms. The correction $-v^j_{\mathrm{edge}}\partial_j v^i_{\mathrm{edge}}$ means the edge terms are not able to advect themselves, but they are able to advect vortices. 
The right hand side involves the familiar combination of pressure, odd viscosity, and magnetic field. Notably, the pressure picks up an  $\mathcal{O}(k^2)$ correction due to density fluctuations. 
We study this equation in detail Appendix~\ref{app:vortex dynamics}, where we derive the density profile around a single vortex in the absence of a magnetic field. We are able to reproduce the result of Ref.~\cite{hoyos_effective_2014} far from the vortex, but additionally obtain the density profile close to its center.

Now taking the curl of Eq.~\eqref{eqn:cons_mom_coherent},  we obtain 
\begin{equation}
\frac{\partial \Omega}{\partial t} + \partial_i(\Omega V^i) - \partial_i(\Omega_{\mathrm{edge}} v^i_{\mathrm{edge}}) = \frac{ \ell}{2m_b}\nabla^2(\partial_i V^i) + \omega_c\partial_iV^i,
\end{equation}
where $\Omega$ denotes the total vorticity and $\Omega_{\mathrm{edge}}$ the edge contribution, 
\begin{align}
\Omega &= \omega_c + \Omega_{\mathrm{phase}} + \Omega_{\mathrm{edge}}\,,\qquad \Omega_{\mathrm{edge}} = \frac{ \ell}{2m_b}\nabla^2 \log(\rho_m)\,.
\end{align}
We now consider an incompressible flow, where $\partial_i V^i = 0$. This will be the case for e.g., a collection of point vortices. Then we can see that the conservation of mass equation becomes
\begin{equation}
\frac{\partial \rho_m}{\partial t} + V^i\partial_i \rho_m = 0,
\end{equation}
so the density is advected with the velocity $V^i$, while the conservation of vorticity equation becomes
\begin{equation}
\frac{\partial \Omega}{\partial t} + V^i\partial_i \Omega = v^i_{\mathrm{edge}} \partial_i \Omega_{\mathrm{edge}},
\end{equation}
so the vorticity is advected with the velocity $V^i$, up to this additional edge correction that ensures the edge terms are not able to advect themselves.

It is challenging to study the above hydrodynamic equations analytically. For a standard superfluid the right hand side would be zero and $\Omega$ would only be supported at pointlike vortices. It can then be argued that such vortices must be advected with the sum of the velocities due to all other point vortices \cite{acheson_elementary_1990}. However, in a superfluid with Hall viscosity, the edge contribution to $\Omega$ is nonzero everywhere and the same argument cannot be made. This necessitates a numerical, rather than analytic, approach to showing the vortices are advected with $V^i$. We showed the results of such a numerical simulation in Section~\ref{sec:LLL_superfluid_results} and discuss it in greater detail in Appendix~\ref{app:vortex dynamics}.

\section{Discussion}
\label{sec: discussion}
In this work, we have developed a microscopic theory of LLL superfluids, in which bosons condense into a single LLL orbital, and we have presented several realistic protocols for observing its hydrodynamic effects in current cold atom systems. LLL superfluids thus represent ideal platforms for the study of Hall viscosity, and we anticipate that they will be the first macroscopic quantum states in which signatures of Hall viscosity will be observed. 

While this work focuses on the observable consequences of Hall viscosity for the motion of vortices and  propagation of sound waves, which should be readily accessible to current imaging techniques, it remains to develop realistic protocols to directly measure Hall viscosity. A major obstacle is the difficulty with measuring transport in cold atomic gases. Overcoming this challenge could lead to a direct measurement of the Hall viscosity through the wave-vector dependence of the Hall conductivity, in particular the dependence on $q^2$, $\partial_{q^2}\sigma_{H}(q)|_{q^2=0}$~\cite{bradlyn_kubo_2012,hoyos_hall_2012}.

It is also of future importance to study the effects of Landau level mixing on the stability of the LLL superfluid. Rather remarkably, in current experiments~\cite{fletcher_geometric_2021,mukherjee_crystallization_2022}, the distribution of bosons in the BEC remains sharply peaked in the LLL despite the energy scales associated with the cyclotron gap, interactions, and trap strength all being comparable. It would be useful to explore both experimentally and theoretically whether this stability continues to hold for general trap geometries such as the ring-shaped trap giving rise to the symmetric gauge LLL superfluid focused on for much of this work.

\section*{Acknowledgements} 

We are especially grateful to R. Fletcher and M. Zwierlein for enlightening discussions. We also thank Jing-Yuan Chen, Eduardo Fradkin, Sriram Ganeshan, Pranav Rao, Colin Scheibner, and Dam Thanh Son for discussions. S.M. was supported by the National Science Foundation Graduate Research Fellowship under Grant No.~1745302. Any opinions, findings, and conclusions or recommendations expressed in this material are those of the author(s) and do not necessarily reflect the views of the National Science Foundation. H.G. was supported by the Gordon and Betty Moore Foundation EPiQS Initiative through Grant No.~GBMF8684 at the Massachusetts Institute of Technology and by the Kadanoff Fellowship from the University of Chicago. T.S. was supported by U.S. Department of Energy Grant No. DE-SC0008739, and partially
through a Simons Investigator Award from the Simons Foundation. This work was also partly supported by the
Simons Collaboration on Ultra-Quantum Matter, which is a grant from the Simons Foundation (Grant No. 651446, T.S.).

\appendix

\section{Pouiselle flow in the presence of Hall viscosity}
\label{app:Pouiselle_flow}

\begin{figure}
    \centering
    \includegraphics[width=0.5\textwidth]{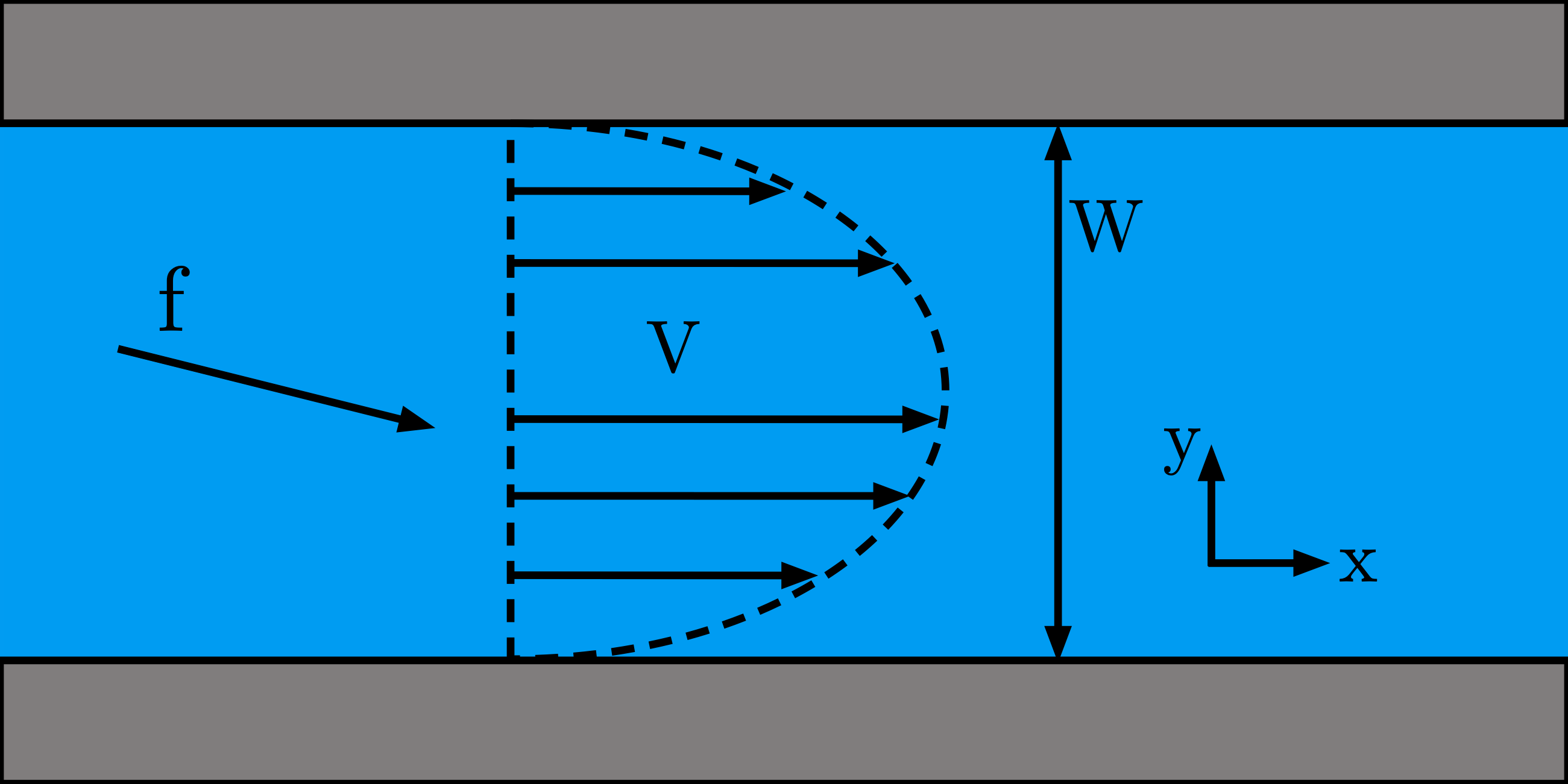}
    \caption{Pouiselle flow in a two dimensional pipe geometry. The pipe has width $W$, the velocity has a parabolic profile which follows the no-slip boundary condition at the edges of the pipe, and the force needed to keep the flow moving is shown with a nonzero $y$-component due to the nonzero magnetic field and Hall viscosity.}
    \label{fig:Pouiselle_flow}
\end{figure}

We solve Eqs.~(\ref{eqn:cons_num1}) and (\ref{eqn:cons_mom_lorentz}) in the two-dimensional pipe geometry shown in Fig.~\ref{fig:Pouiselle_flow}. By symmetry we take the steady-state form of the velocity and density to be given by $\vec{v} = v_x(y) \hat{x}, \rho_m = \rho_m(y)$. With this form we automatically satisfy the conservation of mass equation of motion and the conservation of momentum equation of motion becomes:
\begin{align}
0 =& -\frac{1}{\rho_m}\partial^x\mathfrak{p} + \nu_s \partial_y^2 v^x,\\
0 =& -\frac{1}{\rho_m}\partial^y\mathfrak{p} + \nu_H\partial_y^2 v^x - \omega_c v^x.
\end{align}
The first equation reveals that the pressure must have the form $\mathfrak{p} = f(y)x + g(y)$ for some functions $f$ and $g$, while the second reveals that $f(y)$ must be a constant $-f_0$. Additionally, we may take the density to be constant. Then the first equation reduces to:
\begin{equation}
\partial_y^2 v^x = -\frac{f_0}{\eta_s},
\end{equation}
which is satisfied by $v^x(y) = f_0y(W-y)/2\eta_s$ when the no-slip boundary conditions $v^x(0) = 0 = v^x(W)$ are imposed. This parabolic velocity profile is shown in Fig.~\ref{fig:Pouiselle_flow}. The average velocity in the pipe is then given by $\overline{v} = f_0W^2/12\eta_s$ allowing for replacement of $f_0$ by $\overline{v}$.

Rearranging the above equations of motion reveals that the force per unit area that must be exerted to keep the fluid flowing is given by
\begin{equation}
\vec{f} = -\vec{\nabla}\mathfrak{p} = \frac{12\overline{v}}{W^2}\left[\eta_s \hat{x} + \left(\eta_H + \frac{\rho_m\omega_c}{2} y(W-y)\right)\hat{y}\right].
\end{equation}
At the edges of the pipe, where the is no flow, the Lorentz force does not need to be balanced as it is zero. However, there is still a nonzero contribution from the Hall viscosity, which is constant across the sample. This can be understood as the vorticity correction to the pressure that arises in incompressible flow with Hall viscosity. Here the vorticity is given by $\Omega = -\partial^y v^x = 6\overline{v}(2y-W)/W^2$ leading to a correction to the pressure gradient of $-\eta_H \partial^y \Omega = -12\overline{v}\eta_H/W^2$. In the bulk of the fluid the Lorentz force must be balanced, adding an extra force. The average force per unit area that must be exerted in the $y$-direction is given by:
\begin{equation}
\overline{f}^y = \rho_m\overline{v}\left(\omega_c + 12\frac{\nu^H}{W^2}\right),
\end{equation}
which allows us to see that $\nu_H$ acts as the order $k^2$ correction to $\omega_c$, with $k$ on the order of $1/W$. This is very similar to the situation considered in Ref.~\cite{scaffidi_hydrodynamic_2017}.

\section{Linearized hydrodynamics}
\label{app:lin_hydro}

In this section we will linearize Eqs.~(\ref{eqn:cons_num1}),(\ref{eqn:cons_mom_lorentz}), and (\ref{eqn:vort_EOM}) about a small density fluctuation, $\rho_m(r,t) = \rho_{0,m} + \delta \rho_m(r,t)$. As is relevant in superfluids we will take $\nu_\zeta = 0 = \nu_s$. In the first we will perturb about zero velocity, while in the latter we will perturb about the velocity of a solid body rotation about the origin~\cite{mukherjee_crystallization_2022}. In both cases we will see that the effects of odd viscosity can be removed via a new ``edge'' contribution to the velocity. In particular we will be able to write
\begin{equation}
v^i = -\partial^i \theta + \s{A}^i + \nu_H \varepsilon^{ij}\partial^j \log(\rho_m),
\end{equation}
where $\s{A}^i$ will contain the background vorticity of the fluid \footnote{In the rest frame we will see $\varepsilon_{ij}\partial^i\s{A}^j = \omega_c\delta \rho_m/\rho_{0,m}$, while in the rotating frame it will be equal to $-\omega_c$.} and the dynamics of the field $\theta$ will be independent of Hall viscosity apart from a correction to the sound velocity.

\subsection{Rest frame}

In the rest frame $v^i(r,t) = \delta v^i(r,t)$. Then the linearized equations of motion become:
\begin{align}
\frac{\partial \delta \rho_m}{\partial t} &= -\rho_{0,m}\partial_i \delta v^i \label{eqn:lin_dens_rest}\\
\frac{\partial \delta v^i}{\partial t} &= -\frac{1}{\rho_{0,m}}\partial^i \mathfrak{p} -\nu_H \partial^i\delta \omega - \nu_H\varepsilon^{ij}\partial^j(\partial_k \delta v^k) + \omega_c \varepsilon^{ij}\delta v^j \label{eqn:lin_vel_rest}\\
\frac{\partial \delta \Omega}{\partial t} &= \nu_H \nabla^2(\partial_i \delta v^i) - \omega_c\partial_i \delta v^i \label{eqn:lin_vort_rest},
\end{align}
where we have used the Helmholtz transformation to simplify the Hall viscosity term. If we now insert Eq.~(\ref{eqn:lin_dens_rest}) into Eq.~(\ref{eqn:lin_vort_rest}), then we see that:
\begin{equation}
\frac{\partial \delta \Omega}{\partial t} = - \frac{\nu^H}{\rho_{0,m}}\nabla^2 \frac{\partial \delta \rho_m}{\partial t} + \frac{\omega_c}{\rho_{0,m}}\frac{\partial \delta \rho_m}{\partial t} .
\end{equation}
If we require the perturbation go to zero at infinity, then we see:
\begin{equation}
\delta \Omega = -\frac{\nu^H}{\rho_{0,m}}\nabla^2 \delta \rho_m + \frac{\omega_c}{\rho_{0,m}}\delta \rho_m  \label{eqn:vort_rest},
\end{equation}
so the density fluctuation entirely determines the incompressible piece of the velocity field. This fact is noted in Ref.~\cite{souslov_topological_2019} where the authors point out that it represents a zero energy eigenmode of the equations of motion.

If we now expand the pressure as $\mathfrak{p}(\rho_{0,m} + \delta \rho_m) = \mathfrak{p}(\rho_{0,m}) + c_s^2 \delta \rho_m + \cdots$ and insert Eq.~(\ref{eqn:lin_dens_rest}) and Eq.~(\ref{eqn:vort_rest}) into Eq.~(\ref{eqn:lin_vel_rest}), then we can simplify it as
\begin{equation}
\frac{\partial}{\partial t}\left(\delta v^i - \nu_H \varepsilon^{ij}\partial^j \frac{\delta \rho_m}{\rho_{0,m}}\right) = -(c_s^2 + 2\omega_c\nu_H -\nu_H^2 \nabla^2)\partial^i \frac{\delta \rho_m}{\rho_{0,m}} + \omega_c\varepsilon^{ij}\left(\delta v^i - \nu_H \varepsilon^{jk}\partial^k \frac{\delta \rho_m}{\rho_{0,m}}\right).
\end{equation}
This suggests that we define a new velocity:
\begin{equation}
\tilde{v}^i \equiv v^i - \nu_H \varepsilon^{ij}\partial^j \log(\rho_m).
\end{equation}
Note that $\partial_i \tilde{v}^i = \partial_i v^i$ so when linearized about the density fluctuation the equations of motion in terms of this new velocity will be given by:
\begin{align}
\frac{\partial \delta \rho_m}{\partial t} &= -\rho_{0,m}(\partial_i \delta \tilde{v}^i),\\
\frac{\partial \delta \tilde{v}^i}{\partial t} &= -(c_s^2 + 2\omega_c\nu_H - \nu_H^2 \nabla^2)\partial^i \frac{\delta \rho_m}{\rho_{0,m}} + \omega_c\varepsilon^{ij}\delta \tilde{v}^j.
\end{align}
We can now solve these equations of motion by decomposing $\delta \tilde{v}^i$ into a vorticity free and a divergenceless component i.e., by writing:
\begin{equation}
\delta \tilde{v}^i = -\partial^i \theta + \s{A}^i, \text{ where } \partial_i \s{A}^i = 0.
\end{equation}
We will not endeavor to actually solve the equations of motion, as they are standard equations for a density wave in the presence of a magnetic field. We will, however, point out that:
\begin{align}
\varepsilon_{ij}\partial^i\s{A}^j &= \delta \Omega + \nu_H \nabla^2 \frac{\delta \rho_m}{\rho_{0,m}}\\
&= \omega_c \frac{\delta \rho_m}{\rho_{0,m}} \text{ by Eq.~(\ref{eqn:vort_rest})} \label{eqn:mag_flux},
\end{align}
so the divergence free part of the velocity is entirely determined by the density fluctuation. Indeed, we can view $\s{A}^i$ as a vector potential for a magnetic field which is localized to density fluctuations.
 
In terms of this new velocity the Hall viscosity contribution to the momentum conservation equation of motion, Eq.~(\ref{eqn:lin_vel_rest}), vanished. The price paid for this is a sound velocity with $k^2$ corrections and the fact that $\tilde{v}^i$ no longer describes the momentum current divided by mass density, but rather features an extra ``edge'' current term which points tangentially to any perturbation in the density. This edge current will induces extra vorticity, which cancels the Hall viscosity contribution to the vorticity, as seen in Eq.~(\ref{eqn:mag_flux}). Further, we note that we can write the original physical velocity as:
\begin{equation}
v^i = -\partial^i \theta + \s{A}^i + \nu_H \varepsilon^{ij}\partial^j \log(\rho_m) \text{ where } \partial_i \s{A}^i = 0 \text{ and } \varepsilon_{ij}\partial^i \s{A}^j = \omega_c \frac{\delta \rho_m}{\rho_{0,m}}.
\end{equation}
To lowest order in the density perturbation the dynamics of $\theta$ will then be governed by a system without Hall viscosity, albeit with a correction to the sound velocity.

\subsection{Rotating frame}

We now consider the rotating frame, where $v^i(r,t) = \s{A}^i(r) + \delta v^i(r,t)$. Here $\s{A}^i(r)$ is a vector potential which satisfies:
\begin{equation}
\partial_i \s{A}^i = 0 \text{ and } \varepsilon_{ij}\partial^i \s{A}^j = -\omega_c,
\end{equation}
where $\omega_c$ is the constant cyclotron frequency. We will specifically choose:
\begin{equation}
\s{A}^i = \omega_c\varepsilon^{ij}r^j/2.
\end{equation}
This thus describes a fluid undergoing a solid body rotation about the origin. With this choice the linearized equations become:
\begin{align}
\frac{\partial \delta \rho_m}{\partial t} + \s{A}^i\partial_i \delta \rho_m &=  -\rho_{0,m}\partial_i \delta v^i \label{eqn:lin_dens_rot},\\
\frac{\partial \delta v^i}{\partial t} + \s{A}^j\partial_j \delta v^i &= -\frac{1}{\rho_{0,m}}\partial^i\left(\mathfrak{p} + \frac{\rho_{0,m}}{8}\omega_c^2 r^2\right) -\nu_H\partial^i \delta \Omega - \nu_H \varepsilon^{ij}\partial^j(\partial_k \delta v^k), \nonumber \\
&+ \omega_c \varepsilon^{ij}\delta v^j/2 \label{eqn:lin_vel_rot}\\
\frac{\partial \delta \Omega}{\partial t} + \s{A}^i\partial_i \delta \Omega  &= \nu_H \nabla^2(\partial_i \delta v^i)  \label{eqn:lin_vort_rot},
\end{align}
where we have used the Helmholtz transformation to simplify the Hall viscosity term. We thus see that in the rotating frame fluctuations are advected along by the background solid body rotation described by $\s{A}^i$ and the system behaves as if it is in a confining potential described by
\begin{equation}
U(r) = \frac{1}{8}m\omega_c^2 r^2;
\end{equation}
which is a familiar transformation~\cite{souslov_topological_2019,mukherjee_crystallization_2022}.

We now see from Eq.~(\ref{eqn:lin_dens_rot}) that:
\begin{align}
\partial_i \delta v^i &= -\frac{1}{\rho_{0,m}}\left(\frac{\partial \delta \rho_m}{\partial t} + \s{A}^i\partial_i \delta \rho_m\right),\\
\varepsilon^{ij}\partial^j(\partial_k \delta v^k) &= -\varepsilon^{ij}\frac{1}{\rho_{0,m}}\partial_t\partial^j \delta \rho_m -\frac{1}{\rho_{0,m}}\varepsilon^{ij}\partial^j \s{A}^k \partial_k \delta \rho_m - \frac{1}{\rho_{0,m}}\varepsilon^{ij} \s{A}^k\partial_k\partial^j \delta \rho_m\\
=& -\varepsilon^{ij}\frac{1}{\rho_{0,m}}\partial_t\partial^j \delta \rho_m - \frac{1}{\rho_{0,m}}\varepsilon^{ij} \s{A}^k\partial_k\partial^j \delta \rho_m - \frac{\omega_c}{2\rho_{0,m}}\partial^i \delta \rho_m.
\end{align}
Inserting this back into Eq.~(\ref{eqn:lin_vel_rot}) allows us to see that:
\begin{align}
\frac{\partial}{\partial t}\left(\delta v^i - \frac{\nu_H}{\rho_{0,m}}\varepsilon^{ik}\partial^k \delta \rho_m \right) &+ \s{A}^j\partial_j\left(\delta v^i - \frac{\nu_H}{\rho_{0,m}}\varepsilon^{ik}\partial^k \delta \rho_m \right)\\
 &= -\frac{1}{\rho_{0,m}}\partial^i\left(\mathfrak{p} + \frac{\rho_{0,m}}{8}\omega_c^2 r^2 - \frac{1}{2} \omega_c \nu_H \delta \rho_m\right) - \nu_H \partial^i \delta \Omega + \omega_c \varepsilon^{ij}\delta v^j /2 \nonumber.
\end{align}
Just as in the rest frame, this motivates us to define a new velocity given by:
\begin{equation}
\tilde{v}^i \equiv v^i - \nu_H \varepsilon^{ij}\partial^j \log(\rho_m).
\end{equation}
We note that $\partial_i \tilde{v}^i = \partial_i v^i$ and that $\delta \tilde{\Omega} = \delta \Omega + \nu_H \nabla^2 \delta\rho_m/\rho_{0,m}$. Then rewriting the equations of motion in this form gives:
\begin{align}
\frac{\partial \delta \rho_m}{\partial t} + \s{A}^i\partial_i \delta \rho_m &= -\rho_{0,m} \partial_i \delta \tilde{v}^i\\
\frac{\partial \delta \tilde{v}^i}{\partial t} + \s{A}^j\partial_j \delta \tilde{v}^i &= -\frac{1}{\rho_{0,m}}\partial^i\left(\mathfrak{p} + \frac{\rho_{0,m}}{8}\omega_c^2 r^2\right) - \nu_H \partial^i \left(\delta \tilde{\Omega} - \nu_H \nabla^2 \frac{\delta \rho_m}{\rho_{0,m}}\right) + \omega_c\varepsilon^{ij}\delta \tilde{v}^j/2\\
\frac{\partial \delta \tilde{\Omega}}{\partial t} + \s{A}^j \partial_j \delta \tilde{\Omega} &= 0.
\end{align}

Let us now suppose that there is initially no vorticity present in $\delta \tilde{v}^i$ except for isolated point vortices, as would be the situation in a rotating superfluid. The latter equation then tells us that these point vortices are simply advected along by the background $\s{A}^j$. Then we can write that:
\begin{align}
\frac{\partial \delta \rho_m}{\partial t} + \s{A}^i\partial_i \delta \rho_m &= -\rho_{0,m} \partial_i \delta \tilde{v}^i,\\
\frac{\partial \delta \tilde{v}^i}{\partial t} + \s{A}^j\partial_j \delta \tilde{v}^i &= -(c_s^2 - \nu_H^2 \nabla^2)\frac{\delta \rho_m}{\rho_{0,m}} -\frac{1}{8}\partial^i(\omega_c^2 r^2) + \omega_c\varepsilon^{ij}\delta \tilde{v}^j/2.
\end{align}
We once again see that apart from a correction to the speed of sound we have eliminated Hall viscosity from the equations of motion. We can then once again write 
\begin{equation}
v^i = -\partial^i \theta + \s{A}^i + \nu_H \varepsilon^{ij}\partial^j \log(\rho_m), \text{ where } \s{A}^i = \frac{\omega_c}{2}\varepsilon^{ij}r^j.
\end{equation}
Once again the dynamics of $\theta$ will be governed by a system without Hall viscosity, albeit with a sound velocity with a $k^2$ correction.

\section{Computation of Hall viscosity in LLL superfluids}
\label{app: Hall visc calculations}

\subsection{Basics of the symmetric gauge}
\label{app:LLL_symm_gauge}

Throughout this section we will work in the symmetric gauge; we will also drop the particle labels where unnecessary. In this gauge we make the definitions:
\begin{equation}
\hat{\pi}_i \equiv \hat{p}_i + \frac{1}{2 l_B^2} \varepsilon_{ij}\hat{r}_j \text{ and } \hat{R}_i \equiv \frac{1}{2}\hat{r}_i + l_B^2\varepsilon_{ij}\hat{p}_j.
\end{equation}
Then we see that:
\begin{equation}
[\hat{\pi}_i, \hat{\pi}_j] = \frac{i}{l_B^2} \varepsilon_{ij}, \text{ and } [\hat{R}_i, \hat{R}_j] = -i l_B^2 \varepsilon_{ij},
\end{equation}
while the two are mutually commuting. The first is referred to as the canonical momentum and has units of momentum, while the latter is the guiding center coordinate and has units of length~\cite{park_guiding-center_2014}. We can now define mutually commuting operators:
\begin{equation}
a \equiv \frac{l_B}{\sqrt{2}}\left(\hat{\pi}_x + i\hat{\pi}_y\right) \text{ and } b \equiv \frac{1}{\sqrt{2}l_B}\left(\hat{R}_x - i\hat{R}_y\right),
\end{equation}
where it can be checked that these satisfy the ladder commutation relations:
\begin{equation}
[a,a^\dagger] = 1 = [b,b^\dagger].
\end{equation}
The first is the ladder operator for raising and lowering the Landau level, while the latter is the ladder operator for raising and lowering the guiding center. In particular, we note that:
\begin{equation}
\frac{\hat{\pi}^2}{2m_b} = \omega_c \left(a^\dagger a + \frac{1}{2}\right),
\end{equation}
as it should. As stated in Ref.~\cite{bradlyn_kubo_2012} 
the angular momentum has a particularly simple representation in this basis. It is given by:
\begin{equation}
\hat{L}_z = b^\dagger b - a^\dagger a,
\end{equation}
and thus is the difference between the guiding center and Landau level value in a given state.

Suppose now that we have a potential which is a function of $\hat{r}^2$ i.e., a central potential, and want to express it in terms of the ladder operators $a$ and $b$. We note that:
\begin{align}
\hat{r}^2 &= \left(\hat{R}_i - l_B^2 \varepsilon_{ij}\hat{\pi}_j \right)\left(\hat{R}^i - l_B^2\varepsilon^{ik}\hat{\pi}^k \right)\\
&= \hat{R}^2 + l_B^4\hat{\pi}^2 - 2l_B^2\hat{R}^i\varepsilon^{ij}\hat{\pi}^j\\
&= 2l_B^2\left(a^\dagger a + b^\dagger b + iab - ia^\dagger b^\dagger + 1\right) \label{eqn:full_rsq_ladder}.
\end{align}
Since the $a^\dagger b^\dagger$ ($ab$) term raises (lowers) the Landau level and guiding center values together, it will not change the angular momentum. Thus, it is clear that $[\hat{r}^2, \hat{L}_z] = 0$ as they must. We note in particular that when projected to the lowest Landau level we will have
\begin{equation}
\s{P}_{LLL} \hat{r}^2 \s{P}_{LLL} = \frac{2}{m_b\omega_c}\left(b^\dagger b + 1\right),
\end{equation}
since any terms involving $a, a^\dagger$ cannot contribute. Thus, if we take the limit $\omega_c\rightarrow \infty$ with $l_B$ fixed i.e., assume that the kinetic term is sufficiently large to energetically project to the lowest Landau level, then:
\begin{equation}
V(\hat{r}^2)\approx \s{P}_{LLL}V(\hat{r}^2)\s{P}_{LLL} = V\left(2l_B^2(b^\dagger b + 1)\right).
\end{equation}
In particular, we note that a potential which has a minimum at $\hat{r}^2 = 2l_B^2(\ell + 1)$ will have a minimum when $b^\dagger b = \ell$, where we are taking $\ell \in \bb{Z}$.

We would now like to make the transformation to the second quantized basis. We first take:
\begin{equation}
|n_\alpha, m_\alpha\rangle = \frac{1}{\sqrt{n_\alpha! m_\alpha!}} a^{\dagger n_\alpha}_\alpha b^{\dagger m_\alpha}_\alpha|0,0\rangle \text{ where } \langle z|0,0\rangle \propto e^{-|z|^2/4l_B^2}
\end{equation}
to be the symmetric gauge state which is an eigenstate of both $a^\dagger_\alpha a_\alpha$ and $b^\dagger_\alpha b_\alpha$. Then we express the many-body wave function as
\begin{equation}
|\Psi\rangle = \sum_{n,m} d_{n,m}|n,m\rangle,
\end{equation}
where $d_{n,m}$ is the annihilation operator for a particle in the $|n,m\rangle$ state. We note that this basis is the orbital or Fock state basis rather than the coherent state basis introduced in Section~\ref{sec:coherent_states}. However, the physics we derive will be the same, as we will comment on.

In this basis we see that:
\begin{align}
\hat{N} &= \sum_{n,m} d^\dagger_{n,m}d_{n,m}\\
\hat{L}_z &= \sum_{n,m}(m-n)d^\dagger_{n,m}d_{n,m}.
\end{align}
In particular, we note that if we restrict to the one-dimensional sector where $m = \ell + n$, then $\hat{L}_z = \ell \hat{N}$.

For more complicated expressions we need the fact that
\begin{equation}
a^\dagger |n,m\rangle = \sqrt{n+1}|n+1,m\rangle,
\end{equation}
and likewise with $b^\dagger$. Then we see that, e.g., 
\begin{equation}
\hat{r}^2 = 2l_B^2\sum_{n,m} \left[(n + m + 1)d^\dagger_{n,m}d_{n,m} + \sqrt{(n+1)(m+1)}\left(id^\dagger_{n+1,m+1}d_{n,m}+\mathrm{h.c.}\right)\right].
\end{equation}
Particularly important will be the shear generators expressed in this basis. These are given as Eq.~(5.19) in the symmetric gauge basis~\cite{bradlyn_kubo_2012}.
Expressed in the second quantized basis, they will be given by:
\begin{align}
\hat{J}^{\mathrm{sh}}_{ij} = \sum_{n,m} &\frac{i}{4}\Bigl(\sqrt{(n+2)(n+1)} d^\dagger_{n+2,m}d_{n,m} - \sqrt{n(n-1)} d^\dagger_{n-2,m}d_{n,m}\nonumber \\
&- \sqrt{m(m-1)}d^\dagger_{n,m-2}d_{n,m} + \sqrt{(m+2)(m+1)}d^\dagger_{n,m+2}d_{n,m}\Bigr)\sigma^z_{ij}\nonumber \\
&-\frac{1}{4}\Bigl(\sqrt{(n+2)(n+1)} d^\dagger_{n+2,m}d_{n,m} + \sqrt{n(n-1)} d^\dagger_{n-2,m}d_{n,m}\nonumber \\
&- \sqrt{m(m-1)}d^\dagger_{n,m-2}d_{n,m} - \sqrt{(m+2)(m+1)}d^\dagger_{n,m+2}d_{n,m}\Bigr)\sigma^z_{ij}\nonumber \\
&+\frac{1}{2}(n-m)d^\dagger_{n,m}d_{n,m}\varepsilon_{ij} \label{eqn:shear_2nd_quant}.
\end{align}
The last term is the shear generator corresponding to rotations i.e., angular momentum, and thus will clearly not change the angular momentum. The other terms will change it by $\pm 2$.
 
\subsection{Hall viscosity with interactions in symmetric gauge}
\label{app:Hall_with_interactions}

We now consider adding the $s$-wave interaction term to the Hamiltonian:
\begin{equation}
\hat{H}_{\mathrm{int}} = g\sum_{\alpha,\beta} \delta^{(2)}(r_\alpha - r_\beta).
\end{equation}
We note that $g$ then has units of energy times length squared. In the symmetric gauge second quantized basis we see that this is equal to:
\begin{align}
\hat{H}_{\mathrm{int}} &= g\sum_{\substack{n_i,m_i\\ i=1,\ldots,4}} f(\{n_i,m_i\}) d^\dagger_{n_1,m_1}d^\dagger_{n_2,m_2}d_{n_3,m_3}d_{n_4,m_4}, \text{ where}\\
f(\{n_i,m_i\}) &\equiv \int d^2 r \ \psi^*_{n_1,m_1}(r)\psi^*_{n_2,m_2}(r)\psi_{n_3,m_3}(r)\psi_{n_4,m_4}(r),
\end{align}
and where $\psi_{n,m}(r)$ is equivalent to $\langle r|n,m\rangle$. In particular, if we project to the LLL, then the above integral will take a particularly simple form:
\begin{align}
f(\{0,m_i\}) =& \frac{1}{4\pi l_B^2}\delta_{m_1+m_2,m_3+m_4} \frac{(m_1+m_2)!}{2^{m_1+m_2}\sqrt{m_1!m_2!m_3!m_4!}},
\end{align}
and we see that:
\begin{align}
\hat{H}_{\mathrm{int}} &= \frac{g}{4\pi l_B^2}\sum_{m_0,m_1,m_2}\frac{1}{2^{m_0}}\sqrt{\begin{pmatrix} m_0\\ m_1\end{pmatrix}\begin{pmatrix} m_0\\ m_2\end{pmatrix}}d^\dagger_{0,m_0-m_2}d^\dagger_{0,m_2}d_{0,m_0-m_1}d_{0,m_1}
\end{align}
when projected to the lowest Landau level. The interaction thus annihilates a pair of bosons with total angular momentum $m_0$ and generates another pair with the same angular momentum. It is not surprising that this two-body interaction conserves angular momentum, as its expression in real space clearly does.

We then consider the full Hamiltonian composed of $\hat{H}_0$ given by Eq.~(\ref{eqn:H0_Mexican_hat}) in the second quantized basis and projected to the LLL, plus this term. It will be given by:
\begin{align}
\hat{H} &= U\sum_m (m-\ell)^2 d^\dagger_{m}d_m + \frac{g}{4\pi l_B^2}\sum_{m_0,m_1,m_2}\frac{1}{2^{m_0}}\sqrt{\begin{pmatrix} m_0\\ m_1\end{pmatrix}\begin{pmatrix} m_0\\ m_2\end{pmatrix}}d^\dagger_{m_0-m_2}d^\dagger_{m_2}d_{m_0-m_1}d_{m_1},
\end{align}
where we have dropped the Landau level index for notational simplicity. Let us further suppose that the state we start with is a condensate in the $m=\ell$ state i.e., the ``giant vortex state'' mentioned in the main text. We can then take $d_{\ell+q} = \sqrt{N_0}\delta_{q=0} + \beta_q$ where $N_0$ is the number of particles in the ground state and $q\neq 0$ for $\beta_q$. Then we will have that:
\begin{equation}
\hat{H} \approx \sum_{q\neq 0} Uq^2 \beta^\dagger_q \beta_q  + \frac{c}{2}N_0\left(4\beta^\dagger_q\beta_q + \beta_q\beta_{-q}+ \beta^\dagger_{q}\beta^\dagger_{-q}\right) \text{ where } c=\frac{g}{2l_B^2\sqrt{\pi^3\ell}},
\end{equation}
and where we assume that $\ell \gg 1,q$ \footnote{A similar equation was found in Ref.~\cite{roncaglia_rotating_2011}, but where the $U,g$ and other terms were chosen to scale in a certain way with the particle number to produce a very high angular momentum state.}. Here $c$ will have units of energy. Note that as we increase the angular momentum of the giant vortex state, we will decrease the relevance of the interaction terms. We can solve this via a Bougoliubov transformation:
\begin{align}
\hat{H} &= \sum_q \varepsilon_q \eta^\dagger_q \eta_q \text{ where } \varepsilon_q = \sqrt{\left(Uq^2 + cN\right)^2 - (cN)^2}, \\
\beta_q &= u_q\eta_q - v_q \eta^\dagger_{-q} \text{ where } u_q^2 - v_q^2 = 1, \text{ and}\\
v_q^2 &= \frac{1}{2}\left(\frac{Uq^2 + cN}{\varepsilon_q} - 1\right).
\end{align}

We note a few things. First, the ground state is clearly the vacuum of $\eta_q$ excitations. Next, the interaction scale will be given at leading order by the chemical potential, $E_{\mathrm{int}} = cN + \cdots \sim g\overline{n}$~\cite{roncaglia_rotating_2011}. If we require staying in the LLL limit, then this means that $cN \ll \omega_c$. In the limit $U\ll E_{\mathrm{int}} \ll \omega_c$, we see that these quasiparticles will have a superfluid dispersion given by:
\begin{equation}
\varepsilon_q \approx \sqrt{2cNU}|q| = \sqrt{\frac{gNU}{l_B^2\sqrt{\pi^3\ell}}}|q|.
\end{equation}
We note that this dispersion is exactly the same as would be obtained in Eq.~\eqref{eq: coherent state EFT} for the dispersion of the $\theta$ variables in the strong interaction limit. The speed of sound for that action would be given by $\sqrt{U/m_*}$, which allows us to see that we should identify $E_{\mathrm{int}}$ with the inverse mass scale $1/m^*$. This was commented on in the main text.

This is the familiar phonon dispersion of a superfluid; however we note that $q$ is an integer in this case. Thus, there is a finite gap to the nearest quasiparticle excitation, $\Delta = \varepsilon_{\pm 1}$. We note that since $E_{\mathrm{int}}$ is fixed with $N$, if $U$ is fixed with $N$, then this gap will stay constant as the number of particles increases. Further, if the experiment can control both the quadratic and quartic terms in $V_{\mathrm{eff}}$, then both $U$ and $\ell$ are freely tunable, so that this gap is tunable. Last, we note that the the fraction of particles outside of the condensate will be given by:
\begin{align}
\frac{N-N_0}{N} &= \frac{1}{N}\sum_{q\neq 0}\langle \beta^\dagger_q \beta_q\rangle_0\\
&= \frac{1}{N}\sum_{q\neq 0} v_q^2\\
&= \frac{1}{2N} \sum_{q\neq 0} \frac{q^2 + cN/U}{\sqrt{(q^2+cN/U)^2-(cN/U)^2}} -1\\
&\approx \sqrt{\frac{c}{2NU}}\sum_{q=1}^{\sqrt{cN/U}}\frac{1}{q} \text{ since } q\gg \sqrt{cN/U} \text{ converges},\\
&\approx \frac{1}{2}\sqrt{\frac{c}{2NU}}\log\left(\frac{cN}{U}\right).
\end{align}
Given the scaling of $c\sim 1/N$ and $U$ fixed, we see that this will quickly approach zero as $N$ becomes large.

We now want to evaluate the Hall viscosity in the interacting state. We first want to project the shear generators of Eq.~(\ref{eqn:shear_2nd_quant}) to the LLL and express them in terms of the operators $\beta_q$. When we do this we find that:
\begin{align}
\hat{J}^{\mathrm{sh}}_{ij} = \sum_q &\frac{i}{4}\left(-\sqrt{(\ell + q)(\ell + q-1)}\beta^\dagger_{q-2}\beta_q + \sqrt{(\ell + q + 2)(\ell + q + 1)} \beta^\dagger_{q+2}\beta_q\right)\sigma^z_{ij}\nonumber \\
&+\frac{1}{4}\left(\sqrt{(\ell + q)(\ell + q-1)}\beta^\dagger_{q-2}\beta_q + \sqrt{(\ell + q + 2)(\ell + q + 1)} \beta^\dagger_{q+2}\beta_q\right)\sigma^z_{ij}\nonumber \\
&-\frac{\ell + q}{2}\beta^\dagger_q \beta_q \varepsilon_{ij} \label{eqn:shear_beta}.
\end{align}
In particular, this reveals that
\begin{align}
\langle \hat{J}^{\mathrm{sh}}_{ij}\rangle_0 &= -\frac{1}{2}\langle \hat{L}_z\rangle_0 \varepsilon_{ij} = -\frac{\ell}{2}N\varepsilon_{ij} -\frac{1}{2}\varepsilon_{ij} \sum_{q\neq 0}qv_q^2,
\end{align}
since the first two strain generators change the angular momentum. We note that $v_q^2$ is symmetric with respect to $q\rightarrow -q$. Then since the sum runs from $q=-\ell$ to infinity, we see that:
\begin{align}
\langle \hat{J}^{\mathrm{sh}}_{ij}\rangle_0 &= -\frac{\ell}{2}N\varepsilon_{ij} -\frac{1}{4}\varepsilon_{ij} \sum_{q=\ell+1}^{\infty} q\left[\frac{q^2 + cN/U}{\sqrt{(q^2+cN/U)^2-(cN/U)^2}} -1\right]\\
&= -\frac{\ell}{2}N\varepsilon_{ij} -\frac{1}{4}\frac{E_{\mathrm{int}}}{U}\varepsilon_{ij}\cdot \begin{cases} \sim E_{\mathrm{int}}/U\ell^2 &\mbox{if } E_{\mathrm{int}}\ll U\ell^2, \\ \sim1 &\mbox{if } E_{\mathrm{int}} \gg U\ell^2.\end{cases}
\end{align}
We see that if the confining potential dominates i.e., if $E_{\mathrm{int}}\ll U\ell^2$, then we can expect the angular momentum to be given very closely by its value without interactions i.e., precisely $\ell N$. However, if interactions dominate, then we will have a correction to the angular momentum of order $E_{\mathrm{int}}/U$. Note that since $c$ and $U$ are both positive this can only enhance the angular momentum. Inspection of the above makes it clear why this is; it is only possible for fluctuations to decrease the angular momentum from $\ell$ to zero while they can increase it without bound. Thus, fluctuations will tend to enhance the angular momentum. Also, since $c\sim 1/N$ we see that this correction will not scale with $N$ so that in the large particle limit we will continue to have angular momentum $\ell N$ to a very good approximation.

Then from Ref.~\cite{bradlyn_kubo_2012} in the strong interaction limit we have
\begin{align}
\eta^{\mathrm{sh}}_{ijkl}(\omega) &= \frac{1}{2}\overline{n}\left(\ell + \frac{E_{\mathrm{int}}}{2NU}\right)\left(\delta_{il}\varepsilon_{kj} - \delta_{kj}\varepsilon_{il}\right) \nonumber \\
&+ \frac{i\omega}{A}\sum_{\nu}\left(\frac{\langle 0|\hat{J}_{ij}^{\mathrm{sh}}|\nu\rangle\langle \nu| \hat{J}_{kl}^{\mathrm{sh}}|0\rangle}{\omega - E_0 + E_\nu} - \frac{\langle 0|\hat{J}_{kl}^{\mathrm{sh}}|\nu\rangle\langle \nu| \hat{J}_{ij}^{\mathrm{sh}}|0\rangle}{\omega + E_0 - E_\nu}\right). \label{eqn:Lehmann_shear}
\end{align}
The latter term contains the corrections to adiabatic response, as discussed in \cite{bradlyn_kubo_2012}. We note that only the first two terms of Eq.~(\ref{eqn:shear_beta}) will contribute to this term. Since they both involve changing the angular momentum by two they will pick up a factor of
\begin{equation}
\frac{1}{\omega \pm \varepsilon_{\pm 2}} \text{ where } \varepsilon_{\pm 2} \sim 2\sqrt{2E_{\mathrm{int}}U}
\end{equation}
in the strong interacting limit. As mentioned before, because $E_{\mathrm{int}}$ and $U$ do not scale with $N$, we see that these gaps will stay finite in the large $N$ limit. Thus, the zero-frequency Hall response will not pick up any corrections since the system will continue to possess many-body energy gaps. We conclude that the effect of interactions in this case is merely to renormalize the value of $\ell$ upwards by an amount proportional to $1/N$.

\subsection{Hall viscosity in the Landau gauge}
\label{app:Hall_Landau}

We begin with the Hamiltonian Eq.~\eqref{eqn:BEC_Ham}, but now in the Landau gauge. In this case $\s{A} = 2m_b\omega_{\mathrm{trap}} x \hat{y}$ and $V(r_\alpha) = m_b\varepsilon \omega_{\mathrm{trap}}^2 x^2$, where $\varepsilon$ will control the strength of squeezing. This can be generated by an anisotropic trap as in Ref.~\cite{fletcher_geometric_2021}. The Hamiltonian is therefore translationally invariant in the $y$ direction and its eigenfunctions will be of the form 
\begin{equation}
\psi_{0,k} \propto e^{iky} \exp\left[-\frac{1}{2}\left(\frac{x}{l_B} + kl_B\right)^2\right],
\end{equation}
where we have restricted to the LLL for ease. We can see that the potential will favor states which are centered around $x=0$ i.e., those with $k=0$. Indeed, when expressed in this basis the Hamiltonian possesses a $k^2$ term which energetically favors the $k=0$ state \cite{sinha_two-dimensional_2005,fletcher_geometric_2021}. There is one important difference about this Hamiltonian to note before interactions are added. Unlike the symmetric gauge case, Eq.~\eqref{eqn:H0_Mexican_hat}, there is no gap to excitations in the thermodynamic limit. This is because $k$ is quantized in units of $2\pi/L_y$, so as $L_y\rightarrow \infty$ the momentum can continuously approach zero. This difference means that adiabatic response cannot be used to find the Hall viscosity. We see that it will further cause the Hall viscosity to diverge.

Let us now add a repulsive $s$-wave interaction as a perturbation to the Hamiltonian. We can do this in the second quantized basis as:
\begin{equation}
\psi(x,y) = \sum_{n,k} a_n(k)\psi_{n,k}(x,y), \label{eqn:2_quant_Landau}
\end{equation}
where $a_n(k)$ is the annihilation operator of a particle in the $n$th Landau level with momentum $k$ in the $y$-direction, and $\psi_{n,k}$ is the corresponding wave function. If we insert this into the Hamiltonian and restrict to the LLL, then the Hamiltonian becomes
\begin{align}
H_{LLL} &= \sum_k \frac{k^2}{2\tilde{m}}a^\dagger(k) a(k)\nonumber \\
&+ \frac{g}{\sqrt{2\pi}l_BL_y}\sum_{k,k',q} a^\dagger(k+q)a^\dagger(k'-q)a(k)a(k') \exp\left(-\frac{l_B^2}{2}[(k-k'+q)^2+q^2]\right),
\end{align}
where we are taking $a_n(k) = \delta_{n0}a(k)$ and where $\tilde{m} \propto m_b/\varepsilon$. We thus see that our Hamiltonian has reduced to a one-dimensional Hamiltonian labeled by $y$ momentum and exhibiting a four-boson interaction term \cite{sinha_two-dimensional_2005}. The four-boson term has an odd momentum dependence arising from the fact that the wave functions $\psi_{n,k}$ are shifted along the $x$ axis by a term proportional to $k$ which decreases their overlap with wave functions not sharing their momentum.

We now assume the kinetic term dominates over the interaction term. In this limit the bosons governed by $H_{LLL}$ can be assumed to condense into the $k=0$ state, \footnote{Even though there are no true condensates in the thermodynamic limit in one dimension \cite{sinha_two-dimensional_2005} shows this system can be treated as one for reasonable system sizes.} so we may write that $a(k) = \delta_{k0}\sqrt{N - \delta n} + \tilde{a}(k)$. This is just the statement that we have a LLL superfluid which is squeezed into the $k=0$ Landau gauge state. For self-consistency we take $\tilde{a}(k)$ to be the boson annihilation operator for $k\neq 0$ and thus $\delta n = \sum_{k\neq 0}\tilde{a}^\dagger(k) \tilde{a}(k)$ is second order in $\tilde{a}(k)$. Linearizing $H_{LLL}$ and diagonalizing it via a Bogoliubov substitution gives eigenenergies:
\begin{equation}
\varepsilon(k)^2 = \left[\frac{k^2}{2\tilde{m}} + \frac{2gn}{\sqrt{2\pi}}\left(2e^{-l_B^2k^2/2}-1\right)\right]^2 - \frac{2g^2n^2}{\pi}e^{-2l_B^2k^2},
\end{equation}
where $n = N_b/L_yl_B$ is the two-dimensional number density of the condensate \cite{sinha_two-dimensional_2005}. Here $\varepsilon(k)$ is a universal function of $kl_B$ and $\beta = 2gn\tilde{m}l_B^2 \sim \tilde{m}gnl_B^2$. As we take $\varepsilon$ to zero, $\beta\rightarrow \infty$, $\varepsilon(k)^2$ will develop a zero at a finite $k_c$ and eventually become negative. This represents the presence of a roton excitation and indicates that the $k=0$ ground state is unstable to filling other values of $k$. Physically this is clear as we will no longer be squeezing into the $k=0$ state. We assume that $\varepsilon$ is chosen to be sufficiently large such that we are far from this instability. This will correspond to the limit $\tilde{m}gn l_B^2\ll 1$. In this limit the interaction terms will be irrelevant except for in the region $k\ll 1/l_B$ where they will lead to a linear dispersion, $\varepsilon(k) \sim c_s k$ with $c_s = (2/\pi)^{1/4} \sqrt{gn/\tilde{m}}$. In fact as soon as $k \sim 1/\xi \equiv \sqrt{\tilde{m} gn} \ll 1/l_B$ there will be higher order corrections to this linear dispersion.

Having used the Bogoliubov transformation to put the Hamiltonian in quadratic form we can now address the form the Hall viscosity takes. It can be checked that in the second quantized basis given by Eq.~\eqref{eqn:2_quant_Landau} we have:
\begin{equation}
\frac{\hat{L}_z}{N_b} = \frac{1}{2} - \frac{1}{2}\sum_k \left[(kl_B)^2 a^\dagger(k)a(k) + \frac{1}{l_B^2} \partial_k a^\dagger(k) \partial_k a(k)\right],
\end{equation}
where the constant $1/2$ term arises from the orbital coordinate, while the latter term is due to the guiding center coordinate and responds to the geometry of the system. The appearance of derivatives with respect to $k$ is due to the presence of $y$ in the original expression. The expectation value of these terms can be evaluated using the Bogoliubov analysis and it can be seen that $\langle \hat{L}_z\rangle/N_b \sim  L_y^4/l_B^4$. The Hall viscosity of the strip geometry will thus na\"{i}vely diverge in the thermodynamic limit. This is once again driven by the fact that $k$ can be tuned continuously to zero.

Of course the full Kubo calculation of the Hall viscosity involves more than just computing the expectation value of angular momentum. As Eq.~\eqref{eqn:Lehmann_shear} reveals, there will also be a contribution from any states $|\nu\rangle$ with $E_\nu = E_0$ in the thermodynamic limit. But, as we have noted, in the thermodynamic limit there is no gap in $\varepsilon(k)$. Thus, unlike the symmetric squeezed case there will be contributions from this term. These gapless effects will regularize the divergence of the first term. We do not perform this calculation, but merely note that the Kubo formula is essential in this case to capture the full Hall viscosity regularized by interactions.

\section{The spin connection and the action of Galilean symmetry}
\label{app:galilean_symm}
We first note that for a metric given by $g_{ij} = \delta_{ij} + \delta g_{ij}$ the spin connection will be given by:
\begin{align}
\omega_t&=\frac{1}{2}\varepsilon^{jk}\delta g_{ij}\,\partial_t\,\delta g_{ik}\,,\\
\omega_i&=-\frac{1}{2}\varepsilon^{jk}\partial_j\delta g_{ik}\,.
\end{align}
We can now consider the transformation of the fields under the space and time dependent translation $r^i \rightarrow r^i + \xi^i(r,t)$. Galilean invariance requires that~\cite{hoyos_effective_2014}:
\begin{align}
\delta n &= -\xi^k \partial_k n,\\
\delta \theta &= -\xi^k\partial_k \theta,\\
\delta \s{A}_t &= -\xi^k\partial_k \s{A}_t + \s{A}_k\dot{\xi}^k,\\
\delta \s{A}^i &= -\xi^k \partial_k \s{A}^i + \s{A}_k\partial^i \xi^k + m_b\dot{\xi}^i,\\
\delta g_{ij} &= -\xi^k\partial_k g_{ij} -g_{ik}\partial_j\xi^k - g_{kj}\partial_i\xi^k.
\end{align}
It can be checked that this means $\omega_i$ will transform as a one-form, but $\omega_t$ will not. Instead the combination of $\omega_t + B/2m_b$ will transform as a one-form, where we note that the factor of mass is needed for dimensions to agree. To account for this Ref.~\cite{hoyos_effective_2014} incorporated $B/2m_b$ into their definition of $\omega_t$. However, one can see from Eq.~\eqref{eqn:action_BH_Hl} that replacing $\omega_t$ with $\omega_t + B/2m_b$ will introduce \textit{exactly} the gyromagnetic term written in Eq.~\eqref{eq: gyro}. The physics of this term is clearer if it is separated out, thus our approach.

Having thus introduced the gyromagnetic term, the spin connection $\omega_t$ will transform as a one-form. However, the theory as written is still not invariant under the Galilean transformation written above, as the phase fluctuations carry the mass $m_*$ instead of $m_b$. The introduction of the fluctuating ``velocity field," $\upsilon^i$, will fix this when it is integrated out, as discussed in the main text. The need for this can also be seen, as discussed in the main text, to be due to the need to enforce Eq.~\eqref{eq: correct momentum}. When both of these things are done the full theory will be invariant under Galilean symmetry.

\section{Details on vortex dynamics}
\label{app:vortex dynamics}

\subsection{Density profiles of a single vortex}
\label{app:dens_single_vortex}

We can now use Eq.~(\ref{eqn:cons_mom_coherent}) to find the density profile of a given vortex. We do so in zero magnetic field for ease. A vortex of charge $k$ will have phase velocity given by:
\begin{equation}
v^r_{\mathrm{phase}} = 0 \text{ and } v^\phi_{\mathrm{phase}} = -\frac{k}{m_br} \label{eqn:vel_vortex}.
\end{equation}
By symmetry we can expect that $\rho_m = \rho_m(r)$. We thus note that $\partial_i V^i = 0$ and $V^i\partial_i \rho_m = 0$ so the conservation of mass equation of motion is automatically satisfied. Satisfying the conservation of momentum equation of motion then reduces to:
\begin{align}
\frac{V^{\phi,2}}{r} - \frac{v^{\phi,2}_{\mathrm{edge}}}{r} &= -\frac{c_s^2}{\rho_{0,m}}\partial^r \rho_m + \frac{1}{2m_b m_*}\partial^r\left(\frac{1}{\sqrt{\rho_m}}\nabla^2 \sqrt{\rho_m}\right),
\end{align}
where we have kept the leading speed of sound term in the pressure. If we simplify, then we see that:
\begin{align}
\frac{k^2}{r^3} &= \frac{k\ell}{r^2 \rho_m}\partial^r \rho_m + \frac{c_s^2}{\rho_{0,m}}\partial^r \rho_m - \frac{1}{2m_bm_*}\partial^r\left(\frac{1}{\sqrt{\rho_m}}\nabla^2 \sqrt{\rho_m}\right) \label{eqn:dens_profile}.
\end{align}
Far from the vortex we can approximate $\rho_m(r) = \rho_{0,m} + \delta \rho_m(r)$. Then the last term will be higher order in derivatives and can be neglected. In particular, we note that:
\begin{equation}
\rho_m(r) = \rho_{0,m}\left[1 - \frac{k^2}{2m_b^2c_s^2r^2} + \s{O}\left(\frac{1}{r^4}\right)\right]
\end{equation}
is a consistent solution for $r$ large. We note that this is of the form found in Ref.~\cite{hoyos_effective_2014}.

If we now numerically integrate Eq.~(\ref{eqn:dens_profile}) with the boundary conditions: $\rho(0) = 0, \rho(\infty) = \rho_0, \rho'(\infty) = 0$, then we will produce the full density profiles displayed in Fig.~\ref{fig:dens_profiles}. Closer to $r=0$ we can see that the term proportional to $c_s^2$ will drop out of Eq.~(\ref{eqn:dens_profile}) and the latter term will be necessary to stabilize the density.

\subsection{Simulation of a vortex dipole}
\label{app:vort_dipole_sim}


We would like to simulate Eq.~(\ref{eqn:cons_mom_coherent}) with $\omega_c=0$ along with the conservation of mass equation of motion. It is very challenging to numerically simulate these in the presence of quantized vortices, as their velocity diverges near the core. In the case of a superfluid without Hall viscosity this can be accomplished via a mapping to a complex field $\psi = \sqrt{n} e^{i\theta}$. Quantized vortices can then be expressed as points where there is a nontrivial winding of $\theta$ and the dynamics expressed in terms of $\psi$ will no longer possess singularities \cite{musser_starting_2019}. It is not immediately clear how to accomplish this in the context of Eq.~(\ref{eqn:cons_mom_coherent}), where the Hall viscosity term provides an obstruction to expressing this equation as a dynamical equation for $\theta$ directly.

We will thus use Lamb-Oseen vortices \cite{banerjee_odd_2017} to avoid the issue of velocity divergence altogether. We will specifically take their velocity to be given by:
\begin{equation}
v^r_{\mathrm{phase}} = 0 \text{ and } v^\phi_{\mathrm{phase}} = -\frac{k}{m_br}\left(1 - e^{-r^2/r_0^2}\right) \label{eqn:vel_LO}.
\end{equation}
Comparison to Eq.~(\ref{eqn:vel_vortex}) reveals that for $r \gtrsim r_0$ the velocity around a Lamb-Oseen vortex will be identical to that of a quantized vortex, while for $r \lesssim r_0$ the velocity will approach zero linearly with $r$. The vorticity, $\Omega_{\mathrm{phase}}$, of this vortex will be a Gaussian. As $r_0\rightarrow 0$ this Gaussian will approach the delta function of a quantized vortex. 

\begin{figure}
    \centering
    \includegraphics[width=0.6\textwidth]{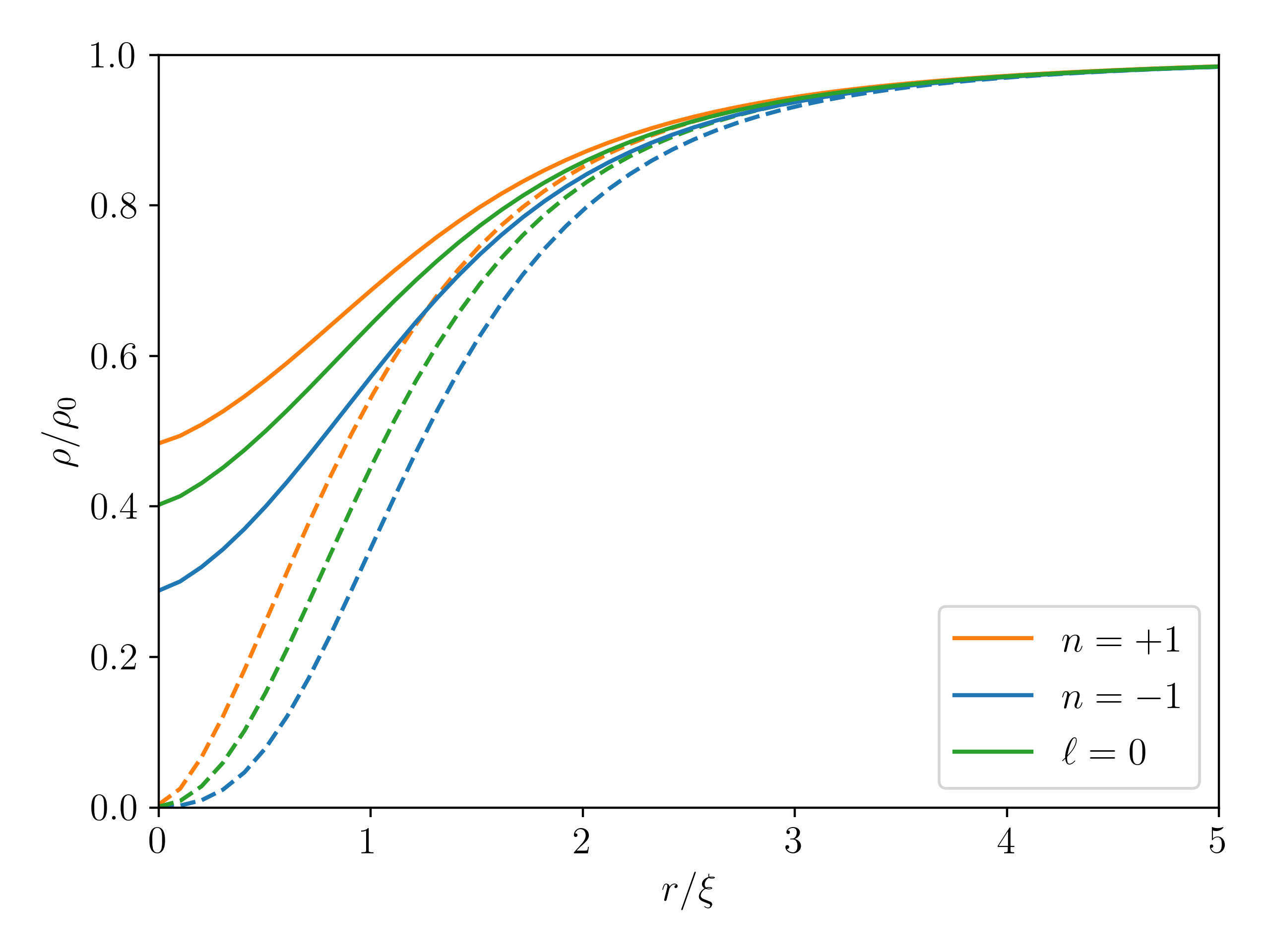}
    \caption{Density profiles of Lamb-Oseen vortices in a superfluid with $\ell = 1$. The solid lines are density profiles where $r_0 = 0.85\xi$, while the dotted lines have $r_0 = 0.01\xi$. The latter are close to those shown in Fig.~\ref{fig:dens_profiles} for the full quantized vortices, while the former avoid the divergence issues caused by low density. For either value of $r_0$ the density of a vortex in a fluid without Hall viscosity interpolates between those with $n = \pm 1$.}
    \label{fig:LO_profiles}
\end{figure}

The density profile of these vortices will need to be reworked using Eq.~(\ref{eqn:dens_profile}) with the new phase velocity. This is shown in Fig.~\ref{fig:LO_profiles}, where we chose boundary conditions $\rho_m'(0) = 0, \rho_m(\infty) = \rho_{0,m}, \rho'_{0,m}(\infty) = 0$. Low values of $r_0$ can be seen to recover the form of the density profiles in Fig.~\ref{fig:dens_profiles}, while higher values will avoid zero density at the core of the vortex. In either case the density profiles are nearly identical for $r\gtrsim 2\xi$. 

In all simulations discussed we took $r_0 = 0.85\xi$ with the two vortices placed a distance $d=5\xi$ apart. As can be seen in Fig.~\ref{fig:LO_profiles} this meant that the vortices were sufficiently well separated that their density did not substantially deviate from the quantized vortex density. The simulation was then initialized by taking:
\begin{equation}
\rho = \rho_{n=-1}(x=-d/2)\rho_{n=+1}(x=+d/2) \text{ and } v_{\mathrm{phase}} = v_{n=-1}(x=-d/2) + v_{n=+1}(x=d/2),
\end{equation}
where these were the density and phase profiles of a single vortex at this location. This provided an approximation of the density and velocity in the presence of the vortex dipole. We note that if the densities multiply initially, then from Eq.~(\ref{eqn:defn_big_V}), the initial velocities $V$ of the vortex dipole will also add. 

With this initial condition we then numerically integrated Eq.~(\ref{eqn:cons_mom_coherent}) and the conservation of mass equation of motion via a second order Runge-Kutta process. The nondimensionalized equations of motion were given by:
\begin{align}
\frac{\partial \rho_m}{\partial t} + \partial_i(\rho_m V^i) &= 0,\\
\frac{\partial V^i}{\partial t} +V^j\partial_j V^i - v^j_{\mathrm{edge}}\partial^j v^i_{\mathrm{edge}} &= -\partial^i \rho_m + \frac{m_b}{2m_*}\partial^i\left(\frac{1}{\sqrt{\rho_m}}\nabla^2\sqrt{\rho_m}\right) \nonumber\\
&- \frac{\ell^2}{4}\partial^i\nabla^2 \log(\rho_m) + \frac{\ell}{2}\varepsilon^{ij}\nabla^2 V^j,
\end{align}
here we have taken $\mathfrak{p} = c_s^2 \rho_m^2/2$. Units of length are in terms of $\xi = 1/m_bc_s$, units of time are in terms of $\xi/c_s$, density is in terms of $\rho_{0,m}$ i.e., the density far from a vortex, and velocity is in terms of $c_s$. For all simulations we took either $\ell = 0$ or $\ell = 1$. The low value of $\ell$ was so that both $n=\pm 1$ vortices had a density at their center sufficiently above zero so as to not cause problems with the simulation. We also note that the parameter $m_b/m_*$ is free and not set by the effective field theory worked out in Section~\ref{sec:coherent_states}. For our simulations we set $m_b = m_*$ for ease. Choosing a different value would have changed the density profile near the center of the vortex, where the term $m_b/m_*$ is larger, but should not affect other physics.

The simulations were performed in a box of size $40\xi \times 40\xi$ with periodic boundary conditions \footnote{In principle, these boundary conditions may require addressing the velocity due to the ``image vortex'' \cite{acheson_elementary_1990}; however, as the box size is increased this effect will become negligible.}. The spatial discretization was taken to be $\Delta r = 0.2\xi$ to resolve vortex cores and the time discretization was chosen to be $\Delta t = 0.001\xi/c_s < 0.1*\Delta r^2$ to ensure numerical stability \cite{musser_starting_2019}. The imperfect initial conditions generated some sound waves as the density equilibrated. It may be possible to remove these via the inclusion of a small amount of shear viscous dissipation. We did not do this, however, as it would cause diffusion of the Lamb-Oseen vortex \cite{acheson_elementary_1990}. Nonetheless, we did not find that the sound waves had a serious effect beyond causing small oscillations of the vortex dipole.

After performing the second order Runge-Kutta integration for a time $t = 50\xi/c_s$ the simulation was halted and the density and velocity values were saved at intervals of $0.1\xi/c_s$. The \verb|locate| function of the python package \verb|trackpy| was then used to track the trajectories of the vortices. In the case of $\ell = 0$ the vortices were tracked using their density depletions. In the case of $\ell = 1$ the density depletion at the core of the $n=+1$ vortex was smaller, as seen in Fig.~\ref{fig:LO_profiles}. This made tracking the $n=+1$ vortex with its density depletion more challenging and lead to greater noise. Instead, we computed the phase vorticity, $|\Omega_{\mathrm{phase}}|$, which was initially two Gaussian peaks at each of the vortices. The phase vorticity stayed sharply peaked at each of the vortices throughout the simulation, allowing for easy tracking. After obtaining the vortex trajectories with this method we compared them to the density and found that the core of the vortex identified this way differed from the core identified via $\Omega_{\mathrm{phase}}$ by less than $2\xi$ at all times. We thus used the trajectories identified via the phase vorticity.

The results of the simulation are shown in Fig.~\ref{fig:dipole_rot_sim} and are discussed in the surrounding text.

\nocite{apsrev41Control}
\bibliographystyle{apsrev4-1}
\bibliography{Draft.bib}

\begin{thebibliography}{61}%
\makeatletter
\providecommand \@ifxundefined [1]{%
 \@ifx{#1\undefined}
}%
\providecommand \@ifnum [1]{%
 \ifnum #1\expandafter \@firstoftwo
 \else \expandafter \@secondoftwo
 \fi
}%
\providecommand \@ifx [1]{%
 \ifx #1\expandafter \@firstoftwo
 \else \expandafter \@secondoftwo
 \fi
}%
\providecommand \natexlab [1]{#1}%
\providecommand \enquote  [1]{``#1''}%
\providecommand \bibnamefont  [1]{#1}%
\providecommand \bibfnamefont [1]{#1}%
\providecommand \citenamefont [1]{#1}%
\providecommand \href@noop [0]{\@secondoftwo}%
\providecommand \href [0]{\begingroup \@sanitize@url \@href}%
\providecommand \@href[1]{\@@startlink{#1}\@@href}%
\providecommand \@@href[1]{\endgroup#1\@@endlink}%
\providecommand \@sanitize@url [0]{\catcode `\\12\catcode `\$12\catcode
  `\&12\catcode `\#12\catcode `\^12\catcode `\_12\catcode `\%12\relax}%
\providecommand \@@startlink[1]{}%
\providecommand \@@endlink[0]{}%
\providecommand \url  [0]{\begingroup\@sanitize@url \@url }%
\providecommand \@url [1]{\endgroup\@href {#1}{\urlprefix }}%
\providecommand \urlprefix  [0]{URL }%
\providecommand \Eprint [0]{\href }%
\providecommand \doibase [0]{http://dx.doi.org/}%
\providecommand \selectlanguage [0]{\@gobble}%
\providecommand \bibinfo  [0]{\@secondoftwo}%
\providecommand \bibfield  [0]{\@secondoftwo}%
\providecommand \translation [1]{[#1]}%
\providecommand \BibitemOpen [0]{}%
\providecommand \bibitemStop [0]{}%
\providecommand \bibitemNoStop [0]{.\EOS\space}%
\providecommand \EOS [0]{\spacefactor3000\relax}%
\providecommand \BibitemShut  [1]{\csname bibitem#1\endcsname}%
\let\auto@bib@innerbib\@empty
\bibitem [{\citenamefont {Avron}\ \emph {et~al.}(1995)\citenamefont {Avron},
  \citenamefont {Seiler},\ and\ \citenamefont {Zograf}}]{avron_viscosity_1995}%
  \BibitemOpen
  \bibfield  {author} {\bibinfo {author} {\bibfnamefont {J.~E.}\ \bibnamefont
  {Avron}}, \bibinfo {author} {\bibfnamefont {R.}~\bibnamefont {Seiler}}, \
  and\ \bibinfo {author} {\bibfnamefont {P.~G.}\ \bibnamefont {Zograf}},\
  }\bibfield  {title} {\enquote {\bibinfo {title} {Viscosity of quantum {Hall}
  fluids},}\ }\href {\doibase 10.1103/PhysRevLett.75.697} {\bibfield  {journal}
  {\bibinfo  {journal} {Physical Review Letters}\ }\textbf {\bibinfo {volume}
  {75}},\ \bibinfo {pages} {697--700} (\bibinfo {year} {1995})}\BibitemShut
  {NoStop}%
\bibitem [{\citenamefont {Avron}(1998)}]{avron_odd_1998}%
  \BibitemOpen
  \bibfield  {author} {\bibinfo {author} {\bibfnamefont {J.~E.}\ \bibnamefont
  {Avron}},\ }\bibfield  {title} {\enquote {\bibinfo {title} {Odd viscosity},}\
  }\href {\doibase 10.1023/A:1023084404080} {\bibfield  {journal} {\bibinfo
  {journal} {Journal of Statistical Physics}\ }\textbf {\bibinfo {volume}
  {92}},\ \bibinfo {pages} {543--557} (\bibinfo {year} {1998})}\BibitemShut
  {NoStop}%
\bibitem [{\citenamefont {Read}(2009)}]{Read2009}%
  \BibitemOpen
  \bibfield  {author} {\bibinfo {author} {\bibfnamefont {N.}~\bibnamefont
  {Read}},\ }\bibfield  {title} {\enquote {\bibinfo {title} {Non-{Abelian}
  adiabatic statistics and {Hall} viscosity in quantum {Hall} states and
  ${p}_{x}+i{p}_{y}$ paired superfluids},}\ }\href {\doibase
  10.1103/PhysRevB.79.045308} {\bibfield  {journal} {\bibinfo  {journal} {Phys.
  Rev. B}\ }\textbf {\bibinfo {volume} {79}},\ \bibinfo {pages} {045308}
  (\bibinfo {year} {2009})}\BibitemShut {NoStop}%
\bibitem [{\citenamefont {Read}\ and\ \citenamefont {Rezayi}(2011)}]{Read2011}%
  \BibitemOpen
  \bibfield  {author} {\bibinfo {author} {\bibfnamefont {N.}~\bibnamefont
  {Read}}\ and\ \bibinfo {author} {\bibfnamefont {E.~H.}\ \bibnamefont
  {Rezayi}},\ }\bibfield  {title} {\enquote {\bibinfo {title} {Hall viscosity,
  orbital spin, and geometry: Paired superfluids and quantum {Hall} systems},}\
  }\href {\doibase 10.1103/PhysRevB.84.085316} {\bibfield  {journal} {\bibinfo
  {journal} {Phys. Rev. B}\ }\textbf {\bibinfo {volume} {84}},\ \bibinfo
  {pages} {085316} (\bibinfo {year} {2011})}\BibitemShut {NoStop}%
\bibitem [{\citenamefont {Hoyos}\ \emph {et~al.}(2014)\citenamefont {Hoyos},
  \citenamefont {Moroz},\ and\ \citenamefont {Son}}]{hoyos_effective_2014}%
  \BibitemOpen
  \bibfield  {author} {\bibinfo {author} {\bibfnamefont {C.}~\bibnamefont
  {Hoyos}}, \bibinfo {author} {\bibfnamefont {S.}~\bibnamefont {Moroz}}, \ and\
  \bibinfo {author} {\bibfnamefont {D.~T.}\ \bibnamefont {Son}},\ }\bibfield
  {title} {\enquote {\bibinfo {title} {Effective theory of chiral
  two-dimensional superfluids},}\ }\href {\doibase 10.1103/PhysRevB.89.174507}
  {\bibfield  {journal} {\bibinfo  {journal} {Physical Review B}\ }\textbf
  {\bibinfo {volume} {89}},\ \bibinfo {pages} {174507} (\bibinfo {year}
  {2014})}\BibitemShut {NoStop}%
\bibitem [{\citenamefont {Moroz}\ \emph {et~al.}(2018)\citenamefont {Moroz},
  \citenamefont {Hoyos}, \citenamefont {Benzoni},\ and\ \citenamefont
  {Son}}]{Moroz2018}%
  \BibitemOpen
  \bibfield  {author} {\bibinfo {author} {\bibfnamefont {S.}~\bibnamefont
  {Moroz}}, \bibinfo {author} {\bibfnamefont {C.}~\bibnamefont {Hoyos}},
  \bibinfo {author} {\bibfnamefont {C.}~\bibnamefont {Benzoni}}, \ and\
  \bibinfo {author} {\bibfnamefont {D.~T.}\ \bibnamefont {Son}},\ }\bibfield
  {title} {\enquote {\bibinfo {title} {{Effective field theory of a vortex
  lattice in a bosonic superfluid}},}\ }\href {\doibase
  10.21468/SciPostPhys.5.4.039} {\bibfield  {journal} {\bibinfo  {journal}
  {SciPost Phys.}\ }\textbf {\bibinfo {volume} {5}},\ \bibinfo {pages} {039}
  (\bibinfo {year} {2018})}\BibitemShut {NoStop}%
\bibitem [{\citenamefont {Moroz}\ and\ \citenamefont {Son}(2019)}]{Moroz2019}%
  \BibitemOpen
  \bibfield  {author} {\bibinfo {author} {\bibfnamefont {S.}~\bibnamefont
  {Moroz}}\ and\ \bibinfo {author} {\bibfnamefont {D.~T.}\ \bibnamefont
  {Son}},\ }\bibfield  {title} {\enquote {\bibinfo {title} {Bosonic superfluid
  on the lowest {Landau} level},}\ }\href {\doibase
  10.1103/PhysRevLett.122.235301} {\bibfield  {journal} {\bibinfo  {journal}
  {Phys. Rev. Lett.}\ }\textbf {\bibinfo {volume} {122}},\ \bibinfo {pages}
  {235301} (\bibinfo {year} {2019})}\BibitemShut {NoStop}%
\bibitem [{\citenamefont {Nie}\ \emph {et~al.}(2020)\citenamefont {Nie},
  \citenamefont {Huang},\ and\ \citenamefont {Yao}}]{Nie2020}%
  \BibitemOpen
  \bibfield  {author} {\bibinfo {author} {\bibfnamefont {W.}~\bibnamefont
  {Nie}}, \bibinfo {author} {\bibfnamefont {W.}~\bibnamefont {Huang}}, \ and\
  \bibinfo {author} {\bibfnamefont {H.}~\bibnamefont {Yao}},\ }\bibfield
  {title} {\enquote {\bibinfo {title} {Edge current and orbital angular
  momentum of chiral superfluids revisited},}\ }\href {\doibase
  10.1103/PhysRevB.102.054502} {\bibfield  {journal} {\bibinfo  {journal}
  {Phys. Rev. B}\ }\textbf {\bibinfo {volume} {102}},\ \bibinfo {pages}
  {054502} (\bibinfo {year} {2020})}\BibitemShut {NoStop}%
\bibitem [{\citenamefont {Rose}\ \emph {et~al.}(2020)\citenamefont {Rose},
  \citenamefont {Golan},\ and\ \citenamefont {Moroz}}]{Rose2020}%
  \BibitemOpen
  \bibfield  {author} {\bibinfo {author} {\bibfnamefont {F.}~\bibnamefont
  {Rose}}, \bibinfo {author} {\bibfnamefont {O.}~\bibnamefont {Golan}}, \ and\
  \bibinfo {author} {\bibfnamefont {S.}~\bibnamefont {Moroz}},\ }\bibfield
  {title} {\enquote {\bibinfo {title} {{Hall viscosity and conductivity of
  two-dimensional chiral superconductors}},}\ }\href {\doibase
  10.21468/SciPostPhys.9.1.006} {\bibfield  {journal} {\bibinfo  {journal}
  {SciPost Phys.}\ }\textbf {\bibinfo {volume} {9}},\ \bibinfo {pages} {006}
  (\bibinfo {year} {2020})}\BibitemShut {NoStop}%
\bibitem [{\citenamefont {Furusawa}\ \emph {et~al.}(2021)\citenamefont
  {Furusawa}, \citenamefont {Fujii},\ and\ \citenamefont
  {Nishida}}]{Furusawa2021}%
  \BibitemOpen
  \bibfield  {author} {\bibinfo {author} {\bibfnamefont {T.}~\bibnamefont
  {Furusawa}}, \bibinfo {author} {\bibfnamefont {K.}~\bibnamefont {Fujii}}, \
  and\ \bibinfo {author} {\bibfnamefont {Y.}~\bibnamefont {Nishida}},\
  }\bibfield  {title} {\enquote {\bibinfo {title} {Hall viscosity in the $a$
  phase of superfluid $^{3}\mathrm{He}$},}\ }\href {\doibase
  10.1103/PhysRevB.103.064506} {\bibfield  {journal} {\bibinfo  {journal}
  {Phys. Rev. B}\ }\textbf {\bibinfo {volume} {103}},\ \bibinfo {pages}
  {064506} (\bibinfo {year} {2021})}\BibitemShut {NoStop}%
\bibitem [{\citenamefont {Berdyugin}\ \emph {et~al.}(2019)\citenamefont
  {Berdyugin}, \citenamefont {Xu}, \citenamefont {Pellegrino}, \citenamefont
  {Krishna~Kumar}, \citenamefont {Principi}, \citenamefont {Torre},
  \citenamefont {Ben~Shalom}, \citenamefont {Taniguchi}, \citenamefont
  {Watanabe}, \citenamefont {Grigorieva}, \citenamefont {Polini}, \citenamefont
  {Geim},\ and\ \citenamefont {Bandurin}}]{berdyugin_measuring_2019}%
  \BibitemOpen
  \bibfield  {author} {\bibinfo {author} {\bibfnamefont {A.~I.}\ \bibnamefont
  {Berdyugin}}, \bibinfo {author} {\bibfnamefont {S.~G.}\ \bibnamefont {Xu}},
  \bibinfo {author} {\bibfnamefont {F.~M.~D.}\ \bibnamefont {Pellegrino}},
  \bibinfo {author} {\bibfnamefont {R.}~\bibnamefont {Krishna~Kumar}}, \bibinfo
  {author} {\bibfnamefont {A.}~\bibnamefont {Principi}}, \bibinfo {author}
  {\bibfnamefont {I.}~\bibnamefont {Torre}}, \bibinfo {author} {\bibfnamefont
  {M.}~\bibnamefont {Ben~Shalom}}, \bibinfo {author} {\bibfnamefont
  {T.}~\bibnamefont {Taniguchi}}, \bibinfo {author} {\bibfnamefont
  {K.}~\bibnamefont {Watanabe}}, \bibinfo {author} {\bibfnamefont {I.~V.}\
  \bibnamefont {Grigorieva}}, \bibinfo {author} {\bibfnamefont
  {M.}~\bibnamefont {Polini}}, \bibinfo {author} {\bibfnamefont {A.~K.}\
  \bibnamefont {Geim}}, \ and\ \bibinfo {author} {\bibfnamefont {D.~A.}\
  \bibnamefont {Bandurin}},\ }\bibfield  {title} {\enquote {\bibinfo {title}
  {Measuring {Hall} viscosity of graphene's electron fluid},}\ }\href {\doibase
  10.1126/science.aau0685} {\bibfield  {journal} {\bibinfo  {journal}
  {Science}\ }\textbf {\bibinfo {volume} {364}},\ \bibinfo {pages} {162--165}
  (\bibinfo {year} {2019})}\BibitemShut {NoStop}%
\bibitem [{\citenamefont {You}\ \emph {et~al.}(2014)\citenamefont {You},
  \citenamefont {Cho},\ and\ \citenamefont {Fradkin}}]{You2014}%
  \BibitemOpen
  \bibfield  {author} {\bibinfo {author} {\bibfnamefont {Y.}~\bibnamefont
  {You}}, \bibinfo {author} {\bibfnamefont {G.~Y.}\ \bibnamefont {Cho}}, \ and\
  \bibinfo {author} {\bibfnamefont {E.}~\bibnamefont {Fradkin}},\ }\bibfield
  {title} {\enquote {\bibinfo {title} {Theory of nematic fractional quantum
  {Hall} states},}\ }\href {\doibase 10.1103/PhysRevX.4.041050} {\bibfield
  {journal} {\bibinfo  {journal} {Phys. Rev. X}\ }\textbf {\bibinfo {volume}
  {4}},\ \bibinfo {pages} {041050} (\bibinfo {year} {2014})}\BibitemShut
  {NoStop}%
\bibitem [{\citenamefont {Son}(2015)}]{Son2015}%
  \BibitemOpen
  \bibfield  {author} {\bibinfo {author} {\bibfnamefont {D.~T.}\ \bibnamefont
  {Son}},\ }\bibfield  {title} {\enquote {\bibinfo {title} {Is the composite
  fermion a {Dirac} particle?}}\ }\href {\doibase 10.1103/PhysRevX.5.031027}
  {\bibfield  {journal} {\bibinfo  {journal} {Phys. Rev. X}\ }\textbf {\bibinfo
  {volume} {5}},\ \bibinfo {pages} {031027} (\bibinfo {year}
  {2015})}\BibitemShut {NoStop}%
\bibitem [{\citenamefont {Levin}\ and\ \citenamefont {Son}(2017)}]{Levin2017}%
  \BibitemOpen
  \bibfield  {author} {\bibinfo {author} {\bibfnamefont {M.}~\bibnamefont
  {Levin}}\ and\ \bibinfo {author} {\bibfnamefont {D.~T.}\ \bibnamefont
  {Son}},\ }\bibfield  {title} {\enquote {\bibinfo {title} {Particle-hole
  symmetry and electromagnetic response of a half-filled {Landau} level},}\
  }\href {\doibase 10.1103/PhysRevB.95.125120} {\bibfield  {journal} {\bibinfo
  {journal} {Phys. Rev. B}\ }\textbf {\bibinfo {volume} {95}},\ \bibinfo
  {pages} {125120} (\bibinfo {year} {2017})}\BibitemShut {NoStop}%
\bibitem [{\citenamefont {Goldman}\ and\ \citenamefont
  {Fradkin}(2018)}]{Goldman2018}%
  \BibitemOpen
  \bibfield  {author} {\bibinfo {author} {\bibfnamefont {H.}~\bibnamefont
  {Goldman}}\ and\ \bibinfo {author} {\bibfnamefont {E.}~\bibnamefont
  {Fradkin}},\ }\bibfield  {title} {\enquote {\bibinfo {title} {Dirac composite
  fermions and emergent reflection symmetry about even-denominator filling
  fractions},}\ }\href {\doibase 10.1103/PhysRevB.98.165137} {\bibfield
  {journal} {\bibinfo  {journal} {Phys. Rev. B}\ }\textbf {\bibinfo {volume}
  {98}},\ \bibinfo {pages} {165137} (\bibinfo {year} {2018})}\BibitemShut
  {NoStop}%
\bibitem [{\citenamefont {Pu}(2020)}]{Pu2020}%
  \BibitemOpen
  \bibfield  {author} {\bibinfo {author} {\bibfnamefont {S.}~\bibnamefont
  {Pu}},\ }\bibfield  {title} {\enquote {\bibinfo {title} {Hall viscosity of
  the composite-fermion {Fermi} seas for fermions and bosons},}\ }\href
  {\doibase 10.1103/PhysRevB.102.165101} {\bibfield  {journal} {\bibinfo
  {journal} {Phys. Rev. B}\ }\textbf {\bibinfo {volume} {102}},\ \bibinfo
  {pages} {165101} (\bibinfo {year} {2020})}\BibitemShut {NoStop}%
\bibitem [{\citenamefont {Banerjee}\ \emph {et~al.}(2017)\citenamefont
  {Banerjee}, \citenamefont {Souslov}, \citenamefont {Abanov},\ and\
  \citenamefont {Vitelli}}]{banerjee_odd_2017}%
  \BibitemOpen
  \bibfield  {author} {\bibinfo {author} {\bibfnamefont {D.}~\bibnamefont
  {Banerjee}}, \bibinfo {author} {\bibfnamefont {A.}~\bibnamefont {Souslov}},
  \bibinfo {author} {\bibfnamefont {A.~G.}\ \bibnamefont {Abanov}}, \ and\
  \bibinfo {author} {\bibfnamefont {V.}~\bibnamefont {Vitelli}},\ }\bibfield
  {title} {\enquote {\bibinfo {title} {Odd viscosity in chiral active
  fluids},}\ }\href {\doibase 10.1038/s41467-017-01378-7} {\bibfield  {journal}
  {\bibinfo  {journal} {Nature Communications}\ }\textbf {\bibinfo {volume}
  {8}},\ \bibinfo {pages} {1573} (\bibinfo {year} {2017})}\BibitemShut
  {NoStop}%
\bibitem [{\citenamefont {Soni}\ \emph {et~al.}(2019)\citenamefont {Soni},
  \citenamefont {Bililign}, \citenamefont {Magkiriadou}, \citenamefont
  {Sacanna}, \citenamefont {Bartolo}, \citenamefont {Shelley},\ and\
  \citenamefont {Irvine}}]{soni_odd_2019}%
  \BibitemOpen
  \bibfield  {author} {\bibinfo {author} {\bibfnamefont {V.}~\bibnamefont
  {Soni}}, \bibinfo {author} {\bibfnamefont {E.~S.}\ \bibnamefont {Bililign}},
  \bibinfo {author} {\bibfnamefont {S.}~\bibnamefont {Magkiriadou}}, \bibinfo
  {author} {\bibfnamefont {S.}~\bibnamefont {Sacanna}}, \bibinfo {author}
  {\bibfnamefont {D.}~\bibnamefont {Bartolo}}, \bibinfo {author} {\bibfnamefont
  {M.~J.}\ \bibnamefont {Shelley}}, \ and\ \bibinfo {author} {\bibfnamefont
  {W.~T.~M.}\ \bibnamefont {Irvine}},\ }\bibfield  {title} {\enquote {\bibinfo
  {title} {The odd free surface flows of a colloidal chiral fluid},}\ }\href
  {\doibase 10.1038/s41567-019-0603-8} {\bibfield  {journal} {\bibinfo
  {journal} {Nature Physics}\ }\textbf {\bibinfo {volume} {15}},\ \bibinfo
  {pages} {1188--1194} (\bibinfo {year} {2019})}\BibitemShut {NoStop}%
\bibitem [{\citenamefont {Souslov}\ \emph {et~al.}(2019)\citenamefont
  {Souslov}, \citenamefont {Dasbiswas}, \citenamefont {Fruchart}, \citenamefont
  {Vaikuntanathan},\ and\ \citenamefont {Vitelli}}]{souslov_topological_2019}%
  \BibitemOpen
  \bibfield  {author} {\bibinfo {author} {\bibfnamefont {A.}~\bibnamefont
  {Souslov}}, \bibinfo {author} {\bibfnamefont {K.}~\bibnamefont {Dasbiswas}},
  \bibinfo {author} {\bibfnamefont {M.}~\bibnamefont {Fruchart}}, \bibinfo
  {author} {\bibfnamefont {S.}~\bibnamefont {Vaikuntanathan}}, \ and\ \bibinfo
  {author} {\bibfnamefont {V.}~\bibnamefont {Vitelli}},\ }\bibfield  {title}
  {\enquote {\bibinfo {title} {Topological waves in fluids with odd
  viscosity},}\ }\href {\doibase 10.1103/PhysRevLett.122.128001} {\bibfield
  {journal} {\bibinfo  {journal} {Physical Review Letters}\ }\textbf {\bibinfo
  {volume} {122}},\ \bibinfo {pages} {128001} (\bibinfo {year}
  {2019})}\BibitemShut {NoStop}%
\bibitem [{\citenamefont {Hargus}\ \emph {et~al.}(2020)\citenamefont {Hargus},
  \citenamefont {Klymko}, \citenamefont {Epstein},\ and\ \citenamefont
  {Mandadapu}}]{hargus_time_2020}%
  \BibitemOpen
  \bibfield  {author} {\bibinfo {author} {\bibfnamefont {C.}~\bibnamefont
  {Hargus}}, \bibinfo {author} {\bibfnamefont {K.}~\bibnamefont {Klymko}},
  \bibinfo {author} {\bibfnamefont {J.~M.}\ \bibnamefont {Epstein}}, \ and\
  \bibinfo {author} {\bibfnamefont {K.~K.}\ \bibnamefont {Mandadapu}},\
  }\bibfield  {title} {\enquote {\bibinfo {title} {Time reversal symmetry
  breaking and odd viscosity in active fluids: {Green-Kubo} and {NEMD}
  results},}\ }\href {\doibase 10.1063/5.0006441} {\bibfield  {journal}
  {\bibinfo  {journal} {The Journal of Chemical Physics}\ }\textbf {\bibinfo
  {volume} {152}},\ \bibinfo {pages} {201102} (\bibinfo {year}
  {2020})}\BibitemShut {NoStop}%
\bibitem [{\citenamefont {Yamauchi}\ \emph {et~al.}(2020)\citenamefont
  {Yamauchi}, \citenamefont {Hayata}, \citenamefont {Uwamichi}, \citenamefont
  {Ozawa},\ and\ \citenamefont {Kawaguchi}}]{yamauchi_chirality-driven_2020}%
  \BibitemOpen
  \bibfield  {author} {\bibinfo {author} {\bibfnamefont {L.}~\bibnamefont
  {Yamauchi}}, \bibinfo {author} {\bibfnamefont {T.}~\bibnamefont {Hayata}},
  \bibinfo {author} {\bibfnamefont {M.}~\bibnamefont {Uwamichi}}, \bibinfo
  {author} {\bibfnamefont {T.}~\bibnamefont {Ozawa}}, \ and\ \bibinfo {author}
  {\bibfnamefont {K.}~\bibnamefont {Kawaguchi}},\ }\href
  {http://arxiv.org/abs/2008.10852} {\enquote {\bibinfo {title}
  {Chirality-driven edge flow and non-{Hermitian} topology in active nematic
  cells},}\ } (\bibinfo {year} {2020}),\ \bibinfo {note} {arXiv:2008.10852
  [cond-mat, physics:hep-th, physics:physics]}\BibitemShut {NoStop}%
\bibitem [{\citenamefont {Han}\ \emph {et~al.}(2021)\citenamefont {Han},
  \citenamefont {Fruchart}, \citenamefont {Scheibner}, \citenamefont
  {Vaikuntanathan}, \citenamefont {de~Pablo},\ and\ \citenamefont
  {Vitelli}}]{han_fluctuating_2021}%
  \BibitemOpen
  \bibfield  {author} {\bibinfo {author} {\bibfnamefont {M.}~\bibnamefont
  {Han}}, \bibinfo {author} {\bibfnamefont {M.}~\bibnamefont {Fruchart}},
  \bibinfo {author} {\bibfnamefont {C.}~\bibnamefont {Scheibner}}, \bibinfo
  {author} {\bibfnamefont {S.}~\bibnamefont {Vaikuntanathan}}, \bibinfo
  {author} {\bibfnamefont {J.~J.}\ \bibnamefont {de~Pablo}}, \ and\ \bibinfo
  {author} {\bibfnamefont {V.}~\bibnamefont {Vitelli}},\ }\bibfield  {title}
  {\enquote {\bibinfo {title} {Fluctuating hydrodynamics of chiral active
  fluids},}\ }\href {\doibase 10.1038/s41567-021-01360-7} {\bibfield  {journal}
  {\bibinfo  {journal} {Nature Physics}\ }\textbf {\bibinfo {volume} {17}},\
  \bibinfo {pages} {1260--1269} (\bibinfo {year} {2021})}\BibitemShut {NoStop}%
\bibitem [{\citenamefont {Hosaka}\ \emph {et~al.}(2021)\citenamefont {Hosaka},
  \citenamefont {Komura},\ and\ \citenamefont
  {Andelman}}]{hosaka_hydrodynamic_2021}%
  \BibitemOpen
  \bibfield  {author} {\bibinfo {author} {\bibfnamefont {Y.}~\bibnamefont
  {Hosaka}}, \bibinfo {author} {\bibfnamefont {S.}~\bibnamefont {Komura}}, \
  and\ \bibinfo {author} {\bibfnamefont {D.}~\bibnamefont {Andelman}},\
  }\bibfield  {title} {\enquote {\bibinfo {title} {Hydrodynamic lift of a
  two-dimensional liquid domain with odd viscosity},}\ }\href {\doibase
  10.1103/PhysRevE.104.064613} {\bibfield  {journal} {\bibinfo  {journal}
  {Physical Review E}\ }\textbf {\bibinfo {volume} {104}},\ \bibinfo {pages}
  {064613} (\bibinfo {year} {2021})}\BibitemShut {NoStop}%
\bibitem [{\citenamefont {Banerjee}\ \emph {et~al.}(2022)\citenamefont
  {Banerjee}, \citenamefont {Souslov},\ and\ \citenamefont
  {Vitelli}}]{banerjee_hydrodynamic_2022}%
  \BibitemOpen
  \bibfield  {author} {\bibinfo {author} {\bibfnamefont {D.}~\bibnamefont
  {Banerjee}}, \bibinfo {author} {\bibfnamefont {A.}~\bibnamefont {Souslov}}, \
  and\ \bibinfo {author} {\bibfnamefont {V.}~\bibnamefont {Vitelli}},\
  }\bibfield  {title} {\enquote {\bibinfo {title} {Hydrodynamic correlation
  functions of chiral active fluids},}\ }\href {\doibase
  10.1103/PhysRevFluids.7.043301} {\bibfield  {journal} {\bibinfo  {journal}
  {Physical Review Fluids}\ }\textbf {\bibinfo {volume} {7}},\ \bibinfo {pages}
  {043301} (\bibinfo {year} {2022})}\BibitemShut {NoStop}%
\bibitem [{\citenamefont {Khain}\ \emph {et~al.}(2022)\citenamefont {Khain},
  \citenamefont {Scheibner}, \citenamefont {Fruchart},\ and\ \citenamefont
  {Vitelli}}]{khain_stokes_2022}%
  \BibitemOpen
  \bibfield  {author} {\bibinfo {author} {\bibfnamefont {T.}~\bibnamefont
  {Khain}}, \bibinfo {author} {\bibfnamefont {C.}~\bibnamefont {Scheibner}},
  \bibinfo {author} {\bibfnamefont {M.}~\bibnamefont {Fruchart}}, \ and\
  \bibinfo {author} {\bibfnamefont {V.}~\bibnamefont {Vitelli}},\ }\bibfield
  {title} {\enquote {\bibinfo {title} {Stokes flows in three-dimensional fluids
  with odd and parity-violating viscosities},}\ }\href {\doibase
  10.1017/jfm.2021.1079} {\bibfield  {journal} {\bibinfo  {journal} {Journal of
  Fluid Mechanics}\ }\textbf {\bibinfo {volume} {934}},\ \bibinfo {pages} {A23}
  (\bibinfo {year} {2022})}\BibitemShut {NoStop}%
\bibitem [{\citenamefont {Reynolds}\ \emph {et~al.}(2022)\citenamefont
  {Reynolds}, \citenamefont {Monteiro},\ and\ \citenamefont
  {Ganeshan}}]{reynolds_hele-shaw_2022}%
  \BibitemOpen
  \bibfield  {author} {\bibinfo {author} {\bibfnamefont {D.}~\bibnamefont
  {Reynolds}}, \bibinfo {author} {\bibfnamefont {G.~M.}\ \bibnamefont
  {Monteiro}}, \ and\ \bibinfo {author} {\bibfnamefont {S.}~\bibnamefont
  {Ganeshan}},\ }\bibfield  {title} {\enquote {\bibinfo {title} {Hele-{Shaw}
  flow for parity odd three-dimensional fluids},}\ }\href {\doibase
  10.1103/PhysRevFluids.7.114201} {\bibfield  {journal} {\bibinfo  {journal}
  {Physical Review Fluids}\ }\textbf {\bibinfo {volume} {7}},\ \bibinfo {pages}
  {114201} (\bibinfo {year} {2022})}\BibitemShut {NoStop}%
\bibitem [{\citenamefont {de~Wit}\ \emph {et~al.}(2024)\citenamefont {de~Wit},
  \citenamefont {Fruchart}, \citenamefont {Khain}, \citenamefont {Toschi},\
  and\ \citenamefont {Vitelli}}]{de_wit_pattern_2024}%
  \BibitemOpen
  \bibfield  {author} {\bibinfo {author} {\bibfnamefont {X.~M.}\ \bibnamefont
  {de~Wit}}, \bibinfo {author} {\bibfnamefont {M.}~\bibnamefont {Fruchart}},
  \bibinfo {author} {\bibfnamefont {T.}~\bibnamefont {Khain}}, \bibinfo
  {author} {\bibfnamefont {F.}~\bibnamefont {Toschi}}, \ and\ \bibinfo {author}
  {\bibfnamefont {V.}~\bibnamefont {Vitelli}},\ }\bibfield  {title} {\enquote
  {\bibinfo {title} {Pattern formation by turbulent cascades},}\ }\href
  {\doibase 10.1038/s41586-024-07074-z} {\bibfield  {journal} {\bibinfo
  {journal} {Nature}\ }\textbf {\bibinfo {volume} {627}},\ \bibinfo {pages}
  {515--521} (\bibinfo {year} {2024})}\BibitemShut {NoStop}%
\bibitem [{\citenamefont {Fruchart}\ \emph {et~al.}(2023)\citenamefont
  {Fruchart}, \citenamefont {Scheibner},\ and\ \citenamefont
  {Vitelli}}]{fruchart_odd_2023}%
  \BibitemOpen
  \bibfield  {author} {\bibinfo {author} {\bibfnamefont {M.}~\bibnamefont
  {Fruchart}}, \bibinfo {author} {\bibfnamefont {C.}~\bibnamefont {Scheibner}},
  \ and\ \bibinfo {author} {\bibfnamefont {V.}~\bibnamefont {Vitelli}},\
  }\bibfield  {title} {\enquote {\bibinfo {title} {Odd viscosity and odd
  elasticity},}\ }\href {\doibase 10.1146/annurev-conmatphys-040821-125506}
  {\bibfield  {journal} {\bibinfo  {journal} {Annual Review of Condensed Matter
  Physics}\ }\textbf {\bibinfo {volume} {14}},\ \bibinfo {pages} {471--510}
  (\bibinfo {year} {2023})}\BibitemShut {NoStop}%
\bibitem [{\citenamefont {Bradlyn}\ \emph {et~al.}(2012)\citenamefont
  {Bradlyn}, \citenamefont {Goldstein},\ and\ \citenamefont
  {Read}}]{bradlyn_kubo_2012}%
  \BibitemOpen
  \bibfield  {author} {\bibinfo {author} {\bibfnamefont {B.}~\bibnamefont
  {Bradlyn}}, \bibinfo {author} {\bibfnamefont {M.}~\bibnamefont {Goldstein}},
  \ and\ \bibinfo {author} {\bibfnamefont {N.}~\bibnamefont {Read}},\
  }\bibfield  {title} {\enquote {\bibinfo {title} {Kubo formulas for viscosity:
  {Hall} viscosity, {Ward} identities, and the relation with conductivity},}\
  }\href {\doibase 10.1103/PhysRevB.86.245309} {\bibfield  {journal} {\bibinfo
  {journal} {Physical Review B}\ }\textbf {\bibinfo {volume} {86}},\ \bibinfo
  {pages} {245309} (\bibinfo {year} {2012})}\BibitemShut {NoStop}%
\bibitem [{\citenamefont {Hoyos}\ and\ \citenamefont
  {Son}(2012)}]{hoyos_hall_2012}%
  \BibitemOpen
  \bibfield  {author} {\bibinfo {author} {\bibfnamefont {C.}~\bibnamefont
  {Hoyos}}\ and\ \bibinfo {author} {\bibfnamefont {D.~T.}\ \bibnamefont
  {Son}},\ }\bibfield  {title} {\enquote {\bibinfo {title} {Hall {Viscosity}
  and {Electromagnetic} {Response}},}\ }\href {\doibase
  10.1103/PhysRevLett.108.066805} {\bibfield  {journal} {\bibinfo  {journal}
  {Physical Review Letters}\ }\textbf {\bibinfo {volume} {108}},\ \bibinfo
  {pages} {066805} (\bibinfo {year} {2012})}\BibitemShut {NoStop}%
\bibitem [{\citenamefont {Delacr\'{e}taz}\ and\ \citenamefont
  {Gromov}(2017)}]{delacretaz_transport_2017}%
  \BibitemOpen
  \bibfield  {author} {\bibinfo {author} {\bibfnamefont {L.~V.}\ \bibnamefont
  {Delacr\'{e}taz}}\ and\ \bibinfo {author} {\bibfnamefont {A.}~\bibnamefont
  {Gromov}},\ }\bibfield  {title} {\enquote {\bibinfo {title} {Transport
  {Signatures} of the {Hall} {Viscosity}},}\ }\href {\doibase
  10.1103/PhysRevLett.119.226602} {\bibfield  {journal} {\bibinfo  {journal}
  {Physical Review Letters}\ }\textbf {\bibinfo {volume} {119}},\ \bibinfo
  {pages} {226602} (\bibinfo {year} {2017})}\BibitemShut {NoStop}%
\bibitem [{\citenamefont {Schine}\ \emph {et~al.}(2016)\citenamefont {Schine},
  \citenamefont {Ryou}, \citenamefont {Gromov}, \citenamefont {Sommer},\ and\
  \citenamefont {Simon}}]{schine_synthetic_2016}%
  \BibitemOpen
  \bibfield  {author} {\bibinfo {author} {\bibfnamefont {N.}~\bibnamefont
  {Schine}}, \bibinfo {author} {\bibfnamefont {A.}~\bibnamefont {Ryou}},
  \bibinfo {author} {\bibfnamefont {A.}~\bibnamefont {Gromov}}, \bibinfo
  {author} {\bibfnamefont {A.}~\bibnamefont {Sommer}}, \ and\ \bibinfo {author}
  {\bibfnamefont {J.}~\bibnamefont {Simon}},\ }\bibfield  {title} {\enquote
  {\bibinfo {title} {Synthetic {Landau} levels for photons},}\ }\href {\doibase
  10.1038/nature17943} {\bibfield  {journal} {\bibinfo  {journal} {Nature}\
  }\textbf {\bibinfo {volume} {534}},\ \bibinfo {pages} {671--675} (\bibinfo
  {year} {2016})}\BibitemShut {NoStop}%
\bibitem [{\citenamefont {Cooper}(2020)}]{Cooper2020}%
  \BibitemOpen
  \bibfield  {author} {\bibinfo {author} {\bibfnamefont {N.}~\bibnamefont
  {Cooper}},\ }\bibfield  {title} {\enquote {\bibinfo {title} {Fractional
  quantum {Hall} states of bosons: Properties and prospects for experimental
  realization},}\ }in\ \href@noop {} {\emph {\bibinfo {booktitle} {Fractional
  Quantum Hall Effects: New Developments}}}\ (\bibinfo  {publisher} {World
  Scientific},\ \bibinfo {year} {2020})\ pp.\ \bibinfo {pages}
  {487--521}\BibitemShut {NoStop}%
\bibitem [{\citenamefont {Fletcher}\ \emph {et~al.}(2021)\citenamefont
  {Fletcher}, \citenamefont {Shaffer}, \citenamefont {Wilson}, \citenamefont
  {Patel}, \citenamefont {Yan}, \citenamefont {Cr\'{e}pel}, \citenamefont
  {Mukherjee},\ and\ \citenamefont {Zwierlein}}]{fletcher_geometric_2021}%
  \BibitemOpen
  \bibfield  {author} {\bibinfo {author} {\bibfnamefont {R.~J.}\ \bibnamefont
  {Fletcher}}, \bibinfo {author} {\bibfnamefont {A.}~\bibnamefont {Shaffer}},
  \bibinfo {author} {\bibfnamefont {C.~C.}\ \bibnamefont {Wilson}}, \bibinfo
  {author} {\bibfnamefont {P.~B.}\ \bibnamefont {Patel}}, \bibinfo {author}
  {\bibfnamefont {Z.}~\bibnamefont {Yan}}, \bibinfo {author} {\bibfnamefont
  {V.}~\bibnamefont {Cr\'{e}pel}}, \bibinfo {author} {\bibfnamefont
  {B.}~\bibnamefont {Mukherjee}}, \ and\ \bibinfo {author} {\bibfnamefont
  {M.~W.}\ \bibnamefont {Zwierlein}},\ }\bibfield  {title} {\enquote {\bibinfo
  {title} {Geometric squeezing into the lowest {Landau} level},}\ }\href
  {\doibase 10.1126/science.aba7202} {\bibfield  {journal} {\bibinfo  {journal}
  {Science}\ }\textbf {\bibinfo {volume} {372}},\ \bibinfo {pages} {1318--1322}
  (\bibinfo {year} {2021})}\BibitemShut {NoStop}%
\bibitem [{\citenamefont {Mukherjee}\ \emph {et~al.}(2022)\citenamefont
  {Mukherjee}, \citenamefont {Shaffer}, \citenamefont {Patel}, \citenamefont
  {Yan}, \citenamefont {Wilson}, \citenamefont {Cr\'{e}pel}, \citenamefont
  {Fletcher},\ and\ \citenamefont
  {Zwierlein}}]{mukherjee_crystallization_2022}%
  \BibitemOpen
  \bibfield  {author} {\bibinfo {author} {\bibfnamefont {B.}~\bibnamefont
  {Mukherjee}}, \bibinfo {author} {\bibfnamefont {A.}~\bibnamefont {Shaffer}},
  \bibinfo {author} {\bibfnamefont {P.~B.}\ \bibnamefont {Patel}}, \bibinfo
  {author} {\bibfnamefont {Z.}~\bibnamefont {Yan}}, \bibinfo {author}
  {\bibfnamefont {C.~C.}\ \bibnamefont {Wilson}}, \bibinfo {author}
  {\bibfnamefont {V.}~\bibnamefont {Cr\'{e}pel}}, \bibinfo {author}
  {\bibfnamefont {R.~J.}\ \bibnamefont {Fletcher}}, \ and\ \bibinfo {author}
  {\bibfnamefont {M.}~\bibnamefont {Zwierlein}},\ }\bibfield  {title} {\enquote
  {\bibinfo {title} {Crystallization of bosonic quantum {Hall} states in a
  rotating quantum gas},}\ }\href {\doibase 10.1038/s41586-021-04170-2}
  {\bibfield  {journal} {\bibinfo  {journal} {Nature}\ }\textbf {\bibinfo
  {volume} {601}},\ \bibinfo {pages} {58--62} (\bibinfo {year}
  {2022})}\BibitemShut {NoStop}%
\bibitem [{\citenamefont {Sinha}\ and\ \citenamefont
  {Shlyapnikov}(2005)}]{sinha_two-dimensional_2005}%
  \BibitemOpen
  \bibfield  {author} {\bibinfo {author} {\bibfnamefont {S.}~\bibnamefont
  {Sinha}}\ and\ \bibinfo {author} {\bibfnamefont {G.~V.}\ \bibnamefont
  {Shlyapnikov}},\ }\bibfield  {title} {\enquote {\bibinfo {title}
  {Two-dimensional {Bose}-{Einstein} condensate under extreme rotation},}\
  }\href {\doibase 10.1103/PhysRevLett.94.150401} {\bibfield  {journal}
  {\bibinfo  {journal} {Physical Review Letters}\ }\textbf {\bibinfo {volume}
  {94}},\ \bibinfo {pages} {150401} (\bibinfo {year} {2005})}\BibitemShut
  {NoStop}%
\bibitem [{\citenamefont {Wiegmann}(2013)}]{wiegmann_anomalous_2013}%
  \BibitemOpen
  \bibfield  {author} {\bibinfo {author} {\bibfnamefont {P.}~\bibnamefont
  {Wiegmann}},\ }\bibfield  {title} {\enquote {\bibinfo {title} {Anomalous
  hydrodynamics of fractional quantum {Hall} states},}\ }\href {\doibase
  10.1134/S1063776113110162} {\bibfield  {journal} {\bibinfo  {journal}
  {Journal of Experimental and Theoretical Physics}\ }\textbf {\bibinfo
  {volume} {117}},\ \bibinfo {pages} {538--550} (\bibinfo {year}
  {2013})}\BibitemShut {NoStop}%
\bibitem [{\citenamefont {Wiegmann}\ and\ \citenamefont
  {Abanov}(2014)}]{wiegmann_anomalous_2014}%
  \BibitemOpen
  \bibfield  {author} {\bibinfo {author} {\bibfnamefont {P.}~\bibnamefont
  {Wiegmann}}\ and\ \bibinfo {author} {\bibfnamefont {A.~G.}\ \bibnamefont
  {Abanov}},\ }\bibfield  {title} {\enquote {\bibinfo {title} {Anomalous
  hydrodynamics of two-dimensional vortex fluids},}\ }\href {\doibase
  10.1103/PhysRevLett.113.034501} {\bibfield  {journal} {\bibinfo  {journal}
  {Physical Review Letters}\ }\textbf {\bibinfo {volume} {113}},\ \bibinfo
  {pages} {034501} (\bibinfo {year} {2014})}\BibitemShut {NoStop}%
\bibitem [{\citenamefont {Abanov}\ \emph {et~al.}(2020)\citenamefont {Abanov},
  \citenamefont {Can}, \citenamefont {Ganeshan},\ and\ \citenamefont
  {Monteiro}}]{abanov_hydrodynamics_2020}%
  \BibitemOpen
  \bibfield  {author} {\bibinfo {author} {\bibfnamefont {A.~G.}\ \bibnamefont
  {Abanov}}, \bibinfo {author} {\bibfnamefont {T.}~\bibnamefont {Can}},
  \bibinfo {author} {\bibfnamefont {S.}~\bibnamefont {Ganeshan}}, \ and\
  \bibinfo {author} {\bibfnamefont {G.~M.}\ \bibnamefont {Monteiro}},\
  }\bibfield  {title} {\enquote {\bibinfo {title} {Hydrodynamics of
  two-dimensional compressible fluid with broken parity: {Variational}
  principle and free surface dynamics in the absence of dissipation},}\ }\href
  {\doibase 10.1103/PhysRevFluids.5.104802} {\bibfield  {journal} {\bibinfo
  {journal} {Physical Review Fluids}\ }\textbf {\bibinfo {volume} {5}},\
  \bibinfo {pages} {104802} (\bibinfo {year} {2020})}\BibitemShut {NoStop}%
\bibitem [{\citenamefont {Markovich}\ and\ \citenamefont
  {Lubensky}(2021)}]{markovich_odd_2021}%
  \BibitemOpen
  \bibfield  {author} {\bibinfo {author} {\bibfnamefont {T.}~\bibnamefont
  {Markovich}}\ and\ \bibinfo {author} {\bibfnamefont {T.~C.}\ \bibnamefont
  {Lubensky}},\ }\bibfield  {title} {\enquote {\bibinfo {title} {Odd viscosity
  in active matter: {Microscopic} origin and 3d effects},}\ }\href {\doibase
  10.1103/PhysRevLett.127.048001} {\bibfield  {journal} {\bibinfo  {journal}
  {Physical Review Letters}\ }\textbf {\bibinfo {volume} {127}},\ \bibinfo
  {pages} {048001} (\bibinfo {year} {2021})}\BibitemShut {NoStop}%
\bibitem [{\citenamefont {Machado~Monteiro}\ \emph {et~al.}(2023)\citenamefont
  {Machado~Monteiro}, \citenamefont {Abanov},\ and\ \citenamefont
  {Ganeshan}}]{machado_monteiro_hamiltonian_2023}%
  \BibitemOpen
  \bibfield  {author} {\bibinfo {author} {\bibfnamefont {G.}~\bibnamefont
  {Machado~Monteiro}}, \bibinfo {author} {\bibfnamefont {A.~G.}\ \bibnamefont
  {Abanov}}, \ and\ \bibinfo {author} {\bibfnamefont {S.}~\bibnamefont
  {Ganeshan}},\ }\bibfield  {title} {\enquote {\bibinfo {title} {Hamiltonian
  structure of {2D} fluid dynamics with broken parity},}\ }\href {\doibase
  10.21468/SciPostPhys.14.5.103} {\bibfield  {journal} {\bibinfo  {journal}
  {SciPost Physics}\ }\textbf {\bibinfo {volume} {14}},\ \bibinfo {pages} {103}
  (\bibinfo {year} {2023})}\BibitemShut {NoStop}%
\bibitem [{\citenamefont {Fetter}(2009)}]{fetter_rotating_2009}%
  \BibitemOpen
  \bibfield  {author} {\bibinfo {author} {\bibfnamefont {A.~L.}\ \bibnamefont
  {Fetter}},\ }\bibfield  {title} {\enquote {\bibinfo {title} {Rotating trapped
  {Bose}-{Einstein} condensates},}\ }\href {\doibase 10.1103/RevModPhys.81.647}
  {\bibfield  {journal} {\bibinfo  {journal} {Reviews of Modern Physics}\
  }\textbf {\bibinfo {volume} {81}},\ \bibinfo {pages} {647--691} (\bibinfo
  {year} {2009})}\BibitemShut {NoStop}%
\bibitem [{\citenamefont {Park}\ and\ \citenamefont
  {Haldane}(2014)}]{park_guiding-center_2014}%
  \BibitemOpen
  \bibfield  {author} {\bibinfo {author} {\bibfnamefont {Y.}~\bibnamefont
  {Park}}\ and\ \bibinfo {author} {\bibfnamefont {F.~D.~M.}\ \bibnamefont
  {Haldane}},\ }\bibfield  {title} {\enquote {\bibinfo {title} {Guiding-center
  {Hall} viscosity and intrinsic dipole moment along edges of incompressible
  fractional quantum {Hall} fluids},}\ }\href {\doibase
  10.1103/PhysRevB.90.045123} {\bibfield  {journal} {\bibinfo  {journal}
  {Physical Review B}\ }\textbf {\bibinfo {volume} {90}},\ \bibinfo {pages}
  {045123} (\bibinfo {year} {2014})}\BibitemShut {NoStop}%
\bibitem [{\citenamefont {Acheson}(1990)}]{acheson_elementary_1990}%
  \BibitemOpen
  \bibfield  {author} {\bibinfo {author} {\bibfnamefont {D.~J.}\ \bibnamefont
  {Acheson}},\ }\href@noop {} {\emph {\bibinfo {title} {Elementary fluid
  dynamics}}},\ Oxford applied mathematics and computing science series\
  (\bibinfo  {publisher} {Clarendon Press ; Oxford University Press},\ \bibinfo
  {address} {Oxford : New York},\ \bibinfo {year} {1990})\BibitemShut {NoStop}%
\bibitem [{\citenamefont {Lucas}\ and\ \citenamefont
  {Fong}(2018)}]{lucas_hydrodynamics_2018}%
  \BibitemOpen
  \bibfield  {author} {\bibinfo {author} {\bibfnamefont {A.}~\bibnamefont
  {Lucas}}\ and\ \bibinfo {author} {\bibfnamefont {K.~C.}\ \bibnamefont
  {Fong}},\ }\bibfield  {title} {\enquote {\bibinfo {title} {Hydrodynamics of
  electrons in graphene},}\ }\href {\doibase 10.1088/1361-648X/aaa274}
  {\bibfield  {journal} {\bibinfo  {journal} {Journal of Physics: Condensed
  Matter}\ }\textbf {\bibinfo {volume} {30}},\ \bibinfo {pages} {053001}
  (\bibinfo {year} {2018})}\BibitemShut {NoStop}%
\bibitem [{\citenamefont {Abanov}\ \emph {et~al.}(2018)\citenamefont {Abanov},
  \citenamefont {Can},\ and\ \citenamefont {Ganeshan}}]{abanov_odd_2018}%
  \BibitemOpen
  \bibfield  {author} {\bibinfo {author} {\bibfnamefont {A.}~\bibnamefont
  {Abanov}}, \bibinfo {author} {\bibfnamefont {T.}~\bibnamefont {Can}}, \ and\
  \bibinfo {author} {\bibfnamefont {S.}~\bibnamefont {Ganeshan}},\ }\bibfield
  {title} {\enquote {\bibinfo {title} {Odd surface waves in two-dimensional
  incompressible fluids},}\ }\href {\doibase 10.21468/SciPostPhys.5.1.010}
  {\bibfield  {journal} {\bibinfo  {journal} {SciPost Physics}\ }\textbf
  {\bibinfo {volume} {5}},\ \bibinfo {pages} {010} (\bibinfo {year}
  {2018})}\BibitemShut {NoStop}%
\bibitem [{\citenamefont {Lingam}(2015)}]{lingam_hall_2015}%
  \BibitemOpen
  \bibfield  {author} {\bibinfo {author} {\bibfnamefont {M.}~\bibnamefont
  {Lingam}},\ }\bibfield  {title} {\enquote {\bibinfo {title} {Hall viscosity:
  {A} link between quantum {Hall} systems, plasmas and liquid crystals},}\
  }\href {\doibase 10.1016/j.physleta.2015.03.014} {\bibfield  {journal}
  {\bibinfo  {journal} {Physics Letters A}\ }\textbf {\bibinfo {volume}
  {379}},\ \bibinfo {pages} {1425--1430} (\bibinfo {year} {2015})}\BibitemShut
  {NoStop}%
\bibitem [{\citenamefont {Roncaglia}\ \emph {et~al.}(2011)\citenamefont
  {Roncaglia}, \citenamefont {Rizzi},\ and\ \citenamefont
  {Dalibard}}]{roncaglia_rotating_2011}%
  \BibitemOpen
  \bibfield  {author} {\bibinfo {author} {\bibfnamefont {M.}~\bibnamefont
  {Roncaglia}}, \bibinfo {author} {\bibfnamefont {M.}~\bibnamefont {Rizzi}}, \
  and\ \bibinfo {author} {\bibfnamefont {J.}~\bibnamefont {Dalibard}},\
  }\bibfield  {title} {\enquote {\bibinfo {title} {From rotating atomic rings
  to quantum {Hall} states},}\ }\href {\doibase 10.1038/srep00043} {\bibfield
  {journal} {\bibinfo  {journal} {Scientific Reports}\ }\textbf {\bibinfo
  {volume} {1}},\ \bibinfo {pages} {43} (\bibinfo {year} {2011})}\BibitemShut
  {NoStop}%
\bibitem [{\citenamefont {{T. W. Neely}}\ \emph {et~al.}(2010)\citenamefont
  {{T. W. Neely}}, \citenamefont {{E. C. Samson}}, \citenamefont {{A. S.
  Bradley}}, \citenamefont {{M. J. Davis}},\ and\ \citenamefont {{B. P.
  Anderson}}}]{t_w_neely_observation_2010}%
  \BibitemOpen
  \bibfield  {author} {\bibinfo {author} {\bibnamefont {{T. W. Neely}}},
  \bibinfo {author} {\bibnamefont {{E. C. Samson}}}, \bibinfo {author}
  {\bibnamefont {{A. S. Bradley}}}, \bibinfo {author} {\bibnamefont {{M. J.
  Davis}}}, \ and\ \bibinfo {author} {\bibnamefont {{B. P. Anderson}}},\
  }\bibfield  {title} {\enquote {\bibinfo {title} {Observation of vortex
  dipoles in an oblate {Bose}-{Einstein} condensate},}\ }\href {\doibase
  https://doi.org/10.1103/PhysRevLett.104.160401} {\bibfield  {journal}
  {\bibinfo  {journal} {Physical Review Letters}\ }\textbf {\bibinfo {volume}
  {104}},\ \bibinfo {pages} {160401} (\bibinfo {year} {2010})}\BibitemShut
  {NoStop}%
\bibitem [{\citenamefont {{Kazuki Sasaki}}\ \emph {et~al.}(2010)\citenamefont
  {{Kazuki Sasaki}}, \citenamefont {{Naoya Suzuki}},\ and\ \citenamefont
  {{Hiroki Saito}}}]{kazuki_sasaki_benard-von_2010}%
  \BibitemOpen
  \bibfield  {author} {\bibinfo {author} {\bibnamefont {{Kazuki Sasaki}}},
  \bibinfo {author} {\bibnamefont {{Naoya Suzuki}}}, \ and\ \bibinfo {author}
  {\bibnamefont {{Hiroki Saito}}},\ }\bibfield  {title} {\enquote {\bibinfo
  {title} {B\'{e}nard-von {K\'{a}rm\'{a}n} vortex street in a {Bose}-{Einstein}
  condensate},}\ }\href
  {https://journals.aps.org/prl/abstract/10.1103/PhysRevLett.104.150404}
  {\bibfield  {journal} {\bibinfo  {journal} {Physical Review Letters}\
  }\textbf {\bibinfo {volume} {104}} (\bibinfo {year} {2010})}\BibitemShut
  {NoStop}%
\bibitem [{\citenamefont {Kwon}\ \emph {et~al.}(2016)\citenamefont {Kwon},
  \citenamefont {Kim}, \citenamefont {Seo},\ and\ \citenamefont
  {Shin}}]{kwon_observation_2016}%
  \BibitemOpen
  \bibfield  {author} {\bibinfo {author} {\bibfnamefont {W.~J.}\ \bibnamefont
  {Kwon}}, \bibinfo {author} {\bibfnamefont {J.~H.}\ \bibnamefont {Kim}},
  \bibinfo {author} {\bibfnamefont {S.~W.}\ \bibnamefont {Seo}}, \ and\
  \bibinfo {author} {\bibfnamefont {Y.}~\bibnamefont {Shin}},\ }\bibfield
  {title} {\enquote {\bibinfo {title} {Observation of von {K\'{a}rm\'{a}n}
  vortex street in an atomic superfluid gas},}\ }\href {\doibase
  10.1103/PhysRevLett.117.245301} {\bibfield  {journal} {\bibinfo  {journal}
  {Physical Review Letters}\ }\textbf {\bibinfo {volume} {117}},\ \bibinfo
  {pages} {245301} (\bibinfo {year} {2016})}\BibitemShut {NoStop}%
\bibitem [{\citenamefont {Musser}\ \emph {et~al.}(2019)\citenamefont {Musser},
  \citenamefont {Proment}, \citenamefont {Onorato},\ and\ \citenamefont
  {Irvine}}]{musser_starting_2019}%
  \BibitemOpen
  \bibfield  {author} {\bibinfo {author} {\bibfnamefont {S.}~\bibnamefont
  {Musser}}, \bibinfo {author} {\bibfnamefont {D.}~\bibnamefont {Proment}},
  \bibinfo {author} {\bibfnamefont {M.}~\bibnamefont {Onorato}}, \ and\
  \bibinfo {author} {\bibfnamefont {W.~T.}\ \bibnamefont {Irvine}},\ }\bibfield
   {title} {\enquote {\bibinfo {title} {Starting flow past an airfoil and its
  acquired lift in a superfluid},}\ }\href {\doibase
  10.1103/PhysRevLett.123.154502} {\bibfield  {journal} {\bibinfo  {journal}
  {Physical Review Letters}\ }\textbf {\bibinfo {volume} {123}},\ \bibinfo
  {pages} {154502} (\bibinfo {year} {2019})}\BibitemShut {NoStop}%
\bibitem [{\citenamefont {Patel}\ \emph {et~al.}(2020)\citenamefont {Patel},
  \citenamefont {Yan}, \citenamefont {Mukherjee}, \citenamefont {Fletcher},
  \citenamefont {Struck},\ and\ \citenamefont
  {Zwierlein}}]{patel_universal_2020}%
  \BibitemOpen
  \bibfield  {author} {\bibinfo {author} {\bibfnamefont {P.~B.}\ \bibnamefont
  {Patel}}, \bibinfo {author} {\bibfnamefont {Z.}~\bibnamefont {Yan}}, \bibinfo
  {author} {\bibfnamefont {B.}~\bibnamefont {Mukherjee}}, \bibinfo {author}
  {\bibfnamefont {R.~J.}\ \bibnamefont {Fletcher}}, \bibinfo {author}
  {\bibfnamefont {J.}~\bibnamefont {Struck}}, \ and\ \bibinfo {author}
  {\bibfnamefont {M.~W.}\ \bibnamefont {Zwierlein}},\ }\bibfield  {title}
  {\enquote {\bibinfo {title} {Universal sound diffusion in a strongly
  interacting {Fermi} gas},}\ }\href {\doibase 10.1126/science.aaz5756}
  {\bibfield  {journal} {\bibinfo  {journal} {Science}\ }\textbf {\bibinfo
  {volume} {370}},\ \bibinfo {pages} {1222--1226} (\bibinfo {year}
  {2020})}\BibitemShut {NoStop}%
\bibitem [{\citenamefont
  {Rogel-Salazar}(2013)}]{rogel-salazar_grosspitaevskii_2013}%
  \BibitemOpen
  \bibfield  {author} {\bibinfo {author} {\bibfnamefont {J.}~\bibnamefont
  {Rogel-Salazar}},\ }\bibfield  {title} {\enquote {\bibinfo {title} {The
  {Gross-Pitaevskii} equation and {Bose-Einstein} condensates},}\ }\href
  {\doibase 10.1088/0143-0807/34/2/247} {\bibfield  {journal} {\bibinfo
  {journal} {European Journal of Physics}\ }\textbf {\bibinfo {volume} {34}},\
  \bibinfo {pages} {247--257} (\bibinfo {year} {2013})}\BibitemShut {NoStop}%
\bibitem [{\citenamefont {Nicolis}\ and\ \citenamefont
  {Son}(2011)}]{Nicolis2011}%
  \BibitemOpen
  \bibfield  {author} {\bibinfo {author} {\bibfnamefont {A.}~\bibnamefont
  {Nicolis}}\ and\ \bibinfo {author} {\bibfnamefont {D.~T.}\ \bibnamefont
  {Son}},\ }\bibfield  {title} {\enquote {\bibinfo {title} {{Hall viscosity
  from effective field theory}},}\ }\href@noop {} {\  (\bibinfo {year}
  {2011})},\ \Eprint {http://arxiv.org/abs/1103.2137} {arXiv:1103.2137
  [hep-th]} \BibitemShut {NoStop}%
\bibitem [{\citenamefont {Abanov}\ and\ \citenamefont
  {Gromov}(2014)}]{Abanov2014}%
  \BibitemOpen
  \bibfield  {author} {\bibinfo {author} {\bibfnamefont {A.~G.}\ \bibnamefont
  {Abanov}}\ and\ \bibinfo {author} {\bibfnamefont {A.}~\bibnamefont
  {Gromov}},\ }\bibfield  {title} {\enquote {\bibinfo {title} {Electromagnetic
  and gravitational responses of two-dimensional noninteracting electrons in a
  background magnetic field},}\ }\href {\doibase 10.1103/PhysRevB.90.014435}
  {\bibfield  {journal} {\bibinfo  {journal} {Phys. Rev. B}\ }\textbf {\bibinfo
  {volume} {90}},\ \bibinfo {pages} {014435} (\bibinfo {year}
  {2014})}\BibitemShut {NoStop}%
\bibitem [{\citenamefont {Gromov}\ and\ \citenamefont
  {Abanov}(2014)}]{Gromov2014}%
  \BibitemOpen
  \bibfield  {author} {\bibinfo {author} {\bibfnamefont {A.}~\bibnamefont
  {Gromov}}\ and\ \bibinfo {author} {\bibfnamefont {A.~G.}\ \bibnamefont
  {Abanov}},\ }\bibfield  {title} {\enquote {\bibinfo {title}
  {Density-curvature response and gravitational anomaly},}\ }\href {\doibase
  10.1103/PhysRevLett.113.266802} {\bibfield  {journal} {\bibinfo  {journal}
  {Phys. Rev. Lett.}\ }\textbf {\bibinfo {volume} {113}},\ \bibinfo {pages}
  {266802} (\bibinfo {year} {2014})}\BibitemShut {NoStop}%
\bibitem [{\citenamefont {Cho}\ \emph {et~al.}(2014)\citenamefont {Cho},
  \citenamefont {You},\ and\ \citenamefont {Fradkin}}]{Cho2014}%
  \BibitemOpen
  \bibfield  {author} {\bibinfo {author} {\bibfnamefont {G.~Y.}\ \bibnamefont
  {Cho}}, \bibinfo {author} {\bibfnamefont {Y.}~\bibnamefont {You}}, \ and\
  \bibinfo {author} {\bibfnamefont {E.}~\bibnamefont {Fradkin}},\ }\bibfield
  {title} {\enquote {\bibinfo {title} {Geometry of fractional quantum {Hall}
  fluids},}\ }\href {\doibase 10.1103/PhysRevB.90.115139} {\bibfield  {journal}
  {\bibinfo  {journal} {Phys. Rev. B}\ }\textbf {\bibinfo {volume} {90}},\
  \bibinfo {pages} {115139} (\bibinfo {year} {2014})}\BibitemShut {NoStop}%
\bibitem [{\citenamefont {Green}\ \emph {et~al.}(2012)\citenamefont {Green},
  \citenamefont {Schwarz},\ and\ \citenamefont
  {Witten}}]{green_superstring_2012}%
  \BibitemOpen
  \bibfield  {author} {\bibinfo {author} {\bibfnamefont {M.}~\bibnamefont
  {Green}}, \bibinfo {author} {\bibfnamefont {J.~H.}\ \bibnamefont {Schwarz}},
  \ and\ \bibinfo {author} {\bibfnamefont {E.}~\bibnamefont {Witten}},\
  }\href@noop {} {\emph {\bibinfo {title} {Superstring {Theory}}}},\ \bibinfo
  {edition} {25th}\ ed.\ (\bibinfo  {publisher} {Cambridge University Press},\
  \bibinfo {address} {Cambridge ; New York},\ \bibinfo {year}
  {2012})\BibitemShut {NoStop}%
\bibitem [{\citenamefont {Greiter}\ \emph {et~al.}(1989)\citenamefont
  {Greiter}, \citenamefont {Wilczek},\ and\ \citenamefont
  {Witten}}]{greiter_hydrodynamic_1989}%
  \BibitemOpen
  \bibfield  {author} {\bibinfo {author} {\bibfnamefont {M.}~\bibnamefont
  {Greiter}}, \bibinfo {author} {\bibfnamefont {F.}~\bibnamefont {Wilczek}}, \
  and\ \bibinfo {author} {\bibfnamefont {E.}~\bibnamefont {Witten}},\
  }\bibfield  {title} {\enquote {\bibinfo {title} {{Hydrodynamic} {relations}
  {in} {superconductivity}},}\ }\href {\doibase 10.1142/S0217984989001400}
  {\bibfield  {journal} {\bibinfo  {journal} {Modern Physics Letters B}\
  }\textbf {\bibinfo {volume} {03}},\ \bibinfo {pages} {903--918} (\bibinfo
  {year} {1989})}\BibitemShut {NoStop}%
\bibitem [{\citenamefont {Scaffidi}\ \emph {et~al.}(2017)\citenamefont
  {Scaffidi}, \citenamefont {Nandi}, \citenamefont {Schmidt}, \citenamefont
  {Mackenzie},\ and\ \citenamefont {Moore}}]{scaffidi_hydrodynamic_2017}%
  \BibitemOpen
  \bibfield  {author} {\bibinfo {author} {\bibfnamefont {T.}~\bibnamefont
  {Scaffidi}}, \bibinfo {author} {\bibfnamefont {N.}~\bibnamefont {Nandi}},
  \bibinfo {author} {\bibfnamefont {B.}~\bibnamefont {Schmidt}}, \bibinfo
  {author} {\bibfnamefont {A.~P.}\ \bibnamefont {Mackenzie}}, \ and\ \bibinfo
  {author} {\bibfnamefont {J.~E.}\ \bibnamefont {Moore}},\ }\bibfield  {title}
  {\enquote {\bibinfo {title} {Hydrodynamic electron flow and hall
  viscosity},}\ }\href {\doibase 10.1103/PhysRevLett.118.226601} {\bibfield
  {journal} {\bibinfo  {journal} {Physical Review Letters}\ }\textbf {\bibinfo
  {volume} {118}},\ \bibinfo {pages} {226601} (\bibinfo {year}
  {2017})}\BibitemShut {NoStop}%
\end{thebibliography}%

\end{document}